\documentclass[a4paper]{article}
\pdfoutput=1
\usepackage{jheppub}

\usepackage[utf8]{inputenc}

\usepackage[table ]{ xcolor}
\usepackage{subcaption}
\usepackage{multirow}

\usepackage{amsmath}

\usepackage{tikz}
\usetikzlibrary{shapes,arrows}
\tikzstyle{startstop} = [rectangle,rounded corners, minimum width=3cm,minimum height=1cm,text centered, draw=black,fill=red!30]
\tikzstyle{io} = [trapezium, trapezium left angle = 70,trapezium right angle=110,minimum width=3cm,minimum height=1cm,text centered,draw=black,fill=blue!30]
\tikzstyle{process} = [rectangle,minimum width=3cm,minimum height=1cm,text centered,text width =3cm,draw=black,fill=orange!30]
\tikzstyle{decision} = [diamond,minimum width=3cm,minimum height=1cm,shape aspect=3,inner sep = 0.4pt,text centered,draw=black,fill=green!30]
\tikzstyle{arrow} = [thick,->,>=stealth]
\tikzstyle{shadow}=[preaction={fill=black,opacity=.5,transform canvas={xshift=0.5mm,yshift=-0.5mm},shading=radial,shading angle=20},fill=red]

\tikzstyle{ellipse}=[draw, rectangle, minimum width=2.8em, rounded corners=6pt,line width=0.5pt]
\tikzstyle{pxsbx}=[trapezium, trapezium left angle=75, trapezium right angle=105, minimum width=3em, text centered, draw = black, fill=white,line width=0.5pt] 
\tikzstyle{lingxing}=[draw,diamond,shape aspect=3,inner sep = 0.4pt,thick,font=\itshape,line width=0.5pt]

\usepackage{comment}

\usepackage{amssymb}
\usepackage{graphicx,color}

\usepackage{amsmath,amsthm}

\usepackage{mathrsfs}

\usepackage{bm}

\newcommand{\Z}{\mathbb{Z}}
\newcommand{\R}{\mathbb{R}}
\newcommand{\C}{\mathbb{C}}

\newcommand{\half}{\frac{1}{2}}

\newcommand{\ccirc}{\kern0.2ex\vcenter{\hbox{$\scriptstyle\circ$}}\kern0.2ex}

\newcommand{\Tr}{\text{Tr}}

\newcommand{\Slc}{\mathrm{SL}(2,\mathbb{C})}

\newcommand{\Su}{\mathrm{SU}(2)}
\newcommand{\su}{\fs\fu_2}

\def\be{\begin{eqnarray}}
\def\ee{\end{eqnarray}}

\newcommand{\ca}{\mathcal A}

\newcommand{\cb}{\mathcal B}
\newcommand{\cc}{\mathcal C}

\newcommand{\ce}{\mathcal E}
\newcommand{\cf}{\mathcal F}
\newcommand{\cg}{\mathcal G}
\newcommand{\ch}{\mathcal H}
\newcommand{\ci}{\mathcal I}

\newcommand{\ck}{\mathcal K}
\newcommand{\cl}{\mathcal L}

\newcommand{\cp}{\mathcal P}
\newcommand{\cq}{\mathcal Q}
\newcommand{\calr}{\mathcal R}
\newcommand{\cs}{\mathcal S}

\newcommand{\cw}{\mathcal W}
\newcommand{\cx}{\mathcal X}
\newcommand{\cy}{\mathcal Y}
\newcommand{\cz}{\mathcal Z}

  \newcommand{\Fa}{\mathfrak{A}}

\newcommand{\fp}{\mathfrak{p}}  \newcommand{\Fp}{\mathfrak{P}}

\newcommand{\fs}{\mathfrak{s}}  
  
\newcommand{\fu}{\mathfrak{u}}

\renewcommand{\a}{\alpha}
\renewcommand{\b}{\beta}
\newcommand{\g}{\gamma}

\newcommand{\eps}{\varepsilon}

\newcommand{\sig}{\sigma}
\newcommand{\Sig}{\Sigma}

\renewcommand{\L }{\Lambda}
\renewcommand{\o}{\omega}

\renewcommand{\t}{\tau}

\newcommand{\rmd}{\mathrm d}

\newcommand{\lt}{\left}
\newcommand{\rt}{\right}

\newcommand{\lag}{\left\langle}
\newcommand{\rag}{\right\rangle}

\newcommand{\tr}{\mathrm{tr}}

\newcommand{\sgn}{\mathrm{sgn}}

\title{Loop Quantum Gravity on Dynamical Lattice and Improved Cosmological Effective Dynamics with Inflaton}

\author[1,2]{Muxin Han}  

\affiliation[1]{Department of Physics, Florida Atlantic University, 777 Glades Road, Boca Raton, FL 33431-0991, USA}

\affiliation[2]{Institut f\"ur Quantengravitation, Universit\"at Erlangen-N\"urnberg, Staudtstr. 7/B2, 91058 Erlangen, Germany}

\author[2]{\ Hongguang Liu}

\emailAdd{hanm(At)fau.edu}
\emailAdd{hongguang.liu(At)gravity.fau.de}

\abstract{In the path integral formulation of the reduced phase space Loop Quantum Gravity (LQG), we propose a new approach to allow the spatial cubic lattice (graph) to change dynamically in the physical time evolution. The equations of motion of the path integral derive the effective dynamics of cosmology from the full LQG, when we focus on solutions with homogeneous and isotropic symmetry. The resulting cosmological effective dynamics with the dynamical lattice improves the effective dynamics obtained earlier from the path integral with fixed spatial lattice: The improved effective dynamics recovers the FLRW cosmology at low energy density and resolves the big-bang singularity with a bounce. The critical density $\rho_c$ at the bounce is Planckian $\rho_c\sim \Delta^{-1}$ where $\Delta$ is a Planckian area serving as certain UV cut-off of the effective theory. The effective dynamics gives the unsymmetric bounce and has the de-Sitter (dS) spacetime in the past of the bounce. The cosmological constant $\L_{eff}$ of the dS spacetime is emergent from the quantum effect $\L_{eff}\sim\Delta^{-1}$. These results are qualitatively similar to the properties of $\bar{\mu}$-scheme Loop Quantum Cosmology (LQC). Moreover, we generalize the earlier path integral formulation of the full LQG by taking into account the coupling with an additional real scalar field, which drives the slow-roll inflation of the effective cosmological dynamics. In addition, we discuss the cosmological perturbation theory on the dynamical lattice, and the relation to the Mukhanov-Sasaki equation.
}

\keywords{}

\begin{document}

\maketitle

\section{Introduction}

Loop Quantum Gravity (LQG) is a candidate for background-independent and non-perturbative theory of quantum gravity \cite{book,review,review1,rovelli2014covariant}. Among successful sub-areas in LQG, applying LQG to cosmology is a fruitful direction in which LQG gives physical predictions and phenomenological impacts. Most LQG-literatures on cosmology is based on Loop Quantum Cosmology (LQC): a LQG-like quantization of symmetry reduced model with homogeneity and isotropy (see e.g. \cite{Ashtekar:2006wn,Bojowald:2001xe,Agullo:2016tjh}). LQC leads to important prediction that the big-bang singularity is resolved with the non-singular bounce. However, the connection between LQC and the full theory of LQG has been a long-term open problem.  

In recent progresses \cite{Han:2019vpw,Han:2019feb,Han:2020iwk}, we develop the top-down derivations of the effective dynamics of homogeneous-and-isotropic cosmology and perturbations from the full theory of LQG. The key tool in our approach is the path integral formulation of the reduced phase space LQG on a fixed spatial cubic lattice (graph) $\g$ (see \cite{Han:2019vpw,Han:2020chr} for details). The semiclassical dynamics from the path integral formulation reproduces the effective dynamics of $\mu_0$-scheme LQC, which recovers the Friedmann-Lema\^itre-Robertson-Walker (FLRW) cosmology at low energy density. Although the $\mu_0$-scheme effective dynamics resolves the big-bang singularity with a bounce, it suffers the problem that the critical density at the bounce depend on the initial condition of the scale factor and is not always Planckian. The cosmic bounce at non-Planckian regime is not physically sound. In LQC, the $\mu_0$-scheme is replaced by the improved $\bar{\mu}$-scheme, which guarantees the critical density to be constant and Planckian.

\begin{figure}[h]
\begin{center}
  \includegraphics[width = 0.7\textwidth]{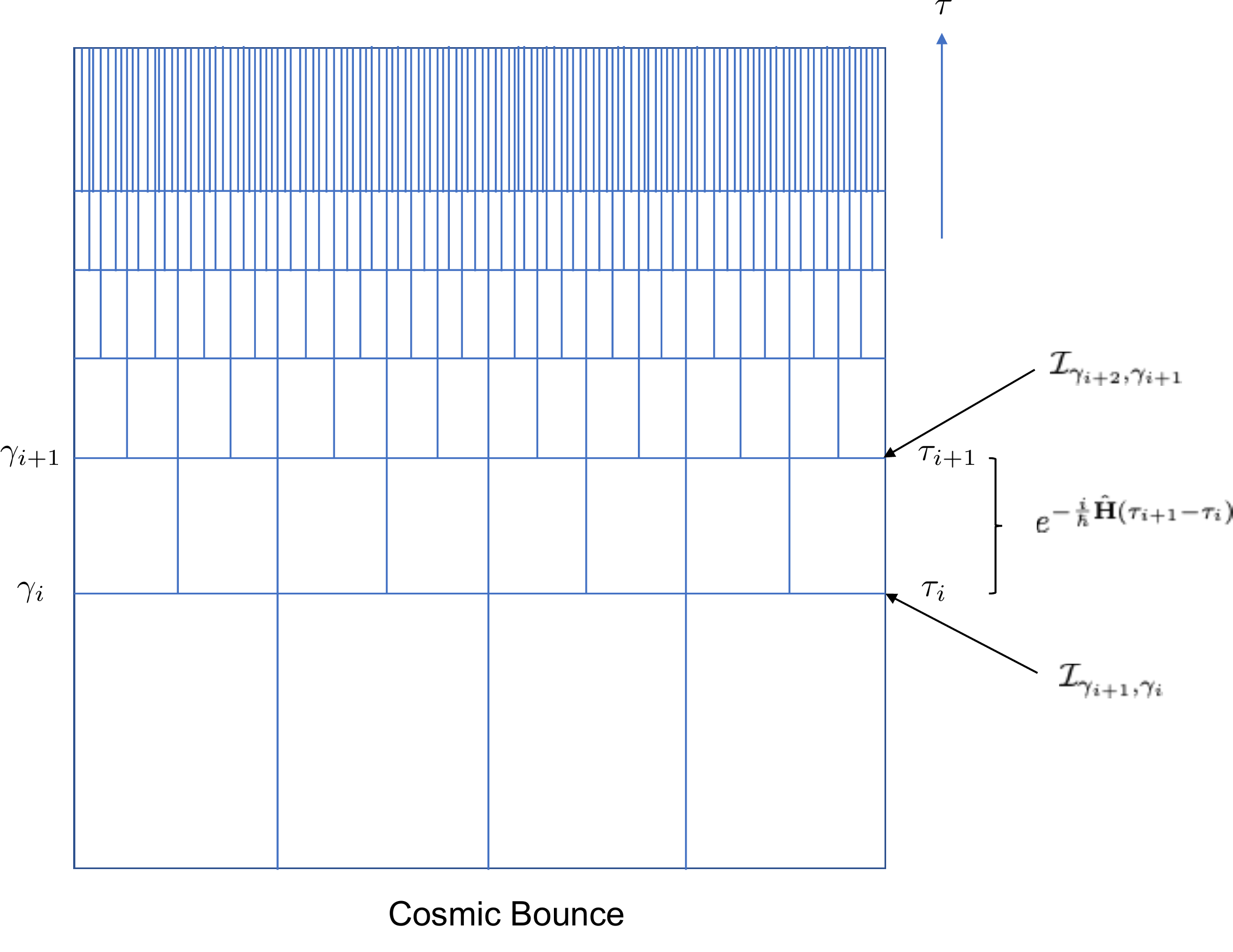} 
    \caption{(1+1)d illustration of the lattice refinement during the time evolution: Each horizontal line is a lattice $\g_i$ partitioning the spatial slice at the time $\t_i$ when the lattice refinement is carried out. Each vertical line is the time evolution of a vertex in the spatial lattice. The unitary evolution is defined in the area between 2 horizontal lines. }
  \label{lattice}
  \end{center}
\end{figure}

In this work, we propose a new strategy in the full theory of the reduced phase space LQG for overcoming the problem of the $\mu_0$-scheme effective dynamics. The key point in our strategy is to allow the spatial cubic lattice (graph) to change in the time evolution. Indeed we consider a large number of discrete time steps $\t_i$ with $i=1,\cdots,m$ in the evolution, such that the spatial lattices $\g_i$ at different time steps may not be the same. We still assume all $\g_i$ are cubic lattices. The LQG Hilbert space $\ch_{\g_i}$ are different if $\g_i$ are different. There are more degrees of freedom (DOFs) on a finer lattice than the DOFs on the coarser lattice. With the coherent states, we define $\ci_{\g_i,\g_{i-1}}:\ch_{\g_{i-1}}\to\ch_{\g_i}$ which is an embedding when $\g_{i-1}$ is coarser than $\g_{i}$, and is a projection otherwise. Inserting $\ci_{\g_i,\g_{i-1}}$ between unitary time evolutions generated by the physical Hamiltonian $\hat{\bf H}$ on $\g_i$ and $\g_{i-1}$, we construct the transition amplitude $\mathcal{A}_{[Z],\left[Z^{\prime}\right]}(\mathcal{K})$ between initial and final semiclassical states $\Psi_{[Z]}^{\hbar},\Psi_{[Z']}^{\hbar}$:  
\be
\mathcal{A}_{[Z],\left[Z^{\prime}\right]}(\mathcal{K})=\left\langle\Psi_{[Z]}^{\hbar}\left|e^{-\frac{i}{\hbar} \hat{\mathbf{H}}\left(T-\tau_{m}\right)} \mathcal{I}_{\gamma_{m}, \gamma_{m-1}} \cdots e^{-\frac{i}{\hbar} \hat{\mathbf{H}}\left(\tau_{i+1}-\tau_{i}\right)} \mathcal{I}_{\gamma_{i}, \gamma_{i-1}} e^{-\frac{i}{\hbar} \hat{\mathbf{H}}\left(\tau_{i}-\tau_{i-1}\right)} \cdots\right| \Psi_{\left[Z^{\prime}\right]}^{\hbar}\right\rangle
\ee
The initial and final states $\Psi_{[Z]}^{\hbar},\Psi_{[Z']}^{\hbar}$ are defined on different spatial lattices. The spatial lattice changes from $\g_{i-1}$ to $\g_i$ at each instance $\t_i$ ($i=1,\cdots,m$), while the unitary time evolution in every time interval $[\t_{i},\t_{i+1}]$ is on the fixed spatial lattice $\g_i$. Furthermore, we follow the coherent state path integral method in \cite{Han:2019vpw} to express $\mathcal{A}_{[Z],\left[Z^{\prime}\right]}(\mathcal{K})$ as a path integral formula. In contrast to our earlier path integral formulation which is defined on hypercubic lattice, $\mathcal{A}_{[Z],\left[Z^{\prime}\right]}(\mathcal{K})$ is defined on the spacetime lattice $\ck$ whose spatial lattices change in time, similar to Figure \ref{lattice}. $\mathcal{A}_{[Z],\left[Z^{\prime}\right]}(\mathcal{K})$ may be viewed as an analog of the spinfoam model.

Based on the lattice Fourier transform of the semiclassical data $Z_{\g_i}$ on $\g_i$, $\ci_{\g_i,\g_{i-1}}$ maps the coherent state $\Psi_{[Z_{\g_{i-1}}]}^{\hbar}\in \ch_{\g_{i-1}}$ to the coherent state $\Psi_{[Z_{\g_{i}}]}^{\hbar}\in\ch_{\g_i}$, such that the set of nonvanishing Fourier modes in $Z_{\g_{i}}$ are the same as $Z_{\g_{i-1}}$ (see Section \ref{Dynamical Lattice Refinement} for details).

By the path integral formula of $\mathcal{A}_{[Z],\left[Z^{\prime}\right]}(\mathcal{K})$, the dominant contribution to $\mathcal{A}_{[Z],\left[Z^{\prime}\right]}(\mathcal{K})$ comes from the trajectory satisfying the semiclassical equations of motion (EOMs) from the stationary phase approximation. We assume each time interval $[\t_{i-1},\t_i]$ are sufficiently small so that the EOMs can be approximated by differential equations with smooth time $\t$. We look for solutions corresponding to the homogeneous and isotropic cosmology. We find that fixing the initial condition, different solutions are determined by different choices of the spacetime lattices $\ck$. The effect of $\ck$ turns out to be an analog of an external force in the effective EOMs of cosmology.

Among choices of the spacetime lattices $\ck$, we propose 2 preferred choices, and call the resulting cosmological effective dynamics the $\mu_{min}$-scheme effective dynamics and the average effective dynamics respectively. Firstly the $\mu_{min}$-scheme effective dynamics is resulting from choosing the finest lattice $\ck_{min}$ such that an UV cut-off $\Delta$ where $\Delta\sim (\text{length})^2$ is saturated at all time steps (see Section \ref{UV Cut-off and effective dynamics}). The UV cut-off $\Delta$ validates the $\hbar$-expansion of the coherent state expectation value of the physical Hamiltonian\footnote{This $\hbar$-expansion is firstly proposed in \cite{Giesel:2006um} and is computed explicitly in \cite{Zhang:2020mld} to the 1st order in $\hbar$.}, so that the derivation of the effective dynamics from the path integral is valid throughout the evolution. $\Delta$ is a Planckian area when the Barbero-Immirzi parameter $\beta$ is relatively small. Secondly, the average effective dynamics is resulting from taking $\ck$ to be the random lattice (see Section \ref{Random Lattice and Average Effective Dynamics}). We perform the disorder average over all $\ck$ which are coarser than $\ck_{min}$. The $\mu_{min}$-scheme and average effective dynamics have the following remarkable features (see Section \ref{Properties of Minimum and Average Effective Dynamics} for details):

\begin{itemize}

\item Both effective dynamics reduce to the classical FLRW cosmology at low energy density.

\item Both effective dynamics resolve the problem of the $\mu_0$-scheme dynamics, they both resolve the big-bang singularity and lead to bounces where the critical density $\rho_c\sim \frac{1}{16\pi G\Delta}$ is Planckian when $\Delta$ is set to be an Planckian area scale. In particular, the critical density of the $\mu_{min}$-scheme dynamics coincides with the prediction from the LQC with unsymmetric bounce \cite{Assanioussi:2019iye} if $\Delta$ is identified with the minimal area gap used in LQC.

\item Both effective dynamics give unsymmetric bounces and have the asymptotic de-Sitter (dS) spacetime on the other side of the bounce, similar to \cite{Assanioussi:2019iye}. The asymptotic dS spacetime has the emergent cosmological constant $\L_{eff}\sim \Delta^{-1}$.

\end{itemize}

In Section \ref{Effective Hamiltonian and Poisson Bracket}, we extract the effective cosmological Hamiltonian and Poisson bracket for homogeneous and isotropic DOFs from the $\mu_{min}$-scheme effective dynamics. 

Another new aspect of this paper is taking into account the coupling to a real scalar field in the path integral formulation\footnote{see e.g. \cite{Kisielowski:2018oiv,Lewandowski:2015xqa,Domagala:2010bm,Han:2006iqa,Sahlmann:2002qj,Thiemann:1997rt} for some earlier works on coupling scalar field to LQG).}, in contrast to the earlier works \cite{Han:2019vpw,Han:2020iwk} where we only consider pure gravity coupling to clock fields. The scalar field with a suitable potential drives the slow-roll inflation in our effective cosmological dynamics\footnote{See \cite{Giesel:2020raf} for recent results on the inflaton in the reduced phase space LQC.}. The inflation provides us another motivation for letting the spatial lattice dynamical: Suppose we use a fixed cubic lattice $\gamma$, the geometrical lengths of the lattice edges are dynamical and describe the scale factor in cosmology. The inflation causes both the scale factor and the extrinsic curvature $K_0$ to grow exponentially. The large $K_0$ leads to the failure in the approximation of the $\mu_0$-scheme effective dynamics to the FLRW cosmology unless the lattice $\gamma$ is very fine. However, fixing a very fine $\gamma$ would cause the geometrical lengths of the lattice edges to be very small before the inflation, so that the $\hbar$-expansion of the coherent state expectation value of $\hat{\bf H}$ became invalid. To resolve this tension, we have to let the spatial lattice dynamical such that we have the fine lattice during the inflation and coarser lattice at early time.    

We generalize our discussion to include perturbations on the $\mu_{min}$-scheme or average effective cosmological background in Section \ref{Perturbations}. The perturbations are derived from the path integral formulation as the first principle. The effective dynamics of the cosmological perturbations are obtained by linearizing the EOMs of the full theory on the effective cosmological background. Our analysis mostly focuses on the scalar-mode perturbation on the $\mu_{min}$-scheme background. In particular, we obtain the consistency result that the scalar-mode perturbation recovers the standard Mukhanov-Sasaki equation at late time, e.g. at the pivot time and later.

Here are our conventions of constants frequently used in this paper: $\kappa=16\pi G$, $\ell_P^2=\hbar\kappa$, $l_P^2=\hbar G$, $m_P=\sqrt{\hbar/G}$.

This paper is organized as follows: Section \ref{Preliminaries} reviews some preliminaries on the reduced phase space LQG of gravity-scalar-dust, the coherent state of the coupled system, and the physical Hamiltonian operator. Section \ref{Coherent State Path Integral of Gravity-Scalar-Dust} extends the coherent state path integral formulation to the reduced phase space LQG coupled to the scalar field. Section \ref{Semiclassical Equations of Motion} derives the EOMs from the path integral on the fixed lattice and discusses the cosmological solution. Section \ref{Dynamical Lattice Refinement} generalizes the formalism to allow the spatial lattice to change in time and applies the formalism to the cosmological effective dynamics. Section \ref{UV Cut-off and effective dynamics} derives the $\mu_{min}$-scheme effective dynamics of cosmology. Section \ref{Random Lattice and Average Effective Dynamics} derives the average effective dynamics of cosmology. Section \ref{Properties of Minimum and Average Effective Dynamics} discusses the properties of the $\mu_{min}$-scheme and average effective dynamics, and compares them to the $\bar{\mu}$-scheme LQC. Section \ref{Effective Hamiltonian and Poisson Bracket} extracts the effective cosmological Hamiltonian and Poisson bracket from the $\mu_{min}$-scheme effective dynamics. Section \ref{Perturbations} derives the cosmological perturbation theory on the effective background from the path integral formulation, and compares the late-time behavior to the Mukhanov-Sasaki equation.

\section{Preliminaries} \label{Preliminaries}

\subsection{Reduced Phase Space Formulation}\label{Reduced Phase Space Formulation}

The reduced phase space formulation couples gravity to clock fields at classical level. In this paper, we mainly focus on the scenario of gravity coupled to Gaussian dust \cite{Kuchar:1990vy,Giesel:2012rb} and a real scalar field. The Gaussian dust serves as the clock fields. The total action is given by 
\be
    S  = S_{\rm GR} + S_{\rm GD}+ S_{\rm Scalar},
\ee
where $S_{\rm GR}$ is the Host action of gravity \cite{holst}
\be 
S_{\rm GR}\lt[e^\mu_{I},\Omega_{\mu \nu}^{IJ}\rt] =\frac{1}{16\pi G} \int_{M} d^{4}x\, \sqrt{|\operatorname{det}(g)|}\lt[ e_{I}^{\mu} e_{J}^{\nu}\left(\Omega_{\mu \nu}^{IJ}+\frac{1}{2 \beta} \epsilon_{\ \ \ KL}^{IJ} \Omega_{\mu \nu}^{KL}\right)+2\Lambda\rt]
\ee 
where the tetrad $e^\mu_{I}$ determines the 4-metric by $g_{\mu\nu}=\eta_{IJ}e^\mu_{I}e^\nu_{J}$, and $\Omega_{\mu \nu}^{IJ}$ is the curvature of the so(1,3) connection $\o_\mu^{IJ}$. $\b$ is the Barbero-Immirzi parameter. The scalar field action reads
\be
S_{\rm Scalar}[\phi,e]=\frac{1}{2}\int_{M}d^{4}x\sqrt{|\det(g)|}\left[g^{\mu\nu}(\partial_{\mu}\phi)\partial_{\nu}\phi+U(\phi)\right].
\ee
where the scalar potential $U(\phi)$ is specified later. $S_{\rm GD}$ is the action of the Gaussian dust:
\be
S_{\rm GD}\left[\rho_{dust}, g_{\mu \nu}, T, S^{j}, W_{j}\right]=-\int_M \mathrm{d}^{4}x \sqrt{|\operatorname{det}(g)|}\left[\frac{\rho_{dust}}{2}\left(g^{\mu \nu} \partial_{\mu} T \partial_{\nu} T+1\right)+g^{\mu \nu} \partial_{\mu} T\left(W_{j} \partial_{\nu} S^{j}\right)\right]
\ee
where $T,S^{j=1,2,3}$ are clock fields and define time and space coordinates in the dust reference frame. $\rho_{dust},W_{j}$ are Lagrange multipliers. The energy-momentum tensor of the Gaussian dust is 
\be
T_{\mu\nu}=\rho_{dust} U_\mu U_\nu-U_{(\mu}  W_{\nu)},\quad U_\mu=-\partial_{\mu} T,\quad W_{\nu}=W_{j}\partial_{\nu} S^{j}.
\ee
which indicates that $\rho_{dust}$ is the energy density and $W_\mu$ relates to the heat-flow \cite{Kuchar:1990vy}.

We assume $M\simeq \R\times \Sig$ and and make Legendre transform of dust variables:
\be
\begin{aligned}
P &:=\frac{\delta{S}_{\mathrm{GD}}}{\delta \dot{T}}=\sqrt{\operatorname{det}(q)}\left\{\rho_{dust}\left[\cl_{n} T\right]+W_{j}\left[\cl_{n} S^{j}\right]\right\} \\
P_{j} &:=\frac{\delta{S}_{\mathrm{GD}}}{\delta \dot{S}^{j}}=\sqrt{\operatorname{det}(q)} W_{j}\left[\cl_{n} T\right] \\
\pi &:=\frac{\delta S_{\mathrm{GD}}}{\delta \dot{\rho}_{dust}}=0 \\
\pi^{j} &:=\frac{\delta S_{\mathrm{GD}}}{\delta \dot{W}_{j}}=0
\end{aligned}
\ee
where $q_{\a\b}$ ($\a,\b=1,2,3$) is the 3-metric and $\cl_n$ denotes the Lie derivative along the normal to the hypersurface $\Sig$. The constraint analysis \cite{Kuchar:1990vy,Giesel:2012rb} results in Hamiltonian and diffeomorphism constraints $\cc^{\rm tot},\cc^{\rm tot}_\a$ which are first-class constraints, and 8 second-class constraints $z,z^j,\zeta_1,\zeta_2,s,K$:
\be
z&=&\pi,\quad z^{j}=\pi^{j},\quad\zeta_{1}=W_{1}P_{2}-{W_{2}} P_{1},\quad \zeta_{2}:=W_{1}P_{3}-W_{3} P_{1},\\
s&=&-\frac{1}{\sqrt{\det(q)}}P_{1}{}^{2}+\sqrt{\det(q)}\left(q^{\a\b} T_{, \a} T_{, \b}+1\right)W_{1}{}^2,\\
K&=&-\frac{P P_{1}{}^2W_{1}}{ \sqrt{\det(q)}}+\frac{\rho_{dust}}{\sqrt{\det(q)}} P_{1}{}^{3}+\sqrt{\det(q)} W_1{}^3 q^{\a \b} T_{, \a}\left(P_{j} S_{, \b}^{j}\right)
\ee
where $T_{,\a}\equiv\partial_\a T$. Solving second-class constraints gives 
\be
W_j&=&\frac{P_j}{\sqrt{\det(q)}\left(q^{\alpha\beta}T_{,\alpha}T_{,\beta}+1\right)^{1/2}}\\
\rho_{dust}&=&\frac{P}{\sqrt{\det(q)}\left(q^{\alpha\beta}T_{,\alpha}T_{,\beta}+1\right)^{1/2}}-\frac{q^{\alpha\beta}T_{,\alpha}\left(P_{j}S_{,\beta}^{j}\right)}{\sqrt{\det(q)}\left(q^{\alpha\beta}T_{,\alpha}T_{,\beta}+1\right)^{3/2}}\label{rho111}
\ee
by a choice of sign in the ratio between $W_j,P_j$. These relations simplifies $\cc^{\rm tot},\cc^{\rm tot}_\a$ to equivalent forms:
\be
\cc^{\mathrm{tot}}&=&P+ h,\quad h= \cc \sqrt{1+q^{\a \b} T_{, \a} T_{, \b}}-q^{\a \b} T_{, \a} \cc_{\b},\label{cc111}\\
\cc_{\a}^{\mathrm{tot}}&=&\cc_{\a}+P T_{, \a}+P_{j} S_{, \a}^{j}
\ee
where $\cc,\cc_a$ are the gravity-scalar Hamiltonian and diffeomorphism constraints from $S_{\rm GR}+S_{\rm Scalar}$. Note that it is possible to have a negative $\rho_{dust}$. However we always guarantee that the total energy density $\rho_{dust}+\rho_s$ ($\rho_s$ is the energy density of the scalar field) must be nonnegative, in order that the energy condition is satisfied.

We construct the Dirac observables based on the fields in $S_{\rm GR}$ and $S_{\rm Scalar}$ with the help of the clock fields:

\begin{description}

\item[Gravity:] We use $A^a_\a(x),E^\a_a(x)$ to be canonical variables of gravity, where $A^a_\a(x)$ is the real Ashtekar-Barbero connection with gauge group SU(2) and $E^\a_a(x)=\sqrt{\det q}\, e^\a_a(x)$ is the densitized triad. $a=1,2,3$ is the Lie algebra index of $\su$\footnote{$[\frac{\t^a}{2},\frac{\t^b}{2}]=-\frac{1}{4}[\sig^a,\sig^b]=-i\eps^{abc}\sig^c/2=\eps^{abc}\frac{\t^c}{2}$ and $\Tr(\frac{\t^a}{2}\,\frac{\t^b}{2})=-\frac{1}{2}\delta^{ab}$}. We choose basis $\t^a=-i\sig^a$ ($\vec{\sig}$ are Pauli matrices) in $\su$. Dirac observables are constructed relationally by parametrizing $(A,E)$ with values of dust fields $T(x)\equiv\t,\ S^j(x)\equiv\sig^j$, i.e. $A_j^a(\sig,\t)=A_j^a(x)|_{T(x)\equiv\t,\,S^j(x)\equiv\sig^j}$ and $E^j_a(\sig,\t)=E^j_a(x)|_{T(x)\equiv\t,\,S^j(x)\equiv\sig^j}$, where $\sig,\t$ are physical space and time coordinates of the dust reference frame. Here $j=1,2,3$ is the dust coordinate index (e.g. $A_j=A_\a S^\a_j$).

\item[Real scalar:] Canonical conjugate variables of real scalar field are $\phi(x),\pi(x)$. Corresponding Dirac observables are  $\phi(\sig,\t)=\phi(x)|_{T(x)\equiv\t,\,S^j(x)\equiv\sig^j}$ and $\pi(\sig,\t)=\pi(x)|_{T(x)\equiv\t,\,S^j(x)\equiv\sig^j}$.

\end{description}

All above fields are Dirac observables weakly Poisson commutative with diffeomorphism and Hamiltonian constraints. They satisfy the standard Poisson bracket in the dust frame 
\be
\{E^i_a(\sig,\t),A_j^b(\sig',\t)\}&=&\frac{1}{2}\kappa \b\ \delta^{i}_j\delta^b_a\delta^{3}(\sig,\sig'),\\
\{\pi(\sig,\t), \phi(\sig,\t)\}&=&\delta^{3}(\sig,\sig')
\ee 
where $\b$ is the Barbero-Immirzi parameter, $\kappa=16\pi G_{\text{Newton}}$. Above conjugate pairs and Poisson brackets define the reduced phase space $\mathscr{P}$.

The evolution in physical time $\t$ is generated by the physical Hamiltonian ${\bf H}_0$ given by integrating $h$ on the constant $T(x)=\t$ slice $\cs$. The constant $T$ slice $\cs$ is coordinated by the value of dust scalars $S^j=\sig^j$ thus is referred to as the dust space \cite{Giesel:2007wn,Giesel:2012rb}. $T_{, \a}=0$ on $\cs$ leads to 
\be
{\bf H}_0=\int_\cs\rmd^3\sig\, \cc(\sig)
\ee 
${\bf H}_0$ formally coincides with smearing the gravity-scalar Hamiltonian $\cc$ with the unit lapse, while here $\cc(\sig)$ is in terms of Dirac observables. 
\be
&&\cc=\cc^{GR}+\cc^{S},\label{ccclass}\\
\text{Gravity:}\quad
&& \cc^{GR}=\frac{1}{\kappa}\lt[F^a_{jk}-\lt({\b^2+1}\rt)\eps_{ade}K^d_{j}K^e_{k}\rt]\eps^{abc}\frac{E^j_bE^k_c}{\sqrt{\det(q)}}+\frac{2\L}{\kappa}\sqrt{\det(q)}\\
\text{Scalar:}\quad
&& \cc^S=\frac{\pi^2}{2\sqrt{\det\left(q\right)}}+\frac{1}{2}\sqrt{\det\left(q\right)}\,q^{jk}(\partial_{j}\phi)(\partial_{k}\phi)+\sqrt{\det\left(q\right)}\,U(\phi)
\ee 
The $\tau$-evolution is governed by the Hamilton's equation
\be
\frac{\rmd f}{\rmd \t}=\{{\bf H}_0, \, f\}.
\ee 
for all functions $f$ on $\mathscr{P}$. This evolution is formally the same as the evolution of gravity-scalar with unit lapse function and zero shift vector. 


In the gravity-scalar-dust model, we resolve the Hamiltonian and diffeomorphism constraints classically, while the SU(2) Gauss constraints
\be
\cg_a&=&\frac{1}{\beta\kappa}\mathcal{D}_{j}E_{a}^{j}=0, \label{GaussGR}
\ee
still has to be imposed to the phase space. 
The time evolution preserves Gauss constraint since $\lt\{\cg_a(\sig,\t),\,{\bf H}_0\rt\}=0$ by the gauge invariance of ${\bf H}_0$. Secondly,
$\cc_j(\sig,\t)$ is conserved on the Gauss constraint surface \cite{Giesel:2007wn}:
\be
\frac{\rmd \cc_j(\sig,\t)}{\rmd \t}=\lt\{{\bf H}_0, \, \cc_j(\sig,\t) \rt\}=0.
\ee
where
\be
\cc_j&=&\cc_j^{GR}+\cc_j^{S},\quad 
\cc^{GR}_j=\frac{2}{\kappa\b}F^b_{jk} {E^k_b},\quad \cc^S_j=\pi \partial_{j}\phi.\label{ccjclass}
\ee
replaces the gravity-scalar diffeomorphism constraint by the corresponding Dirac observable.

\subsection{Quantization}\label{review_quantum}

\begin{figure}[t]
  \includegraphics[width = 0.9\textwidth]{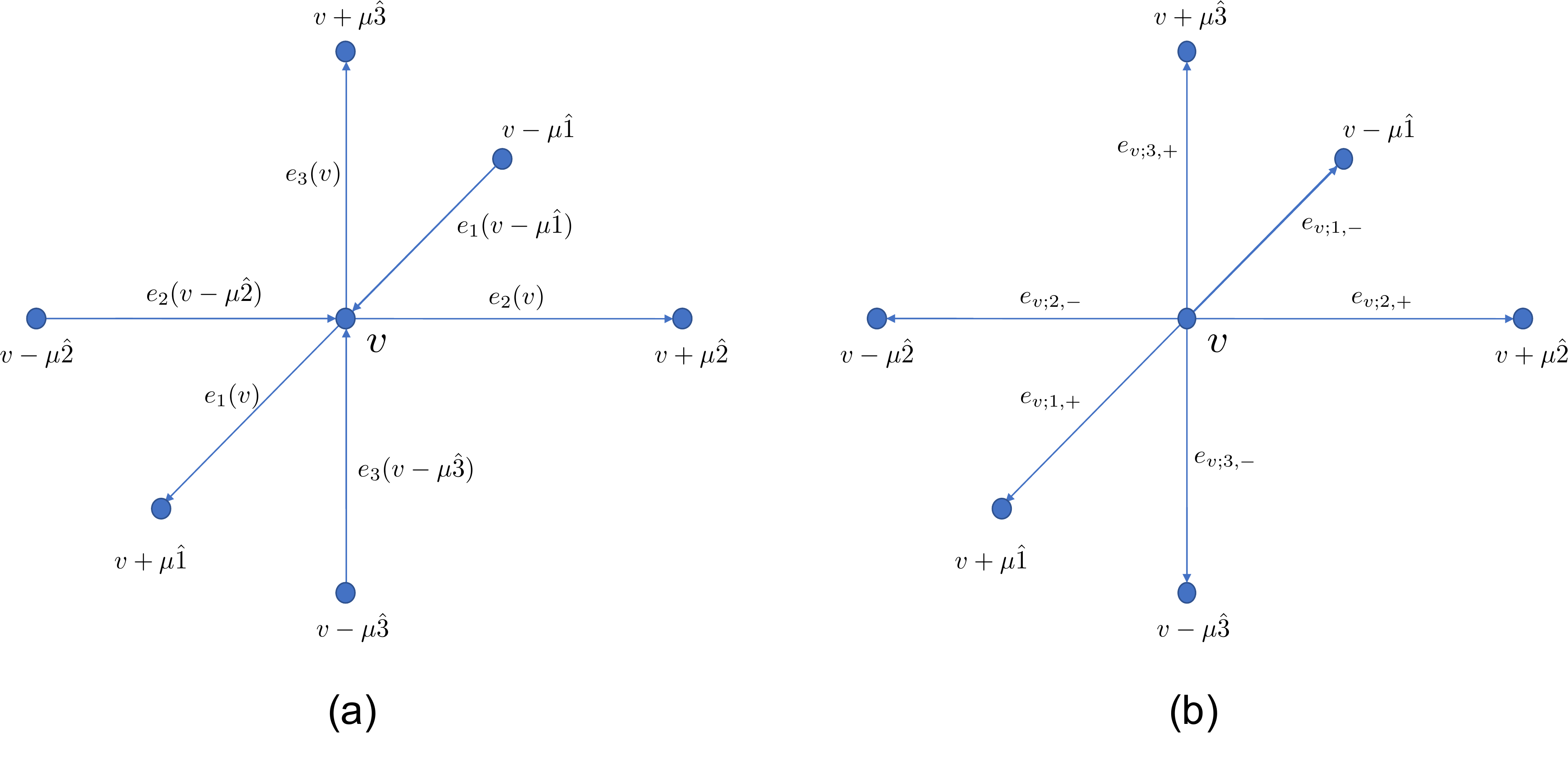} 
    \caption{(a) Notations of edges and vertices when all 6 edges are oriented toward positive directions of coordinates. (b) Notations of edge and vertices when all 6 edge are oriented outgoing from $v$.}
  \label{frame}
\end{figure}

We fix $\g$ to be a cubic lattice which partitions the dust space $\cs$. In this work, we assume $\cs\simeq T^3$, and $\g$ is a finite lattice. We denote by $E(\g)$ and $V(\g)$ sets of (oriented) edges and vertices in $\g$. By the dust coordinate $\sig^j$ on $\cs$, we assign every edge a constant coordinate length $\mu$ in the dust frame. $\mu\to 0,\ |V(\g)|\to\infty$ keeping $\mu^3|V(\g)|$ fixed is the lattice continuum limit. Every vertex $v\in V(\g)$ is 6-valent. At $v$ there are 3 outgoing edges $e_i(v)$ ($i=1,2,3$) and 3 incoming edges $e_i(v-\mu\, \hat{i})$ where $\mu\,\hat{i}$ is the lattice vector along the $i$-th direction. It is often convenient to orient all 6 edges at $v$ to be outgoing from $v$, and denote 6 edges by $e_{v;i,s}$ ($s=\pm$):
\be
e_{v;i,+}=e_i(v),\quad e_{v;i,-}=e_I(v-\mu\, \hat{i})^{-1}.
\ee
These notations are illustrated in FIG.\ref{frame}.

Canonical Dirac observables of gravity and matters can be discretized on the lattice $\g$ and quantized as follows: 

\subsubsection*{Gravity:}

Discretizations of gravity canonical pair $A^a_j(\sig,\t),E^j_a(\sig,\t)$ gives holonomy $h(e)$ and gauge covariant flux $p^a(e)$ at every $e\in E(\g)$ \cite{Thiemann:2000bv}:
\be
h(e)&:=&\cp \exp \int_{e}\rmd\sig^j A_j^a\t^a/2,\nonumber\\
p^a(e)&:=&-\frac{1}{2\b a^2}\tr\lt[\t^a\int_{S_e}\eps_{ijk}\rmd \sig^i\wedge\rmd \sig^j\ h\lt(\rho_e(\sig)\rt)\, E_b^k(\sig)\t^b\, h\lt(\rho_e(\sig)\rt)^{-1}\rt],\label{hpvari}
\ee 
where recall $\t^a=-i(\text{Pauli matrix})^a$. $S_e$ is a 2-face intersecting $e$ in the dual lattice $\g^*$. $\rho_e$ is a path starting at the source of $e$ and traveling along $e$ until $e\cap S_e$, then running in $S_e$ until $\vec{\sig}$. $a$ is a length unit for making $p^a(e)$ dimensionless. Note that because $p^a(e)$ is gauge covariant flux, we have
\be
p^{a}\left(e_{v ; I,-}\right)=\frac{1}{2} \operatorname{Tr}\left[\tau^{a} h\left(e_{v-\hat{I} ; I,+}\right)^{-1} p^{b}\left(e_{v-\hat{I} ; I,+}\right) \tau^{b} h\left(e_{v-\hat{I} ; I,+}\right)\right].
\ee
The Poisson algebra of $h(e)$ and $p^a(e)$ are called the holonomy-flux algebra:
\be
\left\{h(e), h\left(e^{\prime}\right)\right\} &=&0 ,\label{handh}\\
\left\{p^{a}(e), h\left(e^{\prime}\right)\right\} &=&\frac{\kappa}{a^{2}} \delta_{e, e^{\prime}} \frac{\tau^{a}}{2} h\left(e^{\prime}\right) ,\label{pandtheta}\\
\left\{p^{a}(e), p^{b}\left(e^{\prime}\right)\right\} &=&-\frac{\kappa}{a^{2}} \delta_{e, e^{\prime}} \varepsilon_{a b c} p^{c}\left(e^{\prime}\right).\label{pandp}
\ee
The LQG quantization defines the Hilbert space of square-integrable (complex valued) functions of all $h(e)$'s on $\g$, ${}^0\ch_\g^{GR}= L^2(\Su,\rmd\mu_H)^{\otimes|E(\g)|}$, where $d\mu_H$ is the Haar measure. $\hat{h}(e)$ becomes multiplication operators on functions in ${}^0\ch_\g^{GR}$. $\hat{p}^a(e)=i t\,\hat{R}_e^a/2$ where $\hat{R}_e^a$ is the right invariant vector field on SU(2) associated to the edge $e$: $\hat{R}^a f(h)=\frac{\rmd}{\rmd \eps}\big|_{\eps=0} f(e^{\eps\t^a}h)$. $t=\ell^2_p/a^2$ is a dimensionless semiclassicality parameter ($\ell^2_p=\hbar\kappa$). $\hat{h}(e),\hat{p}^a(e)$ satisfy the commutation relations:
\be
\lt[\hat{h}(e),\hat{h}(e')\rt] &=&0\nonumber\\
\lt[\hat{p}^a(e),\hat{h}(e')\rt] &=&i t \delta_{e,e'} \frac{\t^a}{2} {h}(e')\nonumber\\
\lt[\hat{p}^a(e),\hat{p}^b(e')\rt]&=&-it \delta_{e,e'} \eps_{abc} {p}^c(e'), \label{ph}
\ee
as quantization of the holonomy-flux algebra. Imposing the Gaussian constraint at the quantum level reduces ${}^0\ch_\g^{GR}$ to the Hilbert space $\ch_\g^{GR}$ of SU(2) gauge invariant functions of $h(e)$.

\subsubsection*{Real Scalar:}

Lattice scalars $\phi(v)$ and $\pi(v)=\int\rmd^3x\chi_\mu(x,v)\pi(x)$ are located at vertices and satisfy the Poisson bracket
\be
\lt\{\pi(v),{\phi}(v')\rt\}=\delta_{v,v'}.
\ee
The quantization defines $\ch^S_\g=\otimes_{v\in V(\g)}\ch_v$ where $\ch_v\simeq L^2(\R,\rmd\phi (v))$ is spanned by squared-integrable functions of $\phi(v)$ with the Lebesgue measure $\rmd\phi (v)$. Quantization of scalar fields give $\hat{\phi}(v),\hat{\pi}(v)$ whose actions on $\ch_\g^S$ are 
\be
&&\hat{\phi} (v)f(\phi)={\phi} (v)f(\phi),\quad
\hat{\pi} (v)f(\phi)=i\hbar\lt[\partial/\partial{\phi} (v)\rt]f(\phi)
\ee
for all functions $f(\phi)\in\ch^S_\g$. Both $\hat{\phi}(v)$ and $\hat{\pi}(v)$ are self-adjoint operators satisfying
\be
\lt[\hat{\pi}(v),\hat{\phi}(v')\rt]= i\hbar\delta_{v,v'}.
\ee

\vspace{1cm}

The reduced phase space on the lattice, denoted by $\mathscr{P}_\g$, has coordinates $h(e),p^a(e),\phi(v),\pi(v)$. As a result from the quantization of $\mathscr{P}_\g$, the LQG Hilbert space of gravity coupled to the scalar is given by the tensor product:
\be
\ch_\g=\ch^{GR}_\g\otimes\ch^S_\g.
\ee
States in $\ch_{\g}$ are SU(2) gauge invariant since the scalar is SU(2) invariant. $\ch_\g$ is the physical Hilbert space on $\g$ free of constraints because it comes from quantizing Dirac observables.

\subsection{Coherent states}

\subsubsection*{Gravity:}

The coherent state for gravity, $\psi_g^t\in \ch^{GR}_\g$, is defined by
\be
\psi_g^t=\prod_{e\in E(\g)}\psi^{t}_{g(e)},\quad \psi^{t}_{g(e)}\lt(h(e)\rt)=\sum_{j_e\in\Z_+/2\cup\{0\}}(2j_e+1)\ e^{-tj_e(j_e+1)/2}\chi_{j_e}\lt(g(e)h(e)^{-1}\rt),\label{coherent}
\ee
Here $\chi_j$ is the SU(2) character of the spin-$j$ irrep. $g(e)\in\Slc$ is the complex parametrization of the gravity sector in the reduced phase space and relates to the classical flux $p^a(e)$ and holonomy $e^{\theta^a(e)\t^a/2}$ by
\be
g(e)=e^{-ip^a(e)\t^a/2}e^{\theta^a(e)\t^a/2}, \quad p^a(e),\ \theta^a(e)\in\R^3.\label{gthetap0}
\ee
The coherent state $\psi_g^t$ is labelled by the dimensionless semiclassicality parameter 
\be
t=\frac{\ell_P^2}{a^2}
\ee
which is the value of $\ell_P^2$ measured by the length unit $a^2$. The semiclassical limit $\hbar\to0$ implies $t\to0$ or $\ell_P^2\ll a^2$.

The above coherent state is not normalized, the normalized coherent state (on a single edge) is denoted by
\be
\tilde{\psi}^t_{g(e)}=\frac{\psi^{t}_{g(e)}}{||\psi^{t}_{g(e)}||}.
\ee
It is useful to review the overlap of two normalized coherent states $\tilde{\psi}^t_{g_2(e)},\tilde{\psi}^t_{g_1(e)}$ \cite{book,Thiemann:2000ca}:
\be
\langle \tilde{\psi}^t_{g_{2}(e)}|\tilde{\psi}^t_{g_{1}(e)}\rangle&=& e^{\frac{K\lt(g_2(e),g_1(e)\rt)}{t}}\lt\{\frac{\xi_{21}(e)}{\sinh(\xi_{21}(e))}\sqrt{\frac{\sinh\lt(p_1(e)\rt)\sinh\lt(p_2(e)\rt)}{p_1(e)p_2(e)}}+O(t^\infty)\rt\},\label{overlap}\\
 K\lt(g_2(e),g_1(e)\rt)&=&\xi_{21}(e)^2-\half p_2(e)^2-\half p_1(e)^2, \quad \xi_{21}(e)=\mathrm{arccosh}\lt( \half\tr\lt[g_2(e)^\dagger g_1(e)\rt]\rt),\label{K12}
\ee
where $O(t^\infty)$ stands for contributions that are suppressed exponentially as $t\to0$. $p_{1,2}(e)=\sqrt{p_{1,2}^a(e)p_{1,2}^a(e)}= \text{arccosh}(\frac{1}{2}\Tr[g_{1,2}(e)^{\dagger} g_{1,2}(e)])$. $\langle \tilde{\psi}^t_{g_{2}(e)}|\tilde{\psi}^t_{g_{1}(e)}\rangle$ is invariant under $\xi_{21}\mapsto -\xi_{21}$ which relates to the Weyl refection of SU(2). We fix the sign ambiguity of $\xi_{21}$ by using the inverse hyperbolic cosine function, so $\mathrm{Re}(\xi_{21})\geq 0$. Our convention for the inverse hyperbolic cosine is $\mathrm{arccosh}(x)=\ln(x+\sqrt{x+1}\sqrt{x-1})$. $|\langle \tilde{\psi}^t_{g_{2}(e)}|\tilde{\psi}^t_{g_{1}(e)}\rangle|$ behaves as a Gaussian sharply peaked at $g_1(e)=g_2(e)$ with the width given by $| p^a_1(e)-p^a_2(e)|\sim| \theta^a_1(e)-\theta^a_2(e)|\sim\sqrt{t}$ \cite{Thiemann:2000ca}.

It is important that at every edge $e$, the normalized coherent states form an over-complete basis in $L^2(\Su)$ \cite{Thiemann:2000ca}:
\be
\int_{\Slc}\rmd g(e)\ |\tilde{\psi}^{t}_{g(e)}\rangle\langle\tilde{\psi}^{t}_{g(e)}|=1_{L^2(\Su)},\quad \rmd g(e)=\frac{c}{t^3}\rmd\mu_H(h(e))\,\rmd^3p(e),\quad c=\frac{2}{\pi}+O(t^\infty)\label{overcomplete}
\ee
where $\rmd\mu_H(h)$ is the Haar measure on SU(2).

\subsubsection*{Scalar:}

Coherent states in $\ch^S_\g$ are similar to coherent states of simple harmonic oscillator. We define annihilation and creation operators
\be
\hat{\Fa} (v)=\frac{a}{\sqrt{2\hbar}}\lt[\hat{\phi} (v)-\frac{i}{a^2}\hat{\pi} (v)\rt],\quad \hat{\Fa} (v)^\dagger=\frac{a}{\sqrt{2\hbar}}\lt[\hat{\phi} (v)+\frac{i}{a^2}\hat{\pi} (v)\rt]
\ee
where $a^2$ appears to balance the dimensions between $\phi (v)\sim (\text{length})^{-1}$ and $\pi (v)\sim (\text{length})^{1}$. We have following commutation relations of $\hat{ \Fa} (v)$, $\hat{ \Fa} (v)^\dagger$
\be
\lt[\hat{ \Fa} (v),\hat{ \Fa}(v')^\dagger\rt]=\delta_{v,v'},
\ee
which give harmonic oscillators at all $v$. Annihilation operators $\hat{ \Fa} (v)$ define a ``ground stat'' $|0\rangle$ in $\ch_\g^S$ by $\hat{ \Fa} (v)|0\rangle=0$. Coherent states are defined by
\be
\hat{ \Fa} (v)|\psi_z^\hbar\rangle=\frac{z (v)}{\sqrt{\hbar}}|\psi_z^\hbar\rangle,\quad |\psi_z^\hbar\rangle=\prod_{v,r}e^{\frac{1}{\sqrt{\hbar}}z (v)\hat{ \Fa} (v)^\dagger}|0\rangle,
\ee
where $z (v),\bar{z} (v)$ gives the complex parametrization of the scalar sector in the reduced phase space:
\be
z (v)=\frac{a}{\sqrt{2}}\lt[{\phi} (v)-\frac{i}{a^2}{\pi} (v)\rt],\quad \bar{z} (v)=\frac{a}{\sqrt{2}}\lt[{\phi} (v)+\frac{i}{a^2}{\pi} (v)\rt]\label{zzbarphipi}
\ee 
$|\psi_z^\hbar\rangle$ can be expressed as a function of ${\phi} (v)$:
\be
\psi_z^\hbar(\phi)=\prod_{v,r}\lt(\frac{a^2}{\hbar}\rt)^\half e^{\frac{1}{2\hbar}z (v)^2-\frac{1}{2\hbar}\lt[a\phi (v)-\sqrt{2}z (v)\rt]^2}
\ee
The inner product between two coherent states $|\psi_z^\hbar\rangle,|\psi_{z'}^\hbar\rangle$ is given by
\be
\langle \psi_{z'}^\hbar|\psi_z^\hbar\rangle&=&e^{\frac{1}{\hbar}\sum_{v}\bar{z}'(v) z(v)}
\label{overlapscalar}
\ee
The normalization $|\tilde{\psi}_z^\hbar\rangle=|{\psi}_z^\hbar\rangle/\|{\psi}_z^\hbar\|$ satisfies the over-completeness relation
\be
\int \prod_{v,r}\frac{\rmd^2 z (v)}{\pi\hbar}|\tilde{\psi}_z^\hbar\rangle\langle \tilde{\psi}_z^\hbar|=1_{\ch_\g^S},\quad \rmd^2 z (v)=\rmd \mathrm{Re}[z (v)]\rmd \mathrm{Im}[z (v)]
\ee
For any normal ordered polynomial operator $O(\hat{ \Fa}(v)^\dagger,\hat{ \Fa}(v))$, 
\be
\lag \psi_{z'}^\hbar \lt|O\lt(\hat{ \Fa} (v)^\dagger,\hat{ \Fa} (v)\rt)\rt|\psi_z^\hbar\rag=O\lt(\bar{z} (v),{z} (v)\rt)\langle \psi_{z'}^\hbar |\psi_z^\hbar\rangle.
\ee

\vspace{1cm}

Coherent states in the Hilbert space $\ch_\g^0={}^0\ch^{GR}_\g\otimes\ch_\g^S$ are given by the tensor product:
\be
\psi^\hbar_{Z}=\psi^t_g\otimes\psi^\hbar_z,\quad Z\equiv(g,z)
\ee
$Z$ is the complex parametrization of the reduced phase space on the lattice $\g$. The normalization $\tilde{\psi}^\hbar_{Z}={\psi}^\hbar_{Z}/\|{\psi}^\hbar_{Z}\|$ satisfies the over-completeness relation
\be
\int\rmd Z |\tilde{\psi}^\hbar_{Z}\rangle\langle\tilde{\psi}^\hbar_{Z}|=1_{\ch^0_\g},\quad \rmd Z=\prod_{e}\rmd g(e)\prod_{v}\lt(\frac{\rmd^2 z(v)}{\pi\hbar}\rt).\label{overcompleteness}
\ee

The gauge transformation $u_v\in \Su$ transforms coherent states as
\begin{align}
\bm{u}:\ \psi^t_g&\to\prod_e \sum_{j_{e} \in \mathbb{Z}_{+} / 2 \cup\{0\}}\left(2 j_{e}+1\right) e^{-t j_{e}\left(j_{e}+1\right) / 2} \chi_{j_{e}}\left(g(e) u_{t(e)} h(e) ^{-1}u_{s(e)}^{-1}\right)\nonumber\\
&=\prod_e\sum_{j_{e} \in \mathbb{Z}_{+} / 2 \cup\{0\}}\left(2 j_{e}+1\right) e^{-t j_{e}\left(j_{e}+1\right) / 2} \chi_{j_{e}}\left(u_{s(e)}^{-1}g(e) u_{t(e)} h(e) ^{-1}\right)\nonumber\\
&=\psi^t_{g^u},\quad\text{where}\quad  g^u(e)=u_{s(e)}^{-1}g(e) u_{t(e)}.\label{SU2gaugatransf}\\
\psi^\hbar_z
&\to \psi^\hbar_{z}.
\end{align}
Gauge invariant coherent states $\Psi^\hbar_{[Z]}\in\ch_\g$ are defined by group averaging
\be
\Psi^\hbar_{[Z]}=\int_{\Su^{|V(\g)|}}\prod_{v\in V(\g)}\rmd\mu_H(u_v)\, \psi^t_{g^u}\otimes \psi^\hbar_{z},\quad\text{where}\quad [Z]\equiv([g],z).\label{gaugeinv}
\ee
We denote by $[g]$ the gauge equivalence class of $g\sim g^u$.

\subsection{Physical Hamiltonian Operator}

We quantize the physical Hamiltonian ${\bf H}_0$ to be a non-graph-changing Hamiltonian operators $\hat{\bf H}$ on the Hilbert space $\ch_\g$ of gauge invariant states \cite{Giesel:2007wn,Giesel:2006uj}:
\be
\hat{\mathbf{H}}=\half\sum_{v\in V(\g)}\lt(\hat{C}_v+\hat{C}_v^\dagger\rt)
\ee
There exist self-adjoint extensions of $\hat{\mathbf{H}}$ \cite{Thiemann:2020cuq,private}, so we choose an extension and define the self-adjoint Hamiltonian, which is still denoted by $\hat{\mathbf{H}}$. $\hat{C}_v$ sums the contributions from gravity and scalar.  
\be
\hat{C}_v&=&\hat{C}^{GR}_v+\hat{C}^S_v,
\ee
$\hat{C}^{GR}_v$ and $\hat{C}^S_v$ are listed below (see Appendix \ref{Discretizations and Quantizations of Matter Constraints} for details of $\hat{C}^S_v$):

\subsection*{Gravity:}
 
\be
\hat{C}^{GR}_{0,v}&=&-\frac{2}{i\b\kappa\ell_p^2}\sum_{s_1,s_2,s_3=\pm1}s_1s_2s_3\ \eps^{i_1i_2i_3}\ \mathrm{Tr}\Bigg(\hat{h}(\a_{v;i_1s_1,i_2s_2}) \hat{h}(e_{v;i_3s_3})\Big[\hat{h}(e_{v;i_3s_3})^{-1},\hat{V}_v\Big] \Bigg),\label{C}\\
\hat{C}^{GR}_v&=&\hat{C}^{GR}_{0,v}+{(1+\b^2)}\hat{C}^{GR}_{L,v}+\frac{2\L}{\kappa}\hat{ V}_v,\quad\quad \hat{K}=\frac{i}{\hbar\b^2}\lt[\sum_{v\in V(\g)}\hat{C}_{0,v},\sum_{v\in V(\g)}V_v\rt],\label{operatorK}\\
\hat{C}^{GR}_{L,v}&=&\frac{16}{\kappa\lt(i\b\ell_p^2\rt)^3}\sum_{s_1,s_2,s_3=\pm1}s_1s_2s_3\ \eps^{i_1i_2i_3}\label{HCO}\\
&&\mathrm{Tr}\Bigg( \hat{h}(e_{v;i_1s_1})\Big[\hat{h}(e_{v;i_1s_1})^{-1},\hat{K}\Big]\ \hat{h}(e_{v;i_2s_2})\Big[\hat{h}(e_{v;i_2s_2})^{-1},\hat{K}\Big]\ \hat{h}(e_{v;i_3s_3})\Big[\hat{h}(e_{v;i_3s_3})^{-1},\hat{V}_v\Big]\ \Bigg).\nonumber
\ee
where directions $i_1,i_2,i_3=1,2,3$ are summed in above formulae. $\hat{C}^{GR}_{0,v}$ and $\hat{C}^{GR}_{L,v}$ are the Euclidean and Lorentzian terms in the Hamiltonian constraint operator $\hat{C}^{GR}_v$ by Giesel and Thiemann \cite{QSD,Giesel:2006uj}. $\frac{2\L}{\kappa}\hat{V}_v$ quantizes the cosmological constant term. $\hat{V}_v$ is the volume operator at $v$:
\be
\hat{V}_v&=&\lt(\hat{Q}_v^2\rt)^{1/4},\\ 
\hat{Q}_v
&=&-i\lt(\frac{\b\ell_P^2}{4}\rt)^3\eps_{abc}\frac{R^a_{e_{v;1+}}-R^a_{e_{v;1-}}}{2}\frac{R^b_{e_{v;2+}}-R^b_{e_{v;2-}}}{2}\frac{R^c_{e_{v;3+}}-R^c_{e_{v;3-}}}{2}\nonumber\\
&=&\b^3a^6\eps_{abc}\frac{\hat{p}^a({e_{v;1+}})-\hat{p}^a({e_{v;1-}})}{4}\frac{\hat{p}^b({e_{v;2+}})-\hat{p}^b({e_{v;2-}})}{4}\frac{\hat{p}^c({e_{v;3+}})-\hat{p}^c({e_{v;3-}})}{4}.\label{Qv}
\ee 


$\hat{C}^{GR}_{v}|_{\L=0}$ is the quantization of $\sgn(e)\cc^{GR}|_{\L=0}$ where $\sgn(e)$ is the sign of $\det(e^a_j)$ \cite{book}, because the quantization uses Thiemann's trick:
\be
\mathrm{sgn}(e)\int_e\frac{\left[E^{j},E^{k}\right]}{\sqrt{\det(q)}}\simeq\frac{8}{\kappa\beta}{h}(e)\{{h}(e)^{-1},{V}_v\} \varepsilon^{ijk}.
\ee 
The right-hand side is quantized to be $\hat{h}(e)\{\hat{h}(e)^{-1},\hat{V}_v\}$. Eq.\eqref{HCO} quantizes the cosmological constant term $\frac{2\L}{\kappa}\sqrt{\det(q)}$ to be the volume operator $\frac{2\L}{\kappa}\hat{V}_v$ without involving $\sgn(e)$. Therefore flipping $\sgn(e)$ effectively flips the cosmological constant from $\L$ to $-\L$ in $\hat{\bf H}$. As a result, even if we fix $\L>0$ in the definition of $\hat{C}^{GR}_{v}$ and $\hat{\bf H}$, both positive and negative cosmological constants can appear from the theory \cite{Han:2019feb}.

\subsection*{Scalar}

We consider the following real scalar field contribution in $\cc$
\be
\mathcal{C}^{S}=\frac{\pi^2}{2\sqrt{\det(q)}}+\frac{1}{2}\sqrt{\det(q)}q^{jk}\left(\partial_{j}\phi\right)\lt(\partial_{k}\phi\rt))+\sqrt{\det(q)}U_1(\phi)+\det(e^a_j)U_2(\phi)
\ee
where we take into account both parity-even and parity-odd potential terms $\sqrt{\det(q)}U_1(\phi)$ and $\det(e^a_j)U_2(\phi)$. 

We define a family of essentially self-adjoint operators parametrized by $r>0$ \cite{Sahlmann:2002qj}:
\be
\hat{\cq}_{r}^{a}(e)=i\mathrm{Tr}\left(\tau^{a}\hat{h}(e)\left[\hat{h}(e)^{-1},\hat{V}_{v}^{r}\right]\right),\quad \hat{\mathcal{Q}}_{r}(e)=\hat{\mathcal{Q}}_{r}^{a}(e)\frac{\tau^{a}}{2}=-i\hat{h}(e)\left[\hat{h}(e)^{-1},\hat{V}_{v}^{r}\right].
\ee
We define the quantization of $\frac{\sgn(e)}{\sqrt{\det(q)}}$:
\be
\widehat{\lt(\frac{\mathrm{sgn}(e)}{V}\rt)}_v=-\left(\frac{18\times64}{\ell_{P}^{6}\beta^{3}}\right)\sum_{s_{1}s_{2}s_{3}}s_{1}s_{2}s_{3}\sum_{i,j,k}\epsilon^{ijk}\mathrm{Tr}\left[\mathcal{\hat{Q}}_{1/3}(e_{v;is_{1}})\mathcal{\hat{Q}}_{1/3}(e_{v;js_{2}})\mathcal{\hat{Q}}_{1/3}(e_{v;ks_{3}})\right].
\ee
$\hat{C}^S_v$ is the quantization of $\sgn(e)\mathcal{C}^{S}$: 
\be
\hat{C}^S_v & =&\half\widehat{\left(\frac{\mathrm{sgn}(e)}{V}\right)}_v\hat{\pi}(v)^2+\half\widehat{\left(\frac{\mathrm{sgn}(e)}{V}\right)}_v{{\frac{a^{4}\beta^{2}}{8}}}\sum_{s_{1}s_{2}s_{3}}\sum_{j,k}s_jX^j_{a}(v)s_kX^k_{a}(v)\left(\delta_{j,s_{j}}\hat{\phi}(v)\right)\lt(\delta_{k,s_{k}}\hat{\phi}(v)\rt)\nonumber\\
 &&-\frac{2}{3}\frac{8^{2}}{(\ell_{P}^{2}\beta)^{3}}\sum_{s_{1}s_{2}s_{3}}s_{1}s_{2}s_{3}\epsilon^{ijk}\mathrm{Tr}\left[\hat{\mathcal{Q}}_{1}(e_{v;is_{1}})\hat{\mathcal{Q}}_{1}(e_{v;js_{2}})\mathcal{\hat{Q}}_{1}(e_{v;ks_{3}})\right]U_{1}(\hat{\phi})+\hat{V}_{v}U_{2}(\hat{\phi})
\ee
where $\delta_{j,s_j}\hat{\phi}(v)$ is the lattice derivative:
\be
\delta_{j,s_j}\hat{\phi}(v)=\hat{\phi}\left(t(e_{v;js_j})\right)-\hat{\phi}\left(v\right).
\ee
\be
X^k_{a}=\frac{\hat{p}^a({e_{v;k+}})-\hat{p}^a({e_{v;k-}})}{4}
\ee

\section{Coherent State Path Integral of Gravity-Scalar-Dust}\label{Coherent State Path Integral of Gravity-Scalar-Dust}

An interesting quantity for quantum dynamics is the transition amplitude
\be
A_{[Z],[Z']}=\langle \Psi^\hbar_{[Z]}|\,\exp\lt[\frac{i}{\hbar}T \hat{\bf H}\rt]\,|\Psi^\hbar_{[Z']}\rangle
\ee
For the purpose of semiclassical analysis, we focus on the initial and final gauge invariant coherent states $\Psi^\hbar_{[Z']}, \Psi^\hbar_{[Z]}$. Recall that $[Z]=([g],z)$ where $[g]$ is the SU(2) gauge orbit with
\be
g(e)=e^{-ip_a(e)\t_a/2}h(e)=
e^{-ip^a(e)\t^a/2}e^{\theta^a(e)\t^a/2}, \quad p^a(e),\ \theta^a(e)\in\R^3.\label{gthetap}
\ee

Applying Eq.\eqref{gaugeinv} and a discretization of time $T=N\delta\t$ with large $N$ and infinitesimal $\delta\t$, followed by inserting $N+1$ overcompleteness relations of normalized coherent state $\tilde{\psi}^\hbar_{Z}$ in Eq.\eqref{overcompleteness}:
\be
A_{[Z],[Z']}&=&\int\rmd u\lag\psi^\hbar_{Z}\rt|\lt[e^{ \frac{i}{\hbar}\delta\t \hat{\mathbf{H}}}\rt]^N |{\psi}^\hbar_{{Z'}^{u}}\rangle,\\
&=&\int\rmd u\prod_{i=1}^{N+1}\mathrm{d}Z_{i}\,\langle\psi^\hbar_{Z}|\tilde{\psi}^\hbar_{Z_{N+1}}\rangle\langle \tilde{\psi}^\hbar_{Z_{N+1}}\big|e^{ \frac{i\delta\t}{\hbar} \hat{\mathbf{H}}}\big|\tilde{\psi}^\hbar_{Z_{N}}\rangle
\langle \tilde{\psi}^\hbar_{Z_{N}}\big|e^{ \frac{i\delta\t}{\hbar}\hat{\mathbf{H}}}\big|\tilde{\psi}^\hbar_{Z_{N-1}}\rangle\cdots\nonumber\\
&&\quad \cdots\ 
\langle \tilde{\psi}^\hbar_{Z_2}\big|e^{ -\frac{i\delta\t}{\hbar}\hat{\mathbf{H}}}\big|\tilde{\psi}^\hbar_{Z_1}\rangle\langle\tilde{\psi}^\hbar_{Z_1}|{\psi}^\hbar_{Z'{}^{u}}\rangle \label{smallsteps}
\end{eqnarray}
where $\int \rmd u\equiv \prod_{v \in V(\gamma)} \int_{\mathrm{SU}(2)}\mathrm{d} \mu_{H}\left(u_{v}\right)$ and $Z'^u\equiv (g'^u,z)$.

Following the standard coherent state functional integral method, we let $N$ arbitrarily large thus $\delta\t$ arbitrarily small. $U(\delta\t)$ is a strongly continuous unitary group and ${[U(\delta\t)|\psi\rangle-|\psi\rangle]}/{\delta\t}\to \frac{i}{\hbar} \hat{\mathbf{H}}\, |\psi\rangle$, so $\hat{\eps}\lt(\frac{\delta\t}{\hbar}\rt):=\frac{\hbar}{\delta\t}[U(\delta\t)-1-\frac{i {\delta \tau} }{\hbar} \hat{\mathbf{H}}]$ satisfies the strong limit $\hat{\eps}\lt(\frac{\delta\t}{\hbar}\rt)|\psi\rangle\to0$ as $\delta\t\to0$ for all $\psi$ in the domain of $\hat{\mathbf{H}}$. The coherent state $\tilde{\psi}^\hbar_{Z}$ belongs to the domain of $\hat{\mathbf{H}}$, thus $\eps_{i+1,i}\lt(\frac{\delta \t}{\hbar}\rt)=\langle \tilde{\psi}^\hbar_{Z_{i+1}}|\hat{\eps}\lt(\frac{\delta\t}{\hbar}\rt)| \tilde{\psi}^\hbar_{Z_{i}}\rangle$ satisfies $\lim\limits_{\delta\t\to0}\eps_{i+1,i}\lt(\frac{\delta \t}{\hbar}\rt)=0$. We obtain the following relation
\be
&&\langle \tilde{\psi}^\hbar_{Z_{i+1}}\big|\exp \lt( \frac{i}{\hbar}\delta\t \hat{\mathbf{H}}\rt)\big|\tilde{\psi}^\hbar_{Z_{i}}\rangle
=\langle \tilde{\psi}^\hbar_{Z_{i+1}}\big|1+ \frac{i\delta\t}{\hbar} \hat{\mathbf{H}} \big|\tilde{\psi}^\hbar_{Z_{i}}\rangle+\frac{\delta \t}{\hbar}\eps_{i+1,i}\lt(\frac{\delta \t}{\hbar}\rt)\nonumber\\
&=&\langle \tilde{\psi}^\hbar_{Z_{i+1}}\big|\tilde{\psi}^\hbar_{Z_{i}}\rangle\lt[1+\frac{i\delta\t}{\hbar}\frac{\langle \tilde{\psi}^\hbar_{Z_{i+1}}\big| \hat{\mathbf{H}}\big|\tilde{\psi}^\hbar_{Z_{i}}\rangle}{\langle \tilde{\psi}^\hbar_{Z_{i+1}}\big|\tilde{\psi}^\hbar_{Z_{i}}\rangle}\rt] +\frac{\delta \t}{\hbar}\eps_{i+1,i}\lt(\frac{\delta \t}{\hbar}\rt)\nonumber\\
&=&\langle \tilde{\psi}^\hbar_{Z_{i+1}}\big|\tilde{\psi}^\hbar_{Z_{i}}\rangle\,e^{\frac{i\delta\t}{\hbar}\frac{\langle \tilde{\psi}^\hbar_{Z_{i+1}}\lt| \hat{\mathbf{H}}\rt|\tilde{\psi}^\hbar_{Z_{i}}\rangle}{\langle \tilde{\psi}^\hbar_{Z_{i+1}}|\tilde{\psi}^\hbar_{Z_{i}}\rangle} +\frac{\delta \t}{\hbar}\tilde{\eps}_{i+1,i}\lt(\frac{\delta \t}{\hbar}\rt)}\label{smallstep}
\ee
where 
\be
\frac{\delta \t}{\hbar}\tilde{\eps}_{i+1,i}\lt(\frac{\delta \t}{\hbar}\rt)=\ln\lt[1+\frac{i\delta\t}{\hbar}\frac{\langle \tilde{\psi}^\hbar_{Z_{i+1}}\big| \hat{\mathbf{H}}\big|\tilde{\psi}^\hbar_{Z_{i}}\rangle}{\langle \tilde{\psi}^\hbar_{Z_{i+1}}\big|\tilde{\psi}^\hbar_{Z_{i}}\rangle} +\frac{\delta \t}{\hbar}\frac{\eps_{i+1,i}\lt({\delta \t}/{\hbar}\rt)}{\langle \tilde{\psi}^\hbar_{Z_{i+1}}\big|\tilde{\psi}^\hbar_{Z_{i}}\rangle}\rt]-\frac{i\delta\t}{\hbar}\frac{\langle \tilde{\psi}^\hbar_{Z_{i+1}}\big| \hat{\mathbf{H}}\big|\tilde{\psi}^\hbar_{Z_{i}}\rangle}{\langle \tilde{\psi}^\hbar_{Z_{i+1}}|\tilde{\psi}^\hbar_{Z_{i}}\rangle}
\ee
is arbitrarily small and satisfies $\lim\limits_{\delta\t\to0}\tilde{\eps}_{i+1,i}\lt(\frac{\delta \t}{\hbar}\rt)=0$. 

By Eq.\eqref{smallstep} and expressions of overlaps between coherent states Eqs.\eqref{overlap} and \eqref{overlapscalar}, a path integral formula can be derived for $A_{[g],[g']}$ 
\be
{A_{[Z],[Z']}}= {\Big\|\psi_{Z}^{\hbar}\Big\|\,\Big\|\psi_{Z^{\prime}}^{\hbar}\Big\|}\int \mathrm{d} u \prod_{i=1}^{N+1} \mathrm{d} Z_{i}\, \nu[Z]\, e^{S[Z, u] / t}\label{Agg}
\ee 
where the action $S[Z,u]$ is given by
\be
S[Z, u]&=&\sum_{i=0}^{N+1} \ck\left(Z_{i+1}, Z_{i}\right) - \frac{i \kappa}{a^{2}} \sum_{i=1}^{N} \delta \tau\left[- \frac{\langle\psi_{Z_{i+1}}^{\hbar}|\hat{\mathbf{H}}| \psi_{Z_{i}}^{\hbar}\rangle}{\langle\psi_{Z_{i+1}}^{\hbar} | \psi_{Z_{i}}^{\hbar}\rangle} + i \tilde{\varepsilon}_{i+1, i}\left(\frac{\delta \tau}{\hbar}\right)\right],\label{Sgh}
\ee
The ``kinetic term'' $\ck\left(Z_{i+1}, Z_{i}\right)$ reads
\be
\ck\left(Z_{i+1}, Z_{i}\right)&=&\sum_{e \in E(\gamma)}\left[\xi_{i+1,i}(e)^{2}-\frac{1}{2} p_{i+1}(e)^{2}-\frac{1}{2} p_{i}(e)^{2}\right]\\
&&\  +\frac{ \kappa}{a^{2}}\sum_{v\in V(\g)}\left[\bar{z}_{i+1}(v)z_i(v)-\frac{1}{2}\bar{z}_{i+1}(v)z_{i+1}(v)-\frac{1}{2}\bar{z}_i(v)z_i(v)\right],\label{ckZZ}
\ee
where $Z_{0} \equiv Z^{\prime u},\ Z_{N+2} \equiv Z$, and $\xi_{i+1,i}(e)$ are given by 
\be
\xi_{i+1,i}(e)=\mathrm{arccosh}\lt(x_{i+1,i}(e)\rt),\quad x_{i+1,i}(e)=\half\tr\lt[g_{i+1}(e)^\dagger g_{i}(e)\rt].
\ee
$\nu[Z]$ is a measure factor from Eq.\eqref{overlap}
\be
\nu[Z]=\nu[g]=\prod_{i=0}^{N+1} \prod_{e \in E(\gamma)}\lt[\frac{\mathrm{arccosh}\lt(x_{i+1,i}(e)\rt)}{\sinh \left(\mathrm{arccosh}\lt(x_{i+1,i}(e)\rt)\right)}\sqrt{\frac{\sinh \left(p_{i+1}(e)\right)}{p_{i+1}(e)} \frac{\sinh \left(p_{i}(e)\right)}{p_{i}(e)}} +O(t^\infty)\rt]
\ee

The path integral Eq.\eqref{Agg} is constructed with discrete time and space, and is a well-define integration formula for the transition amplitude $A_{[g],[g']}$ as long as $\delta\t$ is arbitrarily small but finite. The time translation of $\g$ with finite $\delta\t$ makes a hypercubic lattice in 4 dimensions, on which the path integral is defined. There is no issue of any divergence in this path integral formulation of LQG, since it is derived from a well-defined transition amplitude. 

\section{Semiclassical Dynamics on Fixed Lattice}\label{Semiclassical Equations of Motion}

\subsection{Equations of Motion of the Full Theory}

The semiclassical limit $\hbar\to0$ (or $t\to0$) of the transition amplitude $A_{[Z],[Z']}$ can be studied by the stationary phase analysis. Dominant contributions to $A_{[Z],[Z']}$ as $\hbar\to0$ come from semiclassical trajectories satisfying the equations of motion (EOMs) $\delta S[Z,u]=0$. We neglect $\tilde{\eps}_{i+1,i}(\delta\t/\hbar)$ in following derivations of the EOMs since we will be interested in the time continuum limit $\delta\t\to0$ of the EOMs, where the contribution of $\tilde{\eps}_{i+1,i}(\delta\t/\hbar)$ is negligible (see Appendix \ref{Corrections of eps in EOMs} for details).

EOMs from $\delta_g S[Z,u]=\delta_u S[Z,u]=0$ has been derived in \cite{Han:2019vpw}

\begin{itemize}
\item The variation with respect to $g_i$ using the holomorphic deformation 
  \be
   g_i(e) \to g_i^\eps(e)=g_i(e) e^{\eps_i^a(e)\t^a} \ , \qquad \eps_i^a(e)\in\C \ 
  \ee
  leads to the following equations from derivatives of $\eps_i^a(e),\bar{\eps}_i^a(e)$ respectively:
\begin{itemize}
\item For $i=1,\cdots,N$, at every edge $e\in E(\g)$, 
\be
&&\frac{1}{\delta\t}\lt[\frac{\mathrm{arccosh}\lt(x_{i+1,i}(e)\rt)\,\tr\lt[\t^a g_{i+1}(e)^\dagger g_i(e)\rt]}{\sqrt{x_{i+1,i}(e)-1}\sqrt{x_{i+1,i}(e)+1}}-\frac{p_i(e)\,\tr\lt[\t^a g_{i}(e)^\dagger g_i(e)\rt]}{\sinh(p_i(e))}\rt]\nonumber\\
&=&- \frac{i\kappa }{a^2}\frac{\partial}{\partial \varepsilon_{i}^{a}(e)} \frac{\langle\psi_{g_{i+1}^{\varepsilon}}^{t}\otimes\psi^\hbar_{z_{i+1}}|\hat{\mathbf{H}}| \psi_{g_{i}^{\varepsilon}}^{t}\otimes\psi^\hbar_{z_i}\rangle}{\langle\psi_{g_{i+1}^{\varepsilon}}^{t}\otimes\psi^\hbar_{z_{i+1}} | \psi_{g_{i}^{\varepsilon}}^{t}\otimes\psi^\hbar_{z_i}\rangle}\Bigg|_{\vec{\eps}=0}\label{eoms1}
\ee 

\item For $i=2,\cdots,N+1$, at every edge $e\in E(\g)$,
\be
&&\frac{1}{\delta\t}\lt[\frac{\mathrm{arccosh}\lt(x_{i,i-1}(e)\rt)\,\tr\lt[\t^a g_{i}(e)^\dagger g_{i-1}(e)\rt]}{\sqrt{x_{i,i-1}(e)-1}\sqrt{x_{i,i-1}(e)+1}}-\frac{p_i(e)\,\tr\lt[\t^a g_{i}(e)^\dagger g_i(e)\rt]}{\sinh(p_i(e))}\rt]\nonumber\\
&=&\frac{i\kappa }{a^2}\frac{\partial}{\partial \bar{\varepsilon}_{i}^{a}(e)} \frac{\langle\psi_{g_{i}^{\varepsilon}}^{t}\otimes\psi_{z_i}^\hbar|\hat{\mathbf{H}}| \psi_{g_{i-1}^{\varepsilon}}^{t}\otimes\psi^\hbar_{z_{i-1}}\rangle}{\langle\psi_{g_{i}^{\varepsilon}}^{t} \otimes\psi^\hbar_{z_i}| \psi_{g_{i-1}^{\varepsilon}}^{t}\otimes\psi^\hbar_{z_{i-1}}\rangle}\Bigg|_{\vec{\eps}=0}.\label{eoms2}
\ee
\end{itemize}
\item The variation with respect to $u_v$ leads to the closure condition at every vertex $v\in V(\g)$ for initial data:
\be
-\sum_{e, s(e)=v}p_1^a(e)+\sum_{e, t(e)=v}\L^a_{\ b}\lt(\vec{\theta}_1(e)\rt)\,p_1^b(e)=0.\label{closure0}
\ee
where $\L^a_{\ b}(\vec{\theta})\in\mathrm{SO}(3)$ is given by $e^{\theta^c\t^c/2}\t^a e^{-\theta^c\t^c/2}=\L^a_{\ b}(\vec{\theta})\t^b$.
\end{itemize}
\noindent
The initial and final conditions for $g_i$ are given by $g_{1}=g'^u$ and $g_{N+1}=g$.

We compute the variation of $S[Z,u]$ with respect to scalar DOFs $z_i(v),\bar{z}_i(v)$:

\begin{itemize}

\item For $i=1,\cdots,N$, at every $v\in V(\g)$,
\be
\frac{\left[\bar{z}_{i+1}(v)-\bar{z}_{i}(v)\right]}{\delta\tau}=- i\frac{\partial}{\partial z_i(v)}\frac{\langle \psi_{Z_{i+1}}^{\hbar}|\hat{\mathbf{H}}|\psi_{Z_{i}}^{\hbar}\rangle }{\langle \psi_{Z_{i+1}}^{\hbar}\mid\psi_{Z_{i}}^{\hbar}\rangle }.\label{eomscalar1}
\ee

\item For $i=2,\cdots,N+1$, at every $v\in V(\g)$,
\be
\frac{\left[z_{i}(v)-z_{i-1}(v)\right]}{\delta\tau}=i\frac{\partial}{\partial \bar{z}_i(v)}\frac{\langle \psi_{Z_{i}}^{\hbar}|\hat{\mathbf{H}}|\psi_{Z_{i-1}}^{\hbar}\rangle }{\langle \psi_{Z_{i}}^{\hbar}\mid\psi_{Z_{i-1}}^{\hbar}\rangle }.\label{eomscalar2}
\ee

\end{itemize}
\noindent
The initial and final conditions for $z_i(v),\bar{z}_i(v)$ are given by $z_{1}(v)=z'(v)$ and $z_{N+1}(v)=z(v)$.

Semiclassical EOMs \eqref{eoms1} - \eqref{closure0} are derived with finite $\delta\t$. We prefer to derive EOMs from the path integral Eq.\eqref{Agg} with discrete time and space, because Eq.\eqref{Agg} is a well-define integration formula for the transition amplitude.

Right-hand sides of Eqs.\eqref{eomscalar1} and \eqref{eomscalar2} can be expressed explicitly by relations $\partial_{z_i(v)}|\psi^\hbar_{z_i}\rangle=\Fa(v)^\dagger|\psi^\hbar_{z_i}\rangle$ and $\partial_{\bar{z}_i(v)}\langle\psi^\hbar_{z_i}|=\langle\psi^\hbar_{z_i}|\Fa(v)$:
\be
\frac{\partial}{\partial z_i(v)}\frac{\langle \psi_{Z_{i+1}}^{\hbar}|\hat{\mathbf{H}}|\psi_{Z_{i}}^{\hbar}\rangle }{\langle \psi_{Z_{i+1}}^{\hbar}\mid\psi_{Z_{i}}^{\hbar}\rangle }&=&\frac{\langle \psi_{Z_{i+1}}^{\hbar}|\hat{\mathbf{H}}\,\mathfrak{\hat{A}}^{\dagger}(v)|\psi_{Z_{i}}^{\hbar}\rangle \langle \psi_{Z_{i+1}}^{\hbar}|\psi_{Z_{i}}^{\hbar}\rangle -\langle \psi_{Z_{i+1}}^{\hbar}|\hat{\mathbf{H}}|\psi_{Z_{i}}^{\hbar}\rangle \langle \psi_{Z_{i+1}}^{\hbar}|\mathfrak{\hat{A}}^{\dagger}(v)|\psi_{Z_{i}}^{\hbar}\rangle }{\langle \psi_{Z_{i+1}}^{\hbar}|\psi_{Z_{i}}^{\hbar}\rangle ^{2}}\quad\\
\frac{\partial}{\partial\bar{z}_{i}(v)}\frac{\langle\psi_{Z_{i}}^{\hbar}|\hat{\mathbf{H}}|\psi_{Z_{i-1}}^{\hbar}\rangle}{\langle\psi_{Z_{i}}^{\hbar}|\psi_{Z_{i-1}}^{\hbar}\rangle}&=&\frac{\langle\psi_{Z_{i}}^{\hbar}|\mathfrak{\hat{A}}(v)\,\hat{\mathbf{H}}|\psi_{Z_{i-1}}^{\hbar}\rangle\langle\psi_{Z_{i}}^{\hbar}|\psi_{Z_{i-1}}^{\hbar}\rangle-\langle\psi_{Z_{i}}^{\hbar}|\hat{\mathbf{H}}|\psi_{Z_{i-1}}^{\hbar}\rangle\langle\psi_{Z_{i}}^{\hbar}|\mathfrak{\hat{A}}(v)|\psi_{Z_{i-1}}^{\hbar}\rangle}{\langle\psi_{Z_{i}}^{\hbar}|\psi_{Z_{i-1}}^{\hbar}\rangle^{2}}\quad
\ee
The time continuous limit $\delta\t\to0$ gives $Z_i\to Z_{i+1}\equiv Z$. By above relations\footnote{$\lim_{z_i\to z}\int\rmd\phi\, \bar{f}(\phi)\psi_{z_i}(\phi)=\int\rmd\phi\, \bar{f}(\phi)\psi_{z}(\phi)$, $\forall f\in \ch^S_\g$ by the dominated convergence, since $|\psi_{z_i}(\phi)|$ is uniformly bounded by an integrable function when $z_i$ is in a finite neighborhood $U$ at $z$. Similarly, $\partial_z \int\rmd\phi\, \bar{f}(\phi)\psi_{z}(\phi)=\int\rmd\phi\, \bar{f}(\phi)\partial_z \psi_{z}(\phi)$ since $|\partial_z\psi_{z}(\phi)|$ is uniformly bounded by an integrable function in a finite neighborhood $U$ at $z$. Note that $\psi_{z}, \partial_z\psi_{z}$ are Schwarz functions on $\R$: $|\psi_{z}| \text{ or } |\partial_z\psi_{z}| \leq C_{k}(z)(1+|x|)^{-k}\leq \mathrm{Max}_{z\in U}(C_{k})(1+|x|)^{-k} \text { for all } k\in\Z_+$.}
\be
\lim_{\delta\t\to0}\frac{\partial}{\partial z_i(v)}\frac{\langle \psi_{Z_{i+1}}^{\hbar}|\hat{\mathbf{H}}|\psi_{Z_{i}}^{\hbar}\rangle }{\langle \psi_{Z_{i+1}}^{\hbar}\mid\psi_{Z_{i}}^{\hbar}\rangle }
&=&\frac{\langle \psi_{Z}^{\hbar}|\hat{\mathbf{H}}\,\mathfrak{\hat{A}}^{\dagger}(v)|\psi_{Z}^{\hbar}\rangle \langle \psi_{Z}^{\hbar}|\psi_{Z}^{\hbar}\rangle -\langle \psi_{Z}^{\hbar}|\hat{\mathbf{H}}|\psi_{Z}^{\hbar}\rangle \langle \psi_{Z}^{\hbar}|\mathfrak{\hat{A}}^{\dagger}(v)|\psi_{Z}^{\hbar}\rangle }{\langle \psi_{Z}^{\hbar}|\psi_{Z}^{\hbar}\rangle ^{2}}\quad\nonumber\\
&=&\frac{\partial}{\partial z(v)}\frac{\langle \psi_{Z}^{\hbar}\mid\hat{\mathbf{H}}\mid\psi_{Z}^{\hbar}\rangle }{\langle \psi_{Z}^{\hbar}\mid\psi_{Z}^{\hbar}\rangle },\label{limitZZZZ1}\\
\lim_{\delta\t\to0}\frac{\partial}{\partial\bar{z}_{i}(v)}\frac{\langle\psi_{Z_{i}}^{\hbar}|\hat{\mathbf{H}}|\psi_{Z_{i-1}}^{\hbar}\rangle}{\langle\psi_{Z_{i}}^{\hbar}|\psi_{Z_{i-1}}^{\hbar}\rangle}
&=&\frac{\langle\psi_{Z}^{\hbar}|\mathfrak{\hat{A}}(v)\,\hat{\mathbf{H}}|\psi_{Z}^{\hbar}\rangle\langle\psi_{Z}^{\hbar}|\psi_{Z}^{\hbar}\rangle-\langle\psi_{Z}^{\hbar}|\hat{\mathbf{H}}|\psi_{Z}^{\hbar}\rangle\langle\psi_{Z}^{\hbar}|\mathfrak{\hat{A}}(v)|\psi_{Z}^{\hbar}\rangle}{\langle\psi_{Z}^{\hbar}|\psi_{Z}^{\hbar}\rangle^{2}}\quad\nonumber\\
&=&\frac{\partial}{\partial\bar{z}(v)}\frac{\langle\psi_{Z}^{\hbar}\mid\hat{\mathbf{H}}\mid\psi_{Z}^{\hbar}\rangle}{\langle\psi_{Z}^{\hbar}\mid\psi_{Z}^{\hbar}\rangle},\label{limitZZZZ2}
\ee
which are finite. The left-hand sides of Eqs.\eqref{eomscalar1} and \eqref{eomscalar2} are finite as $\delta\t\to0$ if and only if $z_i(v),\bar{z}_i(v)$ admit approximations $z(\t, v),\bar{z}(\t, v)$ which are differentiable in $\t$. All solutions $z_i(v),\bar{z}_i(v)$ of Eqs.\eqref{eomscalar1} and \eqref{eomscalar2} must give finite left and right hand sides in Eqs.\eqref{eomscalar1} and \eqref{eomscalar2}. Therefore for all solutions, we can take the time continuous limit:
\be
\frac{\rmd \bar{z}(v)}{\rmd\tau}=- i \frac{\partial}{\partial z(v)}{\langle \tilde{\psi}_{Z}^{\hbar}|\hat{\mathbf{H}}|\tilde{\psi}_{Z}^{\hbar}\rangle }\quad \frac{\rmd z(v)}{\rmd\tau}=i\frac{\partial}{\partial\bar{z}(v)}{\langle\tilde{\psi}_{Z}^{\hbar}|\hat{\mathbf{H}}|\tilde{\psi}_{Z}^{\hbar}\rangle}
\ee  
Recall Eq.\eqref{zzbarphipi}, above continuous-time EOMs can be written as Hamilton's equations with the Hamiltonian $\langle \tilde{\psi}_{Z}^{\hbar}|\hat{\mathbf{H}}|\tilde{\psi}_{Z}^{\hbar}\rangle$:
\be
\frac{\rmd\phi(v)}{\rmd\tau}= \frac{\partial}{\partial\pi(v)}\langle \tilde{\psi}_{Z}^{\hbar}|\hat{\mathbf{H}}|\tilde{\psi}_{Z}^{\hbar}\rangle ,\quad 
\frac{\rmd\pi(v)}{\rmd\tau}= - \frac{\partial}{\partial\phi(v)}\langle \tilde{\psi}_{Z}^{\hbar}|\hat{\mathbf{H}}|\tilde{\psi}_{Z}^{\hbar}\rangle.\label{hamiltonphipi} 
\ee

Similarly \cite{Han:2019vpw,Han:2020chr} proves that Eqs.\eqref{eoms1} - \eqref{eoms2} also admit continuous time approximation and can be expressed in terms of ${\bm p}(e)=(p^1(e),p^2(e),p^3(e))^T$ and ${\bm \theta}(e)=(\theta^1(e),\theta^2(e),\theta^3(e))^T$ and their time derivatives:
\be
\left( \begin{array}{l}  \frac{\rmd {\bm p}(e)}{\rmd \tau} \\ \frac{\rmd \bm{\theta}(e)}{\rmd \tau}  \end{array} \right)  =  - \frac{i\kappa}{a^2}\, {T}\lt({\bm p},{\bm \theta}\rt) ^{-1}\left( \begin{array}{l}  {\frac{\partial}{\partial {\bm p} (e)} \langle \tilde{\psi}_{Z}^{\hbar}|\hat{\mathbf{H}}|\tilde{\psi}_{Z}^{\hbar}\rangle} \\ {\frac{\partial}{\partial \bm{\theta} (e)} \langle \tilde{\psi}_{Z}^{\hbar}|\hat{\mathbf{H}}|\tilde{\psi}_{Z}^{\hbar}\rangle } \end{array} \right) .\label{eom0}
\ee
where $T({p}, {\theta})$ (whose explicit express is given in \cite{github0}) is a $6\times6$ matrix satisfying
\be
-\frac{i a^{2}}{\kappa} P(\bm{p}, \bm{\theta}) T(\bm{p}, \bm{\theta})=1_{6 \times 6},\quad P({\bm p},{\bm \theta})=\lt(\begin{array}{cc}
\lt\{p^a(e),p^b(e)\rt\} & \lt\{p^a(e),\theta^b(e)\rt\}\\
\lt\{\theta^a(e),p^b(e)\rt\} & 0
\end{array}\rt).
\ee
Eq.\eqref{eom0} is equivalent to Hamilton's equations: 
\be
\frac{\rmd {h}(e)}{\rmd \t}= \lt\{\langle \tilde{\psi}_{Z}^{\hbar}|\hat{\mathbf{H}}|\tilde{\psi}_{Z}^{\hbar}\rangle , \, {h}(e) \rt\},\quad \frac{\rmd {p}^a(e)}{\rmd \t}= \lt\{ \langle \tilde{\psi}_{Z}^{\hbar}|\hat{\mathbf{H}}|\tilde{\psi}_{Z}^{\hbar}\rangle , \, {p}^a(e)  \rt\}.\label{hamiltonhp}
\ee 

The coherent state expectation value of $\hat{\bf H}$ have the correct semiclassical limit
\be
\langle\tilde{\psi}_{Z}^{\hbar}|\hat{\mathbf{H}}| \tilde{\psi}_{Z}^{\hbar}\rangle={\bf H}[Z,\bar{Z}]+O(\hbar),\label{semiclassH}
\ee
where ${\bf H}[Z,\bar{Z}]$ is the classical discrete Hamiltonian evaluated at $p^a(e),h(e),\pi(v),\phi(v)$ determined by $Z=(g,z)$ in Eqs.\eqref{gthetap} and \eqref{zzbarphipi}. Note that the above semiclassical behavior of $\langle\tilde{\psi}_{g}^{t}|\hat{\mathbf{H}}| \tilde{\psi}_{g}^{t}\rangle$ relies on the following semiclassical expansion of volume operator $\hat{V}_v$ \cite{Giesel:2006um}:
\be
\hat{V}_v=\langle \hat{Q}_v\rangle^{\half}\lt[1+\sum_{n=1}^{2k+1}(-1)^{n+1}\frac{q(1-q)\cdots(n-1+q)}{n!}\lt(\frac{\hat{Q}_v^2}{\langle\hat{Q}_v\rangle^2}-1\rt)^n\rt]_{q=\frac{1}{4}}+O(\hbar^{k+1}),\label{expandvolume}
\ee
where $\langle \hat{Q}_v\rangle=\langle\psi^t_g|\hat{Q}_v|\psi^t_g\rangle$. When we apply this expansion to e.g. the expectation value of $\hat{V}_v$, using $\langle\hat{Q}^N\rangle=\langle\hat{Q}_v\rangle^N[1+\frac{3t}{8p^2}N(N-1)]$ \cite{Dapor:2017gdk}, we can see that the expansion is valid in the regime $p^2\gg t$. When Eq.\eqref{expandvolume} is actually applied to compute perturbatively $\langle\tilde{\psi}_{g}^{t}|\hat{\mathbf{H}}| \tilde{\psi}_{g}^{t}\rangle$, The validation of the expansion and in particular Eq.\eqref{semiclassH} uses the same requirement $p^2\gg t$ (see \cite{Zhang:2020mld} for details). 

The EOMs derived from the semiclassical approximation is not sensitive to $O(\hbar)$. Neglecting $O(\hbar)$ in Eqs.\eqref{hamiltonhp} and \eqref{hamiltonphipi} imply that for function $f$ on the reduces phase space $\mathscr{P}_\g$, its $\t$-evolution is given by the Hamiltonian flow generated by the classical discrete Hamiltonian ${\bf H} $:
\be
\frac{\rmd f}{\rmd \tau}= \lt\{  {\bf H}, \, f\rt\}.\label{hamilton}
\ee
The closure condition is preserved by $\t$-evolution by $\{G^a_v,\,{\bf H}\}=0$.


\subsection{Homogeneous and Isotropic Cosmological Dynamics on the Fixed Lattice }\label{Homogeneous and Isotropic Cosmological Dynamics on Fixed Lattice }

We would like to find the solution of Eq.\eqref{hamilton} describing the homogeneous and isotropic cosmology. For this purpose, we apply the following homogeneous and isotropic ansatz to the semiclassical EOMs
\be
\theta^a(e_i(v))=\theta\delta^a_I=\mu \b K_0\delta^a_i,&& p^a(e_i(v))=p\delta^a_i=\frac{2\mu^2}{\b a^2}P_0\delta^a_i,\label{cosansatz1}\\
\phi(v)=\phi=\phi_0,\quad&&\pi(v)=\pi=\mu^3\pi_0.\label{cosansatz2}
\ee
Here $K_0=K_0(\t)$, $P_0=P_0(\t)$, $\phi_0=\phi_0(\t)$, and $\pi_0=\pi_0(\t)$ are constant on $\g$ but evolve with the dust time $\t$. This ansatz is a simple generalization of the one used in \cite{Han:2019vpw,Dapor:2017rwv}.

Inserting the ansatz, equations in \eqref{eom0} with $e=e_I(v)$ are split into 2 sets: (1) $\rmd p^a(e_i(v))/\rmd \t=\cdots$ and $\rmd \theta^a(e_i(v))/\rmd \t=\cdots$ with $a=i$: Left-hand sides of these 6 equations are proportional to $\dot{P}_0=\rmd P_0/\rmd\t$ and $\dot{K}_0=\rmd K_0/\rmd\t$. They reduces to 
\be
\frac{4 \beta^{2}\left[-2 \mu^{2} \sqrt{P_{0}} \dot{K}_{0}+\sin ^{4}\left(\beta \mu K_{0}\right)+\Lambda \mu^{2} P_{0}\right]-\sin ^{2}\left(2 \beta \mu K_{0}\right)}{\sqrt{P_{0}}}&=&\kappa\b^2\mu^2\sqrt{P_0}\lt( \pi_0^2 P_0^{-3}- U\rt),\label{cos1}\\
\sqrt{P_{0}}\left[2 \beta^{2} \sin \left(2 \beta \mu K_{0}\right)-\left(\beta^{2}+1\right) \sin \left(4 \beta \mu K_{0}\right)\right]+2 \beta \mu \dot{P}_{0}&=&0,\label{cos2}
\ee
where $U=U_1+U_2$, and (2) equations of $\rmd p^a(e_i(v))/\rmd \t=\cdots$ and $\rmd \theta^a(e_i(v))/\rmd \t=\cdots$ with $a\neq i$: Left-hand sides of these 12 equations are zero. We explicitly check that the ansatz also reduces their right-hand sides to zero, so that these equations are trivial. Note that this check and Eqs.\eqref{cos1} and \eqref{cos2} are nontrivial since they involve brute-force computation of right-hand sides of \eqref{eom0} or poisson bracket in \eqref{hamiltonhp} in the full theory before reducing with the ansatz. Detailed computations and Mathematica files are given in \cite{github}. See also \cite{Kaminski:2020wbg} a recent more abstract argument about obtaining Eqs.\eqref{cos1} and \eqref{cos2} from Eqs.\eqref{hamiltonhp}. 

For the scalar field, \eqref{hamiltonphipi} reduces to 
\be
P_0^{3/2} \dot{\phi}_0 -\pi_0=0,\qquad {P_0}^{3/2} {U}'(\phi_0)=-2\dot{\pi}_0.\label{cos3}
\ee
In the following discussion, we set $U(\phi_0)$ to be the Starobinsky inflationary potential
\be
U(\phi_0)=\frac{3 m^2 }{ \kappa }\left[1-\exp \left(-\sqrt{\frac{\kappa }{3}} \phi_0 \right)\right]^2
\ee
where $m$ is the mass parameter.

The physical Hamiltonian ${\bf H}$ is conserved by the time evolution governed by Eqs.\eqref{cos1} - \eqref{cos3},
\be
\frac{\bf H}{|V(\gamma)|}=C_v&=&-\mu^3\Bigg( \frac{3}{ \beta ^2 \kappa \mu^2}  P_0^{1/2} \sin ^2(\beta  \mu  K_0) \left[-\beta ^2+\left(\beta ^2+1\right) \cos (2 \beta  \mu  K_0)+1\right]\nonumber\\
&&-\frac{1}{2 \kappa }  P_0^{3/2} (4 \Lambda +\kappa  U(\phi_0))-\frac{\pi_0^2}{2P_0^{3/2}}\Bigg).\label{H=|C|}
\ee
The dust density $\rho_{dust}$ relates to ${\bf H}$ by (recall Eqs.\eqref{rho111} and \eqref{cc111})
\be
\rho_{dust}=-\frac{C_v}{P_0^{3/2}\mu^3}.\label{rhodustandH}
\ee

$\dot{P}_0=0$ gives the bounce, the matter density $\rho=\rho_{dust} + \rho_s +\rho_\L=\rho_{dust}+\frac{\pi_0^2}{2 P_0^3}+\frac{1}{2} U(\phi_0)+\frac{2 \Lambda }{\kappa }$ at the bounce is the critical density: 
\be
\rho_c=\frac{3}{2\b^2(\b^2+1)\kappa P_0(\text{bounce})\mu^2}.
\ee

When $\beta \mu K_{0}\ll1$, by $\sin(\beta \mu K_{0})\simeq \beta \mu K_{0}$, Eqs.\eqref{cos1} - \eqref{cos3} reduce to the classical cosmological dynamics:
\be
&8 P_0^{5/2} \dot{K}_0+4 K_0^2 P_0^2+\kappa  \pi_0^2=P_0^3 (4 \Lambda +\kappa  U(\phi_0)),\quad 2 K_0 \sqrt{P_0}=\dot{P}_0,&\label{clcos1}\\
&\pi_0=P_0^{3/2} \dot{\phi}_0,\quad P_0^{3/2} U'(\phi_0)+2 \dot{\pi}_0 =0,&\label{clcos2}
\ee
Eqs.\eqref{clcos2} give the classical EOM of $\phi_0$
\be
2\ddot{\phi}_0+{6 H \dot{\phi}_0}+U'(\phi_ 0)=0,\quad \text{where}\  H=\frac{\dot{P}_0}{2P_0}\ \text{is the Hubble parameter w.r.t.}\ \t. \label{clcos3}
\ee
Eq.\eqref{clcos1} can reduce to 
\be
\mathcal{H}^{\prime}=-\frac{4 \pi G}{3} P_0(\rho+3 \cp),\quad \text{where}\ \ch=\frac{P_0'}{2P_0}\ \text{is the Hubble parameter w.r.t.}\ \eta\label{clcosH}
\ee
where $\eta$ is the conformal time ($\rmd\eta=\frac{1}{\sqrt{P_0}}\rmd \t$)
\be
f'=\sqrt{P_0}\dot{f}.
\ee
$\rho, \cp$ are total matter density and pressure (including cosmological constant):
\be
\rho&=&\rho_{dust} + \rho_s +\rho_\L, 
\\
\rho_{dust} &=& -\frac{2 \Lambda }{\kappa }+\frac{6 K_0^2}{\kappa  P_0}-\frac{\pi_0^2}{2 P_0^3}-\frac{1}{2} U(\phi_0) = -\frac{2 \Lambda }{\kappa }+\frac{6 \mathcal{H}^2}{\kappa  P_0}-\frac{(\phi_0')^2}{2 P_0}-\frac{1}{2} U(\phi_0), \\
\rho_s &=& \frac{\pi_0^2}{2 P_0^3}+\frac{1}{2} U(\phi_0) = \frac{(\phi_0')^2}{2 P_0}+\frac{1}{2} U(\phi_0), \\
\rho_\L&=&\frac{2 \Lambda }{\kappa },\\
\cp&=&- \frac{2 \Lambda }{\kappa }+\frac{\pi_0^2}{2 P_0^3}-\frac{1}{2} U(\phi_0) = -\frac{2 \Lambda }{\kappa }-\frac{1}{2} U(\phi_0) +\frac{(\phi_0')^2}{2 P_0}. 
\ee
We have the relation
\be
\mathcal{H}^{2}=\frac{8 \pi G}{3} P_0 \rho.
\ee

\begin{figure}[t]
\begin{center}
  \includegraphics[width = 0.5\textwidth]{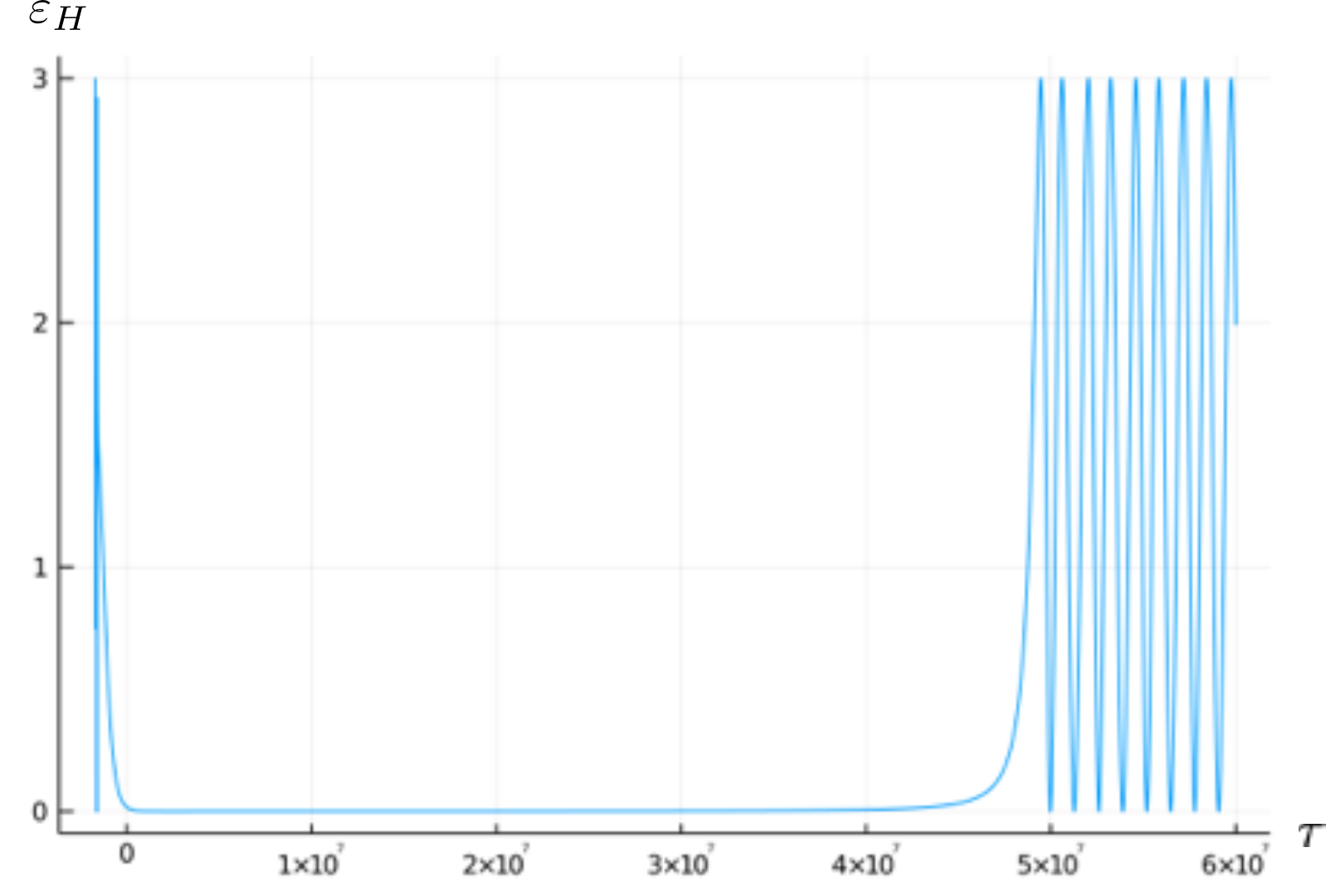} 
    \caption{The inflation is in the period with $0<\eps_H<1$ (before $t=5\times 10^7$). $\t_{pivot}=0$ is the pivot time. Parameters in this solution is $m=2.44\times10^{-6}m_P,\ H(\t_{pivot})=1.21\times 10^{-6}m_P$, $\phi_0(\t_{pivot})=1.07 m_P$, $\pi_0(\t_{pivot})=-5.03\times 10^{-9}m_P^2$, $P_0(\t_{pivot})=1$, $\L=0$. For very small $\mu$, the difference in $\eps_H$ is negligible between solutions of \eqref{cos1} - \eqref{cos3} and of \eqref{clcos1} - \eqref{clcos2}. }
  \label{inflation0}
  \end{center}
\end{figure}

In order to demonstrate the inflation from Eqs.\eqref{cos1} - \eqref{cos3} or classical counter-parts \eqref{clcos1} and \eqref{clcos2}, we define the slow-roll parameters
\be
\eps_H:=-\frac{\dot{H}}{H^{2}}, \quad\delta_H:=\frac{\ddot{H}}{\dot{H} H}. 
\ee
The inflation corresponds to $0<\eps_H<1$. Figure \ref{inflation0} plots $\eps_H$ of a solution and demonstrates the inflation. When practically solving EOMs \eqref{cos1} - \eqref{cos3} or other versions of EOMs to be discussed later, we use values of dynamical variables at pivot time $\t_{pivot}$ to uniquely determine the solution. We require that the dynamics at $\t_{pivot}$ and later must be well approximated by the classical dynamics of cosmology Eqs.\eqref{clcos1}, \eqref{clcos2}, and \eqref{clcos3} (this requirement is always fulfilled by our model as discussed later). The values of ${\phi}_0(\t_{pivot})$ and $H(\t_{pivot})$, as well as the value of the parameter $m$ in $U(\phi_0)$, are determined by the observational values of $A_s$ and $n_s$ (here we use the same data as in \cite{Giesel:2020raf})
\be
A_s=2.10\time 10^{-9},\quad n_s=0.96,
\ee
where $A_s$ is the amplitude of the scalar power spectrum at the pivot mode $k_{pivot}=0.002\, \mathrm{Mpc}^{-1}$, and $n_s$ is the spectral index of scalar perturbations. Following the procedure described in \cite{Ashtekar:2016wpi}, we obtain that
\be
 H(\t_{pivot})=1.21\times 10^{-6}m_P,\quad \phi_0(\t_{pivot})=1.07 m_P,\quad m=2.44\times10^{-6}m_P.
\ee
In principle this derivation is based on zero dust density, but numerical errors in the above numbers give tiny but nonzero dust density. In this paper, we work with nonzero dust density, but we always consider the dust density to be very small in numerical studies. Classically $K_0(\t_{pivot})=H(\t_{pivot})\sqrt{P_0(\t_{pivot})}$, so
\be
K_0(\t_{pivot})=1.21\times 10^{-6}m_P\sqrt{P_0(\t_{pivot})}.
\ee 
The values of ${\phi}_0(\t_{pivot})$ and $H(\t_{pivot})$ determine $\dot{\phi}_0(\t_{pivot})$ by Eq.\eqref{clcos3} ($\ddot{\phi}_0$ is negligible by the slow-roll approximation), and further determine $\pi_0(\t_{pivot})$ by Eq.\eqref{cos3} up to a choice of $P_0(\t_{pivot})$,
\be
\pi_0(\t_{pivot})=-5.03\times 10^{-9}m_P^2\, P_0(\t_{pivot})^{3/2}.
\ee 
$P_0(\t_{pivot})$ is not determined since $P_0$ is the square of scale factor which is defined up to rescaling. 

We are going to take the above values of ${\phi}_0(\t_{pivot})$, $K_0(\t_{pivot})$, and $\pi_0(\t_{pivot})$ to determine the cosmological dynamics, while $P_0(\t_{pivot})$ is left undetermined. The classical cosmological dynamics (of $H$ and $\phi_0$) is free of the ambiguity since Eqs.\eqref{clcos1} and \eqref{clcos2} are invariant under the constant rescaling 
\be
P_0(\t)\to \a P_0(\t),\quad K_0(\t)\to \a^{1/2}K_0(\t),\quad \pi_0(\t)\to\a^{3/2}\pi_0(\t).
\ee 
But this invariance is broken by Eqs.\eqref{cos1} and \eqref{cos2} due to the length scale $\mu$. Consequently, the dynamics of Eqs.\eqref{cos1} and \eqref{cos2} on the fixed spatial lattice is ambiguous due to the dependence on the choice of $P_0(\t_{pivot})$. In particular, the critical density $\rho_c$ at the bounce is ambiguous, and may be even very small if $P_0(\t_{pivot})$ is large. The bounce occurring at low density is not physically sound. This is considered as a problem of the cosmological dynamics from LQG on the fixed spatial lattice.

\section{Dynamical Lattice Refinement}\label{Dynamical Lattice Refinement}

\subsection{Motivation}\label{Motivation Lattice Refinement}

The cosmological dynamics described above coincides with the $\mu_0$-scheme effective dynamics of LQC. However, It is not the popular scheme in LQC. The preferred scheme in LQG is the improved dynamics, or namely the $\bar{\mu}$-scheme, in which $\mu$ is not a constant but set to be dynamical $\mu\to{\mu}(\t)=\sqrt{\Delta/P_0(\t)}$, in other words, the spatial lattice changes during the time evolution. The $\bar{\mu}$-scheme has the advantage that the critical density $\rho_c=\rho(\text{bounce})$ at the bounce only depends on constants $\kappa,\Delta,\b$ and is Planckian if $\Delta\sim\ell_P^2$, in contrast to the $\mu_0$-scheme where
\be
\rho_c=\frac{3}{2\b^2(\b^2+1)\kappa P_0(\text{bounce})\mu^2}.
\ee   
$P_0(\text{bounce})$ depends on the initial/final condition, e.g. $P_0(\t_{pivot})$, which does not guarantee $\rho_c$ to be Planckian.  

It turns out that the existence of inflation leads to a difficulty if the spatial lattice $\g$ is fixed. The reason is the following: First of all, the path integral \eqref{Agg} and the approximation to classical gravity on the continuum rely on 2 requirements: 

\begin{itemize} 

\item[(1)] As a key step in deriving the EOMs \eqref{hamilton}, the semiclassical approximation of $\langle\tilde{\psi}^\hbar_Z|\hat{\bf H}|\tilde{\psi}^\hbar_Z\rangle$ in \eqref{semiclassH} uses the expansion of volume operator as in \eqref{expandvolume}. This expansion requires $p^2\gg t$ or 
\be
4\mu^4 P_0^2\gg\b^2\ell_P^2a^2=\b^2\ell_P^4/t.\label{lbound}
\ee 

\item[(2)] $\theta^a(e)$ has to be sufficiently small in order to approximate the classical theory on the continuum. Namely, the background EOMs \eqref{cos1} and \eqref{cos2}, where $\beta \mu K_{0}=\theta(e)$ has to be small enough to validate $\sin(\theta)\simeq\theta$ and reduce these EOMs to the ones in classical cosmology.

\end{itemize} 
However the requirement (1) can contradict the requirement (2) if we fix the lattice $\g$ throughout the cosmological evolution including the inflation. Indeed, if we construct the lattice $\g$ such that $\mathring{p}^a(e),\mathring{\theta}^a(e)$ satisfying the requirement (1) before the inflation, during the inflation, the classical cosmology gives $P_0=e^{2\chi (t_f-t_i)}\mathring{P}_0$ and $K_0=\chi e^{\chi (t_f-t_i)}\mathring{P}_0^{1/2}$ ($t_i$ or $t_f$ is the time of starting/ending the inflation, and $\mathring{P}_0$ is determined by $\mathring{p}^a(e)$). the classical solution with large $K_0$ cannot approximately satisfy \eqref{cos1} - \eqref{cos3}, unless we set $\mu$ to be extremely small. But an extremely small $\mu$ makes the requirement (1) hard to be satisfied at the early time. For example, if we set $e^{\chi (t_f-t_i)}\sim 10^{24}$ and {$\chi\sim 10^{-6}l_P^{-1} $ ($l_P^2=\hbar G=\frac{1}{16\pi}\ell_P^2$)}, we have to set $\mu$ extremely small such that $\b \mu\mathring{P}_0^{1/2}\sim 10^{-20}l_P$ or less to validate $\sin(\beta \mu K_{0})\simeq\beta \mu K_{0}$ and fulfill the requirement (2). However by the requirement (1) $\mu^4 \mathring{P}_0^2\sim 10^{-80}l_P^4/\b^4 \gg\frac{1}{4}\b^2\ell_P^4/t\ \Leftrightarrow\ t\gg\frac{1}{4(16\pi)^2}10^{80}\b^6$ which violate the semiclassical limit $t\to0$ unless $\b$ is extremely small (or $\b\to 0$ even much faster than $t\to0$). For not so small $\b$, setting $\mu$ extremely small results in that $\mathring{p}^a(e)$ violate the requirement (1) before the inflation. This would not be in favor since the semiclassical approximation is expected to be valid before and during the inflation.

A way to resolve the tension between the requirements (1) and (2) is to refine the spatial lattice during the time evolution, as suggested by the $\bar{\mu}$-scheme in LQC. We are going to ask the lattice to be finer at late time while coarser at early time, so that we have a small enough $\mu$ to satisfy the requirement (2) in the inflation, while having a large enough $\mu$ to satisfy the requirement (1) at early time before the inflation. 


\subsection{Transition Amplitude with Dynamical Lattice Refinement}\label{The Linear Map}

Firstly as the setup, we write the lattice variables $\theta^a(e),p^a(e),\phi(v),\pi(v)$ generally as below
\be
\theta^{a}\left(e_{I}(v)\right)=\mu\left[\beta K_{0} \delta_{I}^{a}+\mathcal{X}^{a}\left(e_{I}(v)\right)\right], && p^{a}\left(e_{I}(v)\right)=\frac{2 \mu^{2}}{\beta a^{2}}P_{0}\left[ \delta_{I}^{a}+\mathcal{Y}^{a}\left(e_{I}(v)\right)\right],\label{perturb1}\\
\phi(v)=\phi_0+\delta\varphi(v),&& \pi(v)=\mu^3\lt[\pi_0+\delta\pi(v)\rt]\label{perturb2}
\ee
where $P_0,K_0,\phi_0,\pi_0$ are the homogeneous and isotropic DOFs. $\mathcal{Y}^{a}\left(e_{I}(v)\right), \mathcal{X}^{a}\left(e_{I}(v)\right),\cw(v),\cz(v)$ are DOFs beyond the homogeneous and isotropic sector. Their dimensions are $\mathcal{X}^{a},\delta\varphi\sim (\text{length})^{-1}$, $\delta\pi\sim (\text{length})^{-2}$, and $\mathcal{Y}^{a}\sim (\text{length})^{0}$. We introduce a vector $V^\rho(v)$ as short-hand notation for non-homogeneous and non-isotropic DOFs:
\be
V^\rho(v)=\left(\mathcal{Y}^{a}\left(e_{I}(v)\right), \mathcal{X}^{a}\left(e_{I}(v)\right),\delta\pi(v),\delta\phi(v)\right)^{T},\quad \rho=1,\cdots 20
\ee
The dictionary between $V^\rho(v)$ and $\cx^a(e_I(v)),\cy^a(e_I(v))$ is given below:
\be
V^1=\cy^1(e_1),\quad &V^2=\cy^2(e_2),&\quad V^3=\cy^3(e_3)\nonumber\\
V^4=\cy^2(e_1),\quad &V^5=\cy^3(e_1),&\quad V^6=\cy^3(e_2)\nonumber\\
V^7=\cy^1(e_2),\quad &V^8=\cy^1(e_3),&\quad V^9=\cy^2(e_3)\nonumber\\
V^{10}=\cx^1(e_1),\quad &V^{11}=\cx^2(e_2),&\quad V^{12}=\cx^3(e_3)\nonumber\\
V^{13}=\cx^2(e_1),\quad &V^{14}=\cx^3(e_1),&\quad V^{15}=\cx^3(e_2)\nonumber\\
V^{16}=\cx^1(e_2),\quad &V^{17}=\cx^1(e_3),&\quad V^{18}=\cx^2(e_3)\nonumber\\
V^{19}=\delta\pi,\quad &V^{20}=\delta\varphi.&
\ee
Eqs.\eqref{perturb1} and \eqref{perturb2} do not lose any generality of the lattice variables. All  $\theta^a(e),p^a(e),\phi(v),\pi(v)$ in $\mathscr{P}_\g$ can be expressed as Eqs.\eqref{perturb1} and \eqref{perturb2}, while $V^\rho$ may be large for phase space points far away from being homogeneous and isotropic. When we discuss cosmological perturbation theory, we are going to assume $V^\rho$ to be small and linearize the EOMs.

Because we work with cubic lattice $\g$ with constant coordinate spacing $\mu$ and periodic boundary, it is convenient to make the following lattice Fourier transformation:
\be
V^\rho(\tau,v)=V^{\rho}(\tau, \vec{\sigma})=\frac{1}{L^3}\sum_{\vec{k}\in(\frac{2\pi }{L}\Z)^3,\ |k^I|\leq \frac{\pi}{\mu}}  e^{i \vec{k} \cdot \vec{\sigma}} \tilde{V}^{\rho}(\tau, \vec{k}), \quad {\sigma}^I \in \mu \mathbb{Z}.\label{fourierksig}
\ee
where both $\vec{\sig}$ and $\vec{k}$ have periodicity ${\sig}^I\sim \sig^I+L$ and $k^I\sim k^I+\frac{2\pi}{\mu}$ ($I=1,2,3$), so the sum $\sum_{\vec{k}}$ has the UV cut-off $|k^I|\leq \frac{\pi}{\mu}$ ($L/\mu$ is assumed to be an interger). Eq.\eqref{fourierksig} can also be expressed as below when we write $\vec{k}=\frac{2\pi }{L}\vec{m}$, ${\sig}^I=\mu{n}^I$, and $L=\mu N$ where $N$ is the total number of vertices along each direction:
\be
V^\rho(\tau,v)=V^{\rho}(\tau, \vec{n})=\frac{1}{L^3}\sum_{\vec{m}\in\Z(N)^3} \prod_{I=1}^3 e^{\frac{2\pi i}{N} {m}^I {n}^I} \tilde{V}^{\rho}(\tau, \vec{m}).\label{fouriermn}\\
\tilde{V}^{\rho}(\tau, \vec{m})=\frac{L^3}{N^3}\sum_{\vec{n}\in\Z(N)^3}   \prod_{I=1}^3 e^{-\frac{2\pi i}{N} {m}^I {n}^I} {V}^{\rho}(\tau, \vec{n})
\ee
where $\Z(N)$ are integers in $[-N/2,N/2-1]$ if $N$ is even, or in $[-(N-1)/2,(N-1)/2]$ if $N$ is odd. 

We may absorb the homogeneous and isotropic DOFs in zero-modes and define
\be
\Phi^\rho(v)&=&\lt(\frac{\b a^2}{2\mu^2}p^{a}\left(e_{I}(v)\right),\frac{1}{\mu}\theta^{a}\left(e_{I}(v)\right),\phi(v),\frac{1}{\mu^3}\pi(v)\rt),\\
&\equiv&\lt(P^{a}_I\left(v\right),C^{a}_I\left(v\right),\phi(v),\Pi(v)\rt)\\
\Phi^\rho(\t,v)&=&\Phi^{\rho}(\tau, \vec{n})=\frac{1}{L^3}\sum_{\vec{m}\in\Z(N)^3} \prod_{I=1}^3 e^{\frac{2\pi i}{N} {m}^I {n}^I}  \tilde{\Phi}^{\rho}(\tau, \vec{m}),\label{fourierZmn}
\ee
where $ \tilde{\Phi}^{\rho}(\tau, \vec{0})$ contains the homogeneous and isotropic DOFs $(P_0(\t),\b K_0(\t),\phi_0(\t),\pi_0(\t))$, e.g.
\be
P^a_I(\t,\vec{0})=P_0(\t)\lt[\delta^a_IL^3+V^{\rho=1,\cdots,9}(\t,\vec{0})\rt]
\ee

We propose a linear map $\ci_{\g_i,\g_{i-1}}:\,\ch_{\g_{i-1}}\to\ch_{\g_i}$ to map states on the coarser lattice $\g_{i-1}$ to a finer lattice $\g_i$. The total number of vertices in $\g_i,\g_{i-1}$ are $N^3_i,N^3_{i-1}$, and here $N_i>N_{i-1}$. $\ci_{\g_i,\g_{i-1}}$ is going to be inserted in the middle of the Hamiltonian evolution by $\hat{\bf H}$, to refine the lattice during the evolution and relate the dynamics on different lattices.  

The formal definition of $\ci_{\g_i,\g_{i-1}}$ is given as the following: Firstly we denote the lattice Fourier transformation Eq.\eqref{fourierZmn} by 
\be
\mathscr{F}_\g:\{ \tilde{\Phi}^{\rho}(\tau, \vec{m})\}_{\vec{m}\in\Z(N)^3}\mapsto \{ \Phi^{\rho}(\tau, \vec{n})\}_{\vec{n}\in\Z(N)^3}, 
\ee
At a given instance $\t_i$ where we apply $\ci_{\g_i,\g_{i-1}}$ to change $\g_{i-1}$ to $\g_i$, The Fourier transformations on $\g_i$ and $\g_{i-1}$ are given respectively by
\be
 \Phi^{\rho}(\tau_i, \vec{n})_{\g_i}&=&\mathscr{F}_{\g_i}   \tilde{\Phi}^{\rho}(\tau_i, \vec{m})_{\g_i}=\frac{1}{L^3}\sum_{\vec{m}\in\Z(N_i)^3} \prod_{I=1}^3 e^{\frac{2\pi i}{N_i} {m}^I {n}^I}   \tilde{\Phi}^{\rho}(\tau_i, \vec{m})_{\g_i} ,\label{fouriernp1}\\
 \Phi^{\rho}(\tau_i, \vec{n})_{\g_{i-1}}&=&\mathscr{F}_{\g_{i-1}}   \tilde{\Phi}^{\rho}(\tau_i, \vec{m})_{\g_{i-1}}=\frac{1}{L^3}\sum_{\vec{m}\in\Z(N_{i-1})^3} \prod_{I=1}^3 e^{\frac{2\pi i}{N_{i-1}} {m}^I {n}^I} \tilde{\Phi}^{\rho}(\tau_i, \vec{m})_{\g_{i-1}} \label{fouriernp2}.
\ee
where we have added the label $\g_i$ to $ \Phi^{\rho}(\tau_i, \vec{n})$ to manifest its corresponding lattice. Recall that the coherent states $\Psi^\hbar_{[Z]}$ are labelled by $Z=(g,z)$ depending on both $\mu$ and $\Phi^\rho$,
\be
\Psi^\hbar_{[Z(\g_{i})]}&=&\Psi^\hbar_{[Z(\mu_i,\Phi_{\g_i})]}=\Psi^\hbar_{[Z(\mu_i,\,\mathscr{F}_{\g_i}\tilde{\Phi}_{\g_i})]},\\
\Psi^\hbar_{[Z(\g_{i-1})]}&=&\Psi^\hbar_{[Z(\mu_{i-1},\Phi_{\g_{i-1}})]}=\Psi^\hbar_{[Z(\mu_{i-1},\,\mathscr{F}_{\g_{i-1}}\tilde{\Phi}_{\g_{i-1}})]}.
\ee
Given $\tilde{\Phi}_{\g_{i-1}}$ on the coarser lattice $\g_{i-1}$, we determines $\tilde{\Phi}_{\g_i}=\tilde{\Phi}_{\g_i}[\tilde{\Phi}_{\g_{i-1}}]$ on the finer lattice $\g_i$ by simple relations
\be
\tilde{\Phi}^{\rho}(\tau_{i}, \vec{m})_{\g_i}&=&\tilde{\Phi}^{\rho}(\tau_{i}, \vec{m})_{\g_{i-1}},\quad \vec{m}\in\Z(N_{i-1})^3, \label{PhiPhiglue1}\\
\tilde{\Phi}^{\rho}(\tau_{i}, \vec{m})_{\g_i}&=&0,\qquad \vec{m}\in\Z(N_{i})^3\setminus\Z(N_{i-1})^3.\label{PhiPhiglue}
\ee
Given $\tilde{\Phi}_{\g_i}[\tilde{\Phi}_{\g_{i-1}}]$ determined by $\tilde{\Phi}_{\g_{i-1}}$ with the above relations, we define the linear embedding map $\ci_{\g_i,\g_{i-1}}$: $\ch_{\g_{i-1}}\to\ch_{\g_i}$ by
\be
\ci_{\g_i,\g_{i-1}}&=&\int\rmd Z(\g_{i-1})\frac{\big|{\Psi}^\hbar_{[Z(\g_i,\g_{i-1})]}\big\rangle \big\langle{\Psi}^\hbar_{[Z(\g_{i-1})]}\big|}{\big\|{\psi}^\hbar_{Z(\g_i,\g_{i-1})}\big\|\,\big\|{\psi}^\hbar_{Z(\g_{i-1})}\big\|},\\
Z(\g_i,\g_{i-1})&=&Z(\mu_i,\,\mathscr{F}_{\g_i}\tilde{\Phi}_{\g_i}[\tilde{\Phi}_{\g_{i-1}}]),\quad Z(\g_{i-1})=Z(\mu_{i-1},\,\mathscr{F}_{\g_{i-1}}\tilde{\Phi}_{\g_{i-1}})
\ee
Given the parameters $\mu_i,\mu_{i-1}$, $Z(\g_i,\g_{i-1})$ is determined by $Z(\g_{i-1})$ since $\tilde{\Phi}_{\g_i}[\tilde{\Phi}_{\g_{i-1}}]$ is determined by $\tilde{\Phi}_{\g_{i-1}}$. 
$Z(\g_i)$ is a representative in the gauge equivalence class $[Z(\g_i)]$. $\|\psi^\hbar_Z\|=\|\psi^\hbar_{Z^u}\|$ is SU(2) gauge invariant. $\ci_{\g_i,\g_{i-1}}$ is densely defined on $\ch_{\g_{i-1}}$ (see Appendix \ref{Properties of I Isometry}).

We assume the Hamiltonian evolution on the fixed lattice $\g_{i-1}$ to occur in the time interval $[\t_{i-1},\t_{i}]$, then we apply $\ci_{\g_i,\g_{i-1}}$ at $\t_i$ at the end of this Hamiltonian evolution to map the state to a refined lattice $\g_i$, followed by the Hamiltonian evolution on $\g_i$ in the time interval $[\t_i,\t_{i+1}]$. By iteration, we define the transition amplitude with dynamically changing lattice:
\be
\ca_{[Z],[Z']}(\ck)&=&\langle\Psi^\hbar_{[Z]}|e^{-\frac{i}{\hbar}\hat{\bf H}(T-\t_m)}\ci_{\g_m,\g_{m-1}}\cdots e^{-\frac{i}{\hbar}\hat{\bf H}(\t_{i+1}-\t_i)}\ci_{\g_i,\g_{i-1}} e^{-\frac{i}{\hbar}\hat{\bf H}(\t_{i}-\t_{i-1})}\cdots|\Psi^\hbar_{[Z']}\rangle\nonumber\\
&=&\big\|\psi^\hbar_{Z}\big\|\,\big\|\psi^\hbar_{Z'}\big\|\int\prod_{i=0}^m\rmd Z(\g_{i-1})\prod_{i=0}^{m+1}\frac{\Big\langle{\Psi}^\hbar_{[Z(\g_i)]}\Big|e^{-\frac{i}{\hbar}\hat{\bf H}(\t_{i+1}-\t_i)}\Big|{\Psi}^\hbar_{[Z(\g_{i},\g_{i-1})]}\Big\rangle}{\big\|{\psi}^\hbar_{Z(\g_i)}\big\|\,\big\|{\psi}^\hbar_{Z(\g_{i},\g_{i-1})}\big\|},\\
&=&\big\|\psi^\hbar_{Z}\big\|\,\big\|\psi^\hbar_{Z'}\big\|\int\prod_{i=0}^m\rmd Z(\g_{i-1})\prod_{i=0}^{m+1}\int  \mathrm{d} Z^{(i)}\mathrm{d} u^{(i)} \, \nu[Z^{(i)}]\, e^{S[Z^{(i)}, u^{(i)}] / t}\label{cagg}
\ee
where $[Z(\g_{m+1})]\equiv[Z]$ and $[Z(\g_0,\g_{-1})]\equiv[Z']$. The initial time is $\t_0$ and the final time is $T$. Each factor in the integrand is the Hamiltonian evolution from $\t_{i}$ to $\t_{i+1}$, and has been expressed as a path integral as in \eqref{Agg}.  The Initial and final conditions of $Z^{(i)}$ are $Z(\g_i)$ and $Z(\g_{i},\g_{i-1})^{u^{(i)}}$, and integrating $u^{(i)}$ implements the SU(2) gauge invariance. The path integral give the EOMs \eqref{hamilton} in every time interval $[\t_{i},\t_{i+1}]$. $\ca_{[Z],[Z']}(\ck)$ can be understood as a discrete path integral formula defined on the spacetime lattice $\ck$ as in Figure \ref{lattice}. $\ca_{[Z],[Z']}(\ck)$ is similar to spinfoam models which are defined on spacetime lattices. The set of $N_i$ or $\mu_{i}={L}/{N_i}$ $(i=0,\cdots,m)$ is determined by the choice of spacetime lattice $\ck$.

Intuitively, in the semiclassical time evolution in each $[\t_{i},\t_{i+1}]$ the initial data $[Z(\g_{i},\g_{i-1})]$ uniquely determines the final data $[Z(\g_i)]$ \cite{Han:2020chr}. The map $\ci_{\g_{i+1},\g_{i}}$ glues the final data $[Z(\g_i)]$ to the initial data $[Z(\g_{i+1},\g_{i})]$ for the time evolution in $[\t_{i+1},\t_{i+2}]$, while the data $[Z(\g_i)]$ and $[Z(\g_{i+1},\g_{i})]$ share the same infrared Fourier modes $\tilde{\Phi}(m)$ with $\vec{m}\in\Z(N_i)^3$. The gluing at all $\t_i$ connects the semiclassical trajectories of Hamiltonian evolutions in all intervals $[\t_{i},\t_{i+1}]$, and make the semiclassical trajectories of the entire time evolution from $\t_0$ to $T$.

Indeed, as is shown in Appendix \ref{Properties of I Equations of Motion}, when we study the variational principle of $\ca_{[Z],[Z']}(\ck)$ by taking into account variations of $Z(\g_{i-1})$ $(i=0,\cdots,m)$ in the definitions of $\ci_{\g_i,\g_{i-1}}$, these variations give some EOMs which are automatically satisfied approximately by solutions of EOMs in every $[\t_{i},\t_{i+1}]$, at least for the homogenous and isotropic cosmological evolution and perturbations. The approximation is up to an arbitrary small error of $O(1/N_i)$ as $N_i$ is arbitrarily large. Connecting by $\ci_{\g_{i+1},\g_{i}}$ the semiclassical trajectories in all $[\t_{i},\t_{i+1}]$ makes the solutions satisfying approximately the variational principle of $\ca_{[Z],[Z']}(\ck)$ up to arbitrarily small error of $O(1/N_i)$. The resulting solutions describe the semiclassical dynamics on the spacetime lattice $\ck$ which relates to the choice of the sequence of spatial lattices $\g_{i=0,\cdots,m}$ and corresponding $\mu_{i=0,\cdots,m}$.


When the initial state $\Psi^\hbar_{[Z']}$ is labelled by the homogeneous and isotropic $[Z']$, both the Hamiltonian evolution and $\ci_{\g_{i},\g_{i-1}}$ preserve the homogeneity and isotropy. $\ci_{\g_{i+1},\g_{i}}$ glues the final data $[Z(\g_i)]$ to the initial data $[Z(\g_{i+1},\g_{i})]$ sharing the same zero modes $P_0,K_0,\phi_0,\pi_0$. 
When the initial state $\Psi^\hbar_{[Z']}$ has cosmological perturbations $V^\rho$, at each step of the lattice refinement, $\ci_{\g_i,\g_{i-1}}$ identifies
\be
&&\tilde{V}^\rho(\t_{i},\vec{m})_{\g_{i}}=\tilde{V}^\rho(\t_{i},\vec{m})_{\g_{i-1}},\quad \vec{m}\in\Z(N_{i-1})^3,\label{VVidentification0}\\
&&\tilde{V}^\rho(\t_{i},\vec{m})_{\g_{i}}=0,\quad \vec{m}\in\Z(N_{i})^3\setminus\Z(N_{i-1})^3,\label{VVidentification01}
\ee
by Eqs.\eqref{PhiPhiglue1} and \eqref{PhiPhiglue}. This prescription freezes ultraviolet modes on the finer lattice, while identifying infrared modes with the ones on the coarser lattice. Our study of cosmological perturbations only focuses on long wavelength perturbations, so Eq.\eqref{VVidentification} is sufficient for our purpose.



$\ca_{[Z],[Z']}$ has the limitation that $\ci_{\g_i,\g_{i-1}}$ only identifies the infrared modes when refining the lattice, while the ultraviolet modes are lost. When we discussion cosmological perturbations $\tilde{V}^\rho(\t,\vec{m})$, the spatial momentum $\vec{k}=\frac{2\pi}{L}\vec{m}$ is conserved, then the EOMs from $\ca_{[Z],[Z']}$ can only describe the dynamics of the modes $\tilde{V}^\rho(\t,\vec{m})$ with $\vec{m}\in\Z(N_0)^3$, i.e. their momenta are bounded by the ultraviolet cut-off on the coarsest lattice $\g_0$ where the initial state is placed, while these modes are infrared at late time in the sense that their momenta are much smaller than the ultraviolet cut-off on the refined lattice. EOMs from $\ca_{[Z],[Z']}$ is not able to predict the dynamics of ultraviolet modes at late time if the initial state is placed at the early time. This feature suggests that $\ca_{[Z],[Z']}$ is likely to be an low energy effective theory, and the early-time dynamics on the coarser lattice is expected to be the coarse-grained model obtained from the full quantum dynamics by integrating out ultraviolet modes (beyond $\Z(N_0)^3$). 
In this work, we only focus on $\ca_{[Z],[Z']}$ understood as the low energy effective theory. 

The time evolution in $\ca_{[Z],[Z']}$ is not unitary. The numbers of DOFs are not equal between early time and late time, and there are ultraviolet modes at late time not predictable by the initial state at early time. But it is possible that the unitarity may be hidden by the coarse-graining, if we view the dynamics on the coarser lattice is the coarse-grain of the dynamics on the finer lattice. Ultimately in more complete treatment, quantum states and dynamics on the coarser lattice should contain the information of ultraviolet modes, although this information is still missing in our effective approach. Standard examples in the Wilsonian renormalization show that when integrating out ultraviolet modes, their effects are not lost but encoded in higher order and higher derivative interaction terms in the effective lagrangian. Here we have ignored the correction to $S[Z,u]$ from coarse-graining since we only focus on long wavelength modes. Our expectation regarding the unitarity in the effective theory is somewhat similar to the one proposed in \cite{Amadei:2019wjp}.

$\ca_{[Z],[Z']}(\ck)$ in Eq.\eqref{cagg} requires $\mu_{i=0,\cdots,m}$ is a monotonically decreasing sequence from early to late time. We can generalize the formulation by relaxing this requirement. When the spacetime lattice $\ck$ is such that at the instance $\t_j$, the lattice $\g_j$ in the future of $\t_j $ is coarser than the lattice $\g_{j-1}$ in the past, i.e. $\mu_j>\mu_{j-1}$, we insert $\ci_{\g_{j-1},\g_j}^\dagger$ in $\ca_{[Z],[Z']}(\ck)$ (recall that $\ci_{\g_{j-1},\g_j}:\ch_{\g_j}\to\ch_{\g_{j-1}}$ from the coarser lattice to the finer)
\be
\ca_{[Z],[Z']}(\ck)=\langle\Psi^\hbar_{[Z]}|e^{-\frac{i}{\hbar}\hat{\bf H}(T-\t_m)}\ci_{\g_m,\g_{m-1}}\cdots e^{-\frac{i}{\hbar}\hat{\bf H}(\t_{j+1}-\t_j)}\ci_{\g_{j-1},\g_{j}}^\dagger e^{-\frac{i}{\hbar}\hat{\bf H}(\t_{j}-\t_{j-1})}\cdots|\Psi^\hbar_{[Z']}\rangle
\ee
where $\ci_{\g_{j-1},\g_{j}}^\dagger:\ch_{\g_{j-1}}\to\ch_{\g_j}$ is defined by
\be
\ci_{\g_{j-1},\g_{j}}^\dagger&=&\int\rmd Z(\g_{j})\frac{\big|{\Psi}^\hbar_{[Z(\g_{j})]}\big\rangle \big\langle{\Psi}^\hbar_{[Z(\g_{j-1},\g_{j})]}\big|}{\big\|{\psi}^\hbar_{Z(\g_{j})}\big\|\,\big\|{\psi}^\hbar_{Z(\g_{j-1},\g_{j})}\big\|}.\\
Z(\g_{j-1},\g_{j})&=&Z(\mu_{j-1},\,\mathscr{F}_{\g_{j-1}}\tilde{\Phi}_{\g_{j-1}}[\tilde{\Phi}_{\g_{j}}]),\quad Z(\g_{j})=Z(\mu_{j},\,\mathscr{F}_{\g_{j}}\tilde{\Phi}_{\g_{j}}).
\ee
Inserting this expression of $\ci_{\g_{j-1},\g_{j}}^\dagger$ in $\ca_{[Z],[Z']}(\ck)$ and denoting $\widetilde{\Psi}^\hbar_{[Z]}={\Psi}^\hbar_{[Z]}/\|{\psi}^\hbar_{Z}\|$, we obtain  
\be
\ca_{[Z],[Z']}(\ck)
=\int \rmd Z(\g_j)\langle\Psi^\hbar_{[Z]}|\cdots e^{-\frac{i}{\hbar}\hat{\bf H}(\t_{j+1}-\t_j)}|\widetilde{\Psi}^\hbar_{[Z(\g_j)]}\rangle\langle\widetilde{\Psi}^\hbar_{[Z(\g_{j-1},\g_j)]}| e^{-\frac{i}{\hbar}\hat{\bf H}(\t_{j}-\t_{j-1})}\cdots|\Psi^\hbar_{[Z']}\rangle\nonumber
\ee
Then the path integral expression of general $\ca_{[Z],[Z']}(\ck)$ is 
\be
\ca_{[Z],[Z']}(\ck)&=&\big\|\psi^\hbar_{Z}\big\|\,\big\|\psi^\hbar_{Z'}\big\|\int\prod_{i=0}^m\rmd Z(\g_{i-1/i})\prod_{i=0}^{m+1}\Big\langle\widetilde{\Psi}^\hbar_{[Z^-(\g_i)]}\Big|e^{-\frac{i}{\hbar}\hat{\bf H}(\t_{i+1}-\t_i)}\Big|\widetilde{\Psi}^\hbar_{[Z^+(\g_{i})]}\Big\rangle,\\
&=&\big\|\psi^\hbar_{Z}\big\|\,\big\|\psi^\hbar_{Z'}\big\|\int\prod_{i=0}^m\rmd Z(\g_{i-1/i})\prod_{i=0}^{m+1}\int  \mathrm{d} Z^{(i)}\mathrm{d} u^{(i)} \, \nu[Z^{(i)}]\, e^{S[Z^{(i)}, u^{(i)}] / t}
\ee
Our notation is: $Z^+(\g_{i})=Z(\g_i,\g_{i-1})$, $Z^-(\g_{i-1})=Z(\g_{i-1})$ and $\rmd Z(\g_{i-1/i})=\rmd Z(\g_{i-1})$ when $\g_i$ is finer than $\g_{i-1}$, but $Z^+(\g_{i})=Z(\g_i)$, $Z^-(\g_{i-1})=Z(\g_{i-1},\g_i)$ and $\rmd Z(\g_{i-1/i})=\rmd Z(\g_{i})$ when $\g_i$ is coarser than $\g_{i-1}$. 

The semiclassical dynamics is still obtained by connecting semiclassical trajectories in $[\t_{i-1},\t_i]$. But we generalize from monotonically decreasing $\mu_{i=0,\cdots,m}$ to arbitrary sequence $\mu_{i=1,\cdots,m}$, in other words, we allow more general spacetime lattice $\ck$. For the cosmological perturbation theory, Eqs.\eqref{VVidentification0} and \eqref{VVidentification01} are generalized to be 
\be
&&\tilde{V}^\rho(\t_{i},\vec{m})_{\g_{i}}=\tilde{V}^\rho(\t_{i},\vec{m})_{\g_{i-1}},\quad \vec{m}\in\Z(\mathrm{Min}(N_i,N_{i-1}))^3,\label{VVidentification}\\
&&\tilde{V}^\rho(\t_{i},\vec{m})_{\g_{i}\ \text{or}\ \g_{i-1}}=0,\quad \vec{m}\in\Z(\mathrm{Max}(N_i,N_{i-1}))^3\setminus\Z(\mathrm{Min}(N_i,N_{i-1}))^3,\label{VVidentification1}
\ee
Eqs.\eqref{PhiPhiglue1} and \eqref{PhiPhiglue} are generalized similarly. The modes captured by the semiclassical dynamics correspond to the ones on the coarsest lattice: 
\be
\vec{m}\in\Z(\mathrm{Min}(\{N_i\}_{i=0,\cdots,m}))^3
\ee 
In general, the coarsest lattice is not necessarily at the initial time $\t_0$.

\subsection{Homogeneous and Isotropic Cosmological Dynamics on Dynamical Lattice}

We impose the initial state $\Psi^\hbar_{[Z']}$ labelled by the homogeneous and isotropic $[Z']$. The Hamiltonian evolution on the fixed lattice $\g_i$ determines the unique semiclassical trajectory from the initial data \cite{Han:2020chr}. The homogeneity and isotropy are preserved by the semiclassical dynamics. $\ci_{\g_{i+1},\g_{i}}$ glues the final data $[Z(\g_i)]$ to the initial data $[Z(\g_{i+1},\g_{i})]$ sharing the same zero-modes $P_0,K_0,\phi_0,\pi_0$. All non-zero Fourier modes vanishes. 

Given a choice of the spacetime lattice $\ck$, or equivalently a sequence of $\g_{i=0,\cdots,m}$ with $\mu_{i=0,\cdots,m}$, the following variables are continuous at each instance $\t_i$ where $\ci_{\g_{i},\g_{i-1}}$ is inserted:
\be
P_0(\t_i)_{\g_i}=P_0(\t_i)_{\g_{i-1}},&& K_0(\t_i)_{\g_i}=K_0(\t_i)_{\g_{i-1}},\label{pthetagamma10}\\
\phi_0(\t_i)_{\g_i}=\phi_0(\t_i)_{\g_{i-1}},&& \pi_0(\t_i)_{\g_i}=\pi_0(\t_i)_{\g_{i-1}}\label{pthetagamma1}
\ee
On the other hand, the semiclassical time evolution within $[\t_i,\t_{i+1}]$ is on a fixed spatial lattice $\g_i$ and is governed by
\be
&&\frac{4 \beta^{2}\left[-2 \mu_i^{2} \sqrt{P_{0}(\t)_{\g_i}} \dot{K}_{0}(\t)_{\g_i}+\sin ^{4}\left(\beta \mu_i K_{0}(\t)_{\g_i}\right)+\Lambda \mu_i^{2} P_{0}(\t)_{\g_i}\right]-\sin ^{2}\left(2 \beta \mu_i K_{0}(\t)_{\g_i}\right)}{\sqrt{P_{0}(\t)_{\g_i}}}\nonumber\\
&=&\kappa\b^2\mu_i^2\sqrt{P_0(\t)_{\g_i}}\lt[ \pi_0(\t)_{\g_i}^2 P_0(\t)_{\g_i}^{-3}- U\lt(\phi_0(\t)_{\g_i}\rt)\rt]\label{eommui1}\\
&&\sqrt{P_{0}(\t)_{\g_i}}\left[2 \beta^{2} \sin \left(2 \beta \mu_i K_{0}(\t)_{\g_i}\right)-\left(\beta^{2}+1\right) \sin \left(4 \beta \mu_i K_{0}(\t)_{\g_i}\right)\right]+2 \beta \mu_i \dot{P}_{0}(\t)_{\g_i}=0.\label{eommui2}\\
&&P_0(\t)_{\g_i}^{3/2} \dot{\phi}_0(\t)_{\g_i} -\pi_0(\t)_{\g_i}=0,\qquad {P_0}(\t)_{\g_i}^{3/2} {U}'\lt(\phi_0(\t)_{\g_i}\rt)=-2\dot{\pi}_0(\t)_{\g_i},\label{eommui3}
\ee
which add lattice labels to Eqs.\eqref{cos1} - \eqref{cos3}.
 
We assume every interval $[\t_i,\t_{i+1}]$ are sufficiently small, and every $N_i\gg 1$ and $N_i-N_{i-1}\sim O(1)$ ($\mu_i/\mu_{i-1}\sim 1$), we can approximate the sequence of $\mu_i$ by a smooth function $\mu(\t)$ (approximating the step function by a smooth function). Moreover, we make following approximations for time derivatives in Eqs.\eqref{eommui1} - \eqref{eommui3}
\be
\dot{P}_0(\t_{i})_{\g_i}\simeq\frac{{P}_0(\t_{i+1})_{\g_i}-{P}_0(\t_{i})_{\g_i}}{\t_{i+1}-\t_{i}},&&  
\dot{K}_0(\t_{i})_{\g_i}\simeq\frac{{K}_0(\t_{i+1})_{\g_i}-{K}_0(\t_{i})_{\g_i}}{\t_{i+1}-\t_{i}}\\
\dot{ \phi}_0(\t_{i})_{\g_i}\simeq\frac{{\phi}_0(\t_{i+1})_{\g_i}-{\phi}_0(\t_{i})_{\g_i}}{\t_{i+1}-\t_{i}},&& \dot{\pi}_0(\t_{i})_{\g_i}\simeq\frac{{\pi}_0(\t_{i+1})_{\g_i}-{\pi}_0(\t_{i})_{\g_i}}{\t_{i+1}-\t_{i}}
\ee
Since ${P}_{0}(\t),{K}_{0}(\t),{\pi}_{0}(\t),{\phi}_{0}(\t)$ are continuous at every $\t_i$ (see Eq.\eqref{pthetagamma1}), when we insert the above approximation in Eqs.\eqref{eommui1} - \eqref{eommui3} and assume ${P}_{0}(\t),{K}_{0}(\t),{\pi}_{0}(\t),{\phi}_{0}(\t),\mu(\t)$ to be smooth functions in $\t$, the resulting EOMs can be approximated by the following differential equations\footnote{Here the intervals $[\t_i,\t_{i+1}]$ has to be coarser than the steps $\delta\tau$ since the path integral formula \eqref{Agg} needs to be valid in each interval. $\delta\tau$ is arbitrarily small so that the intervals can be sufficiently small to validate the approximation.}:
\be
&&\frac{4 \beta^{2}\left[-2 {\mu}(\t)^{2} \sqrt{P_{0}(\t)} \dot{K}_{0}(\t)+\sin ^{4}\left(\beta {\mu}(\t) K_{0}(\t)\right)+\Lambda {\mu}(\t)^{2} P_{0}(\t)\right]}{\sqrt{P_{0}(\t)}}-\frac{\sin ^{2}\left(2 \beta {\mu}(\t)K_{0}(\t)\right)}{\sqrt{P_{0}(\t)}}\nonumber\\
&=&\kappa\b^2{\mu}(\t)^2\sqrt{P_0(\t)}\lt[ \pi_0(\t)^2 P_0(\t)^{-3}-U\lt(\phi_0(\t)\rt)\rt],\label{mueom1}\\
&&\sqrt{P_{0}(\t)}\left[2 \beta^{2} \sin \left(2 \beta {\mu}(\t) K_{0}(\t)\right)-\left(\beta^{2}+1\right) \sin \left(4 \beta {\mu}(\t) K_{0}(\t)\right)\right]+2 \beta {\mu}(\t) \dot{P}_{0}(\t)=0,\label{mueom2}\\
&&P_0(\t)^{3/2} \dot{\phi}_0(\t) -\pi_0(\t)=0,\qquad {P_0}(\t)^{3/2} {U}'(\phi_0(\t))=-2\dot{\pi}_0(\t),\label{mueom3}
\ee
These equations are defined for the entire time evolution from $\t_0$ to $T$. The choice of spacetime lattice $\ck$ relates to the choice of function $\mu(\t)$ in Eqs.\eqref{mueom1} - \eqref{mueom3}. 

It is convenient to make a change of variable $b_0(\t)=K_0(\t)/\sqrt{P_0(\t)}$ since $b_0(\t)$ equals to the Hubble parameter $H(\t)$ in the semiclassical regime. Eqs.\eqref{mueom1} - \eqref{mueom3} become
\be
\dot{b}_0(\tau )&=& -\frac{\sin ^2\left(2 \beta  b_0(\tau ) \mu (\tau ) \sqrt{P_0(\tau )}\right)}{8 \beta ^2 \mu (\tau )^2 P_0(\tau )}+\frac{\sin ^4\left(\beta  b_0(\tau ) \mu (\tau ) \sqrt{P_0(\tau )}\right)}{2 \mu (\tau )^2 P_0(\tau )}\nonumber\\
&&+\frac{\beta  b_0(\tau ) \sin \left(2 \beta  b_0(\tau ) \mu (\tau ) \sqrt{P_0(\tau )}\right)}{2 \mu (\tau ) \sqrt{P_0(\tau )}}-\frac{\beta  b_0(\tau ) \sin \left(4 \beta  b_0(\tau ) \mu (\tau ) \sqrt{P_0(\tau )}\right)}{4 \mu (\tau ) \sqrt{P_0(\tau )}}\nonumber\\
&&-\frac{b_0(\tau ) \sin \left(4 \beta  b_0(\tau ) \mu (\tau ) \sqrt{P_0(\tau )}\right)}{4 \beta  \mu (\tau ) \sqrt{P_0(\tau )}}-\frac{\kappa  \pi_0(\tau )^2}{8 P_0(\tau )^3}+\frac{1}{8} \kappa  U(\phi_0(\tau ))+\frac{\L}{2},\label{bp1}\\
\frac{\dot{P}_0(\tau )}{P_0(\t)}&=& -\frac{\beta   \sin \left(2 \beta  b_0(\tau ) \mu (\tau ) \sqrt{P_0(\tau )}\right)}{\mu (\tau )\sqrt{P_0(\tau )}}+\frac{\beta   \sin \left(2 \beta  b_0(\tau ) \mu (\tau ) \sqrt{P_0(\tau )}\right) \cos \left(2 \beta  b_0(\tau ) \mu (\tau ) \sqrt{P_0(\tau )}\right)}{\mu (\tau )\sqrt{P_0(\tau )}}\nonumber\\
&&+\frac{ \sin \left(2 \beta  b_0(\tau ) \mu (\tau ) \sqrt{P_0(\tau )}\right) \cos \left(2 \beta  b_0(\tau ) \mu (\tau ) \sqrt{P_0(\tau )}\right)}{\beta  \mu (\tau )\sqrt{P_0(\tau )}}\label{bp2}\\
\dot{\phi}_0(\t)&=& \pi_0(\t)/P_0(\t)^{3/2},\qquad {P_0}(\t)^{3/2} {U}'(\phi_0(\t))=-2\dot{\pi}_0(\t),\label{bp3}
\ee
Given the choice of $\mu(\t)$, Eqs.\eqref{bp1} - \eqref{bp3} uniquely determine the solution ${P}_{0}(\t),{b}_{0}(\t),{\pi}_{0}(\t),{\phi}_{0}(\t)$ provided their initial condition at $\t_0$. Since the solution depends on the function $\mu(\t)$, we denote the solution by 
\be
{P}_{0}[\mu],\ {b}_{0}[\mu],\ {\pi}_{0}[\mu],\ {\phi}_{0}[\mu]
\ee
Given the initial condition ${P}_{0}(\t_0),{b}_{0}(\t_0),{\pi}_{0}(\t_0),{\phi}_{0}(\t_0)$ or the final condition ${P}_{0}(T),{b}_{0}(T),{\pi}_{0}(T),{\phi}_{0}(T)$, Eqs.\eqref{bp1} - \eqref{bp3} defines a map $\iota$ from the space $\cf_\mu$ of functions $\mu(\t)$ to the space $\cf_{\cb}$ of solutions ${P}_{0}(\t),{b}_{0}(\t),{\pi}_{0}(\t),{\phi}_{0}(\t)$, and ${P}_{0}[\mu], {b}_{0}[\mu], {\pi}_{0}[\mu], {\phi}_{0}[\mu]$ is the image of this map from a given function $\mu$:
\be
\iota:&\cf_\mu& \to \cf_\cb\nonumber\\
&\mu&\mapsto \lt({P}_{0}[\mu], {b}_{0}[\mu], {\pi}_{0}[\mu], {\phi}_{0}[\mu]\rt)
\ee 

\section{UV Cut-off and $\mu_{min}$-scheme effective dynamics}\label{UV Cut-off and effective dynamics}



First of all, we fix the final condition ${P}_{0}(T),{b}_{0}(T),{\pi}_{0}(T),{\phi}_{0}(T)$ by letting $T=\t_{pivot}$ to be the pivot time. 
\be
{P}_{0}[\mu](T),\ {b}_{0}[\mu](T),\ {\pi}_{0}[\mu](T),\ {\phi}_{0}[\mu](T)
\ee
are independent of $\mu$. Here ${b}_{0}(T),{\pi}_{0}(T),{\phi}_{0}(T)$ have been given in Section \ref{Homogeneous and Isotropic Cosmological Dynamics on Fixed Lattice }:
\be
{b}_{0}(T)=1.21\times 10^{-6}m_P,\quad {\pi}_{0}(T)=-5.03\times 10^{-9}m_P^2\, {P}_{0}(T)^{3/2},\quad {\phi}_{0}(T)=1.07 m_P,\label{finalconditionT}
\ee 
Although ${P}_{0}(T)$ is ambiguous, we fix its value e.g. ${P}_{0}(T)=1$ and proceed, but we are going to show that the effective dynamics obtained at the end of this section is invariant under rescaling ${P}_{0}(T)\to \a {P}_{0}(T)$ for all $\a\in \R$.   

At each time-step, $\mu$ parametrize the discreteness of the theory and $\mu\to 0$ is the continuum limit. We would like to find a choice of $\mu$ to minimize this discreteness while still validating all above discussions. Recall that the validity of the semiclassical dynamics requires $\mu^4P_0^2\gg\frac{1}{4}\b^2\ell_P^4/t$. We impose a UV cut-off $\Delta$ (a small area scale) such that $\Delta^2\gg\frac{1}{4}\b^2\ell_P^4/t$, and we define $\mu_{min}(\t)$ to saturate this UV cut-off 
\be
\mu_{min}(\t)^2P_0[\mu_{min}](\t)= \Delta.\label{saturate}
\ee
The geometrical volume at each lattice vertex is $\mu_{min}(\t)^3P_0[\mu_{min}](\t)^{3/2}=\Delta^{3/2}$ is minimal. As a result, the semiclassical approximation of the dynamics is valid throughout the evolution, while higher order corrections in $\hbar$ of $\left\langle\tilde{\psi}_{Z}^{\hbar}|\hat{\mathbf{H}}| \tilde{\psi}_{Z}^{\hbar}\right\rangle$ only give relatively small corrections to the predictions of the effective dynamics.

$\mu_{min}(\t)$ satisfying Eq.\eqref{saturate} is unique. $P_0[\mu_{min}](\t)$ and $\mu_{min}(\t)$ can be obtained as the following: We firstly recover the original variables in Eqs.\eqref{cosansatz1} and \eqref{cosansatz2} 
\be
p(\t)=\frac{2\mu(\t)^2}{\b a^2}P_0(\t),\quad \theta(\t)=\beta {\mu}(\t) K_{0}(\t),\quad \pi(\t)=\mu(\t)^3\pi_0(\t).
\ee
Eqs.\eqref{mueom1} - \eqref{mueom3} are rewritten as below
\be
\dot{p} &=& \frac{\sqrt{2} \sqrt{\beta  p } \sin (2 \theta  ) \left[\left(\beta ^2+1\right) \cos (2 \theta  )-\beta ^2\right]}{a \beta ^2}+\frac{2 p  \dot{\mu} }{\mu  },\label{eompthmu1}\\
\dot{\theta } &=& \frac{1}{16} \Bigg[4 \sqrt{2} a \beta  \Lambda  \sqrt{\beta  p}-\frac{8 \sqrt{2} \kappa  \pi ^2 \sqrt{\beta  p }}{a^5 \beta ^2 p ^3}-\frac{4 \sqrt{2} p  \sin ^2(\theta  ) \left(-\beta ^2+\left(\beta ^2+1\right) \cos (2 \theta  )+1\right)}{a (\beta  p )^{3/2}}\nonumber\\
&&\ +\sqrt{2} a \beta  \kappa  \sqrt{\beta  p }\, U(\phi_0)+\frac{16 \theta   \dot{\mu } }{\mu  }\Bigg].\label{eompthmu2}\\
\dot{\phi}_0&=&\lt(\frac{2}{a^2\b}\rt)^{3/2} \frac{\pi}{p^{3/2}},\qquad \dot{\pi}=\frac{3 \pi  \dot{\mu} }{\mu}-\frac{a^3 (\beta  p)^{3/2} U'(\phi_0)}{4 \sqrt{2}}. \label{eompthmu3}
\ee
Eq.\eqref{saturate} implies $p(\t)=2\Delta/(\b a^2)$ to be a constant and $\dot{p}=0$. Eqs.\eqref{eompthmu1}, \eqref{eompthmu2}, and \eqref{eompthmu3} become 1st-order ordinary differential equations of $\mu(\t)=\mu_{min}(\t),\theta(\t),\phi_0(\t),\pi(\t)$. The final condition of $\mu_{min}(\t)$ is given by $\mu_{min}(T)=\sqrt{\Delta/P_0(T)}$.

We make the following change of variable according to Eq.\eqref{saturate}
\be
\mu_{min}(\t)=\sqrt{\frac{\Delta}{P_0 [\mu_{min}](\t)}}
\ee
Here $\mu_{min}$ has the same expression as $\bar{\mu}$ in the improved dynamics of LQC, if we identify $\Delta$ to the parameter $\Delta$ (usually set to be the minimal area gap) in LQC. Changing the variable and recovering $b_0,\pi_0$ from Eqs.\eqref{eompthmu1}, \eqref{eompthmu2}, and \eqref{eompthmu3} give the effective EOMs:
\be
\dot{b}_0[\mu_{min}]&=& 
-\frac{\sin ^2\left(2 \sqrt{\Delta}\beta  b_0  [\mu_{min}]\right)}{8 \beta ^2 \Delta }
+\frac{\sin ^4\left(\sqrt{\Delta}\beta  b_0 [\mu_{min}]\right)}{2 \Delta}\nonumber\\
&&+\frac{\beta  b_0  \sin \left(2\sqrt{\Delta} \beta  b_0 [\mu_{min}]\right)}{2 \sqrt{\Delta}}
-\frac{\beta  b_0  \sin \left(4\sqrt{\Delta} \beta  b_0  [\mu_{min}]\right)}{4 \sqrt{\Delta}   }\nonumber\\
&&-\frac{b_0  \sin \left(4\sqrt{\Delta} \beta  b_0   [\mu_{min}]\right)}{4 \sqrt{\Delta}\beta   }-\frac{\kappa  \pi_0^2}{8 P_0^3}+\frac{1}{8} \kappa  U(\phi_0)+\frac{\L}{2},\label{min1}\\
\frac{\dot{P}_0 [\mu_{min}]}{P_0[\mu_{min}]}&=& -\frac{\beta   \sin \left(2 \sqrt{\Delta}\beta  b_0[\mu_{min}] \right)}{\sqrt{\Delta}}
+\frac{\beta   \sin \left(4\sqrt{\Delta} \beta  b_0[\mu_{min}]  \right)}{2\sqrt{\Delta}}\nonumber\\
&&+\frac{ \sin \left(4\sqrt{\Delta} \beta  b_0[\mu_{min}]\right) }{2\sqrt{\Delta}\beta  }\label{min2}\\
\dot{\phi}_0[\mu_{min}]  &=& \pi_0[\mu_{min}]  /P_0[\mu_{min}]  ^{3/2},\qquad {P_0}[\mu_{min}]  ^{3/2} {U}'(\phi_0[\mu_{min}]  )=-2\dot{\pi}_0[\mu_{min}].\label{min3}
\ee
These effective EOMs are equivalent to replacing $\mu(\t)$ by $\mu_{min}(\t)=\sqrt{{\Delta}/{P_0(\t)}}$ in Eqs.\eqref{bp1} - \eqref{bp3}. We coin the name of Eqs.\eqref{min1} - \eqref{min3} as ``$\mu_{min}$-scheme effective dynamics''.

Note that the $\mu_{min}$-scheme effective dynamics is not the same as the $\bar{\mu}$-scheme in LQC (see Fig.\ref{bounce} for the curves labelled by ``Minimum''), although there are several important similarities which are discussed in Section \ref{Properties of Minimum and Average Effective Dynamics}.

Eqs.\eqref{min1} - \eqref{min3} are invariant under the following rescaling:
\be
&P_0[\mu_{min}](\t)\to \a P_0[\mu_{min}](\t),\quad \pi_0[\mu_{min}](\t)\to\a^{3/2}\pi_0[\mu_{min}](\t),&\\
& b_0[\mu_{min}](\t)\to b_0[\mu_{min}](\t),\quad \phi_0[\mu_{min}](\t)\to\phi_0[\mu_{min}](\t).&\label{rescalemin}
\ee
Recall that the final condition $P_0(T)$ is ambiguous (defined up to rescaling). If we rescale $P_0(T)\to \a P_0(T)$ and $\pi_0(T)\to\a^{3/2}\pi_0(T)$ of the final condition ($b_0,\dot{\phi}_0,\phi_0$ are left invariant), the solution of Eqs.\eqref{min1} - \eqref{min3} from the rescaled final condition is the rescaling \eqref{rescalemin} of the solution from the original final condition, since the solution of the equations is uniquely determined by the final condition. The dynamics of Hubble parameter $H$ and scalar field $\phi_0$ is rescaling invariant, thus is ambiguity-free.

\section{Random Lattice and Average Effective Dynamics}\label{Random Lattice and Average Effective Dynamics}


The dynamics of ${P}_{0}(\t_0),{b}_{0}(\t_0),{\pi}_{0}(\t_0),{\phi}_{0}(\t_0)$ depends on the choice of $\mu(\t)$, or equivalently the choice of the spacetime lattice $\ck$. Different spacetime lattices $\ck$ give different definition of transition amplitude $\ca_{[Z],[Z']}(\ck)$, and can be viewed as corresponding to different superselection sectors. When we approximate the discrete $\mu_{i=0,\cdots,m}$ by the smooth function $\mu(\t)$, the supersection sectors are labelled by functions $\mu(\t)$. $\mu(\t)$ behaves as giving an ``external force'' to the cosmological dynamics as shown in Eqs.\eqref{eompthmu1} and \eqref{eompthmu2}.

We propose $\ck$ to be a random lattice such that $\mu(\t)$ is a random function with respect to certain probability distribution on the ensemble $\cf_\mu$. The random $\mu(\t)$ gives a ``random external force'' in Eqs.\eqref{eompthmu1} and \eqref{eompthmu2}. The probability distribution on $\cf_\mu$ is described as the following: 

We again fix the final condition ${P}_{0}(T),{b}_{0}(T),{\pi}_{0}(T),{\phi}_{0}(T)$ at $T=\t_{pivot}$ as \eqref{finalconditionT} and ${P}_{0}(T)=1$ (the effective dynamics obtained at the end of this section is invariant under rescaling ${P}_{0}(T)\to \a {P}_{0}(T)$ for all $\a\in \R$).

We take $\mu_{min}(\t)=\sqrt{\Delta/P_0[\mu_{min}](\t)}$ as the minimal $\mu(\t)$ of the spatial lattice at the instance $\t$. Recall that a general $\mu(\t)$ is an approximation of $L/N(\t)$ (or $L/N_i$ with earlier notations), where the integer $N(\t)$ (or $N_i$) is the number of vertices along each direction on the spatial lattice $\g(\t)$ (or $\g_i$) at $\t$, we have $N(\t)<N_{max}(\t)$ where $N_{max}(\t)$ satisfies $L/N_{max}(\t)\simeq \mu_{min}(\t)$. We let $\g(\t)$ with $N(\t)^3$ vertices to be a random sublattice of the finest one with $N_{max}(\t)^3$ vertices. Randomly selecting $N(\t)$ out of $N_{max}(\t)$ vertices along every direction gives a random sublattice\footnote{We label vertices in the finest lattice by $(i,j,k)$ where integer $i,j,k\in \{1,\cdots,N_{max}\}$. We choose of $N$ vertices in each of 3 directions $i_1,\cdots, i_N$, $j_1,\cdots, j_N$, $k_1,\cdots,k_N$ where $i_a,j_b,k_c\in\{1,\cdots,N_{max}\} $ with $a,b,c\in\{1,\cdots,N\}$. The choices make a $N\times N\times N$ sublattice whose vertices are $(i_a,j_b,k_c)$ with $a,b,c\in\{1,\cdots,N\}$.}. Assuming all sublattices are democratic, the probability $\fp(\mu(\t))$ of having $\g(\t)$ is proportional to the multiplicity \footnote{This probability distribution is different from the one in \cite{Alesci:2016rmn} although the idea is similar.}
\be
\fp(\mu(\t))=\lt[\frac{1}{2^{N_{max}(\t)}}\left(\begin{array}{l}N_{max}(\t) \\ \ \  N(\t)\end{array}\right)\rt]^3
\ee
This is the probability distribution of $\mu(\t)$ at the instance $\t$. The probability of the function $\mu$ is given by $\prod_\t\fp(\mu(\t))$, where the product over $\t$ is essentially a finite product since $\ca_{[Z],[Z']}(\ck)$ assumes a finite number of lattice changes. When $N_{max}(\t)$ are large, $\Fp(\mu(\t))$ can be approximated by a Gaussian function
\be
\fp(\mu(\t))=\lt(\frac{1}{\sqrt{\pi N_{max }(\t) / 2}} e^{-\frac{[N(\t)-N_{max }(\t) / 2]^{2}}{N_{max }(\t) / 2}}\lt[1+O\lt(\frac{1}{\sqrt{N_{max}}}\rt)\rt]\rt)^3.
\ee
It leads to the probability distribution on $\cf_\mu$ as
\be
\Fp(\mu)\simeq
\prod_{\t}\lt(\frac{L}{\mu(\t)^2}\sqrt{\frac{2\mu_{min}(\t)}{\pi L}} e^{-\frac{2L}{\mu_{min }(\t)}\lt(\frac{\mu_{min }(\t)}{\mu(\t)}-\half\rt)^{2}}\rt)^3
\ee
which indicates that the most probable $\mu(\t)$ is $\overline{\mu}(\t)=2\mu_{min}(\t)$. We extend $\Fp(\mu)$ to entire $\cf_\mu$ including those $\mu(\t)$ even smaller than $\mu_{min}(\t)$ since they give negligible probability.

The averaged dynamics of $P_0,K_0,\phi_0,\pi_0$ is given by the ensemble average over $\cf_\mu$, 
\be
&\overline{P}_0(\t)=\int_{\cf_\mu}D\mu\,\Fp(\mu){P}_0[\mu](\t),\quad \overline{b}_0(\t)=\int_{\cf_\mu}D\mu\,\Fp(\mu){b}_0[\mu](\t),&\\
&\overline{\pi}_0(\t)=\int_{\cf_\mu}D\mu\,\Fp(\mu){\pi}_0[\mu](\t),\quad \overline{\phi}_0(\t)=\int_{\cf_\mu}D\mu\,\Fp(\mu){\phi}_0[\mu](\t),&
\ee
where $D\mu=\prod_\t d\mu(\t)$. When $N_{max}(\t)$ is large, we have the following approximation
\be
\lt(\overline{P}_0,\overline{b}_0,\overline{\pi}_0,\overline{\phi}_0\rt)\simeq\lt({P}_0[2\mu_{min}],{b}_0[2\mu_{min}],{\pi}_0[2\mu_{min}],{\phi}_0[2\mu_{min}]\rt),\quad \mu_{min}(\t)=\sqrt{\frac{\Delta}{P_0[\mu_{min}](\t)}}.
\ee
The average effective dynamics is given by the EOMs \eqref{bp1} - \eqref{bp3} with $\overline{\mu}(\t)=2\mu_{min}(\t)$. 
\be
\dot{\overline{b}}_0&=& 
-\frac{\sin ^2\left(4 \sqrt{\Delta}\beta  \overline{b}_0  \sqrt{\overline{P}_0 /P_0[\mu_{min}] }\right)}{32 \beta ^2 \Delta \overline{P}_0 /P_0[\mu_{min}]}
+\frac{\sin ^4\left(2\sqrt{\Delta}\beta  \overline{b}_0  \sqrt{\overline{P}_0 /P_0[\mu_{min}]}\right)}{8 \Delta \overline{P}_0 /P_0[\mu_{min}]}\nonumber\\
&&+\frac{\beta  \overline{b}_0  \sin \left(4\sqrt{\Delta} \beta  \overline{b}_0  \sqrt{P_0  /P_0[\mu_{min}]}\right)}{4 \sqrt{\Delta}\sqrt{\overline{P}_0 /P_0[\mu_{min}] }}
-\frac{\beta  \overline{b}_0  \sin \left(8\sqrt{\Delta} \beta  \overline{b}_0  \sqrt{\overline{P}_0 /P_0[\mu_{min}]}\right)}{8 \sqrt{\Delta}   \sqrt{\overline{P}_0 /P_0[\mu_{min}]}}\nonumber\\
&&-\frac{\overline{b}_0  \sin \left(8\sqrt{\Delta} \beta  \overline{b}_0    \sqrt{\overline{P}_0 /P_0[\mu_{min}]}\right)}{8 \sqrt{\Delta}\beta   \sqrt{\overline{P}_0 /P_0[\mu_{min}] }}-\frac{\kappa  \pi_0^2}{8 P_0^3}+\frac{1}{8} \kappa  U(\overline{\phi}_0)+\frac{\L}{2},\label{bp11}\\
\frac{\dot{\overline P}_0 }{\overline{P}_0}&=& -\frac{\beta   \sin \left(4 \sqrt{\Delta}\beta  \overline{b}_0    \sqrt{\overline{P}_0 /P_0[\mu_{min}] }\right)}{2\sqrt{\Delta}  \sqrt{\overline{P}_0 /P_0[\mu_{min}] }}
+\frac{\beta   \sin \left(8\sqrt{\Delta} \beta  \overline{b}_0  \sqrt{\overline{P}_0 /P_0[\mu_{min}] }\right)}{4\sqrt{\Delta}  \sqrt{\overline{P}_0 /P_0[\mu_{min}]}}\nonumber\\
&&+\frac{ \sin \left(8\sqrt{\Delta} \beta  \overline{b}_0   \sqrt{\overline{P}_0 /P_0[\mu_{min}] }\right) }{4\sqrt{\Delta}\beta  \sqrt{\overline{P}_0 /P_0[\mu_{min}] }}\label{bp22}\\
\dot{\overline{\phi}}_0  &=& \overline{\pi}_0  /\overline{P}_0  ^{3/2},\qquad {\overline{P}_0}  ^{3/2} {U}'(\overline{\phi}_0  )=-2\dot{\overline \pi}_0  .\label{bp33}
\ee

These equations are invariant by rescaling 
\be
&\overline{P}_0(\t)\to\a \overline{P}_0(\t),\quad P_0[\mu_{min}](\t)\to \a P_0[\mu_{min}](\t),\quad \overline{\pi}_0(\t)\to \a^{3/2}\overline{\pi}_0(\t),&\\
& \overline{b}_0(\t)\to \overline{b}_0(\t),\quad \overline{\phi}_0(\t)\to\overline{\phi}_0(\t),&\label{rescaleoverline}
\ee
with constant $\a$. If we rescale the final condition $P_0(T)\to \a P_0(T)$ and $\pi_0(T)\to\a^{3/2}\pi_0(T)$ ($b_0,\dot{\phi}_0,\phi_0$ are left invariant), the solution of Eqs.\eqref{min1} - \eqref{min3} is rescaled as discussed above
\be
&P_0[\mu_{min}](\t)\to \a P_0[\mu_{min}](\t),\quad \pi_0[\mu_{min}](\t)\to\a^{3/2}\pi_0[\mu_{min}](\t),&\\
& b_0[\mu_{min}](\t)\to b_0[\mu_{min}](\t),\quad \phi_0[\mu_{min}](\t)\to\phi_0[\mu_{min}](\t).&
\ee
When we insert this rescaling of $P_0[\mu_{min}]\to \a P_0[\mu_{min}]$ in Eqs.\eqref{bp11} - \eqref{bp33}, and apply the rescaled final condition $P_0(T)\to \a P_0(T)$ and $\pi_0(T)\to\a^{3/2}\pi_0(T)$, the solution of the average effective dynamics is rescaled
\be
&\overline{P}_0(\t)\to \a \overline{P}_0(\t),\quad \overline{\pi}_0(\t)\to\a^{3/2}\overline{\pi}_0(\t)\,&\\
& \overline{b}_0(\t)\to \overline{b}_0(\t),\quad \overline{\phi}_0(\t)\to\overline{\phi}_0(\t),&
\ee 
due to the symmetry and the uniqueness of the solution. The average effective dynamics of Hubble parameter $H$ and scalar field $\phi_0$ is free of the ambiguity of $P_0(T)$.

\section{Properties of $\mu_{min}$-Scheme and Average Effective Dynamics}\label{Properties of Minimum and Average Effective Dynamics}

\begin{figure}[t]
\begin{center}
  \includegraphics[width = 0.6\textwidth]{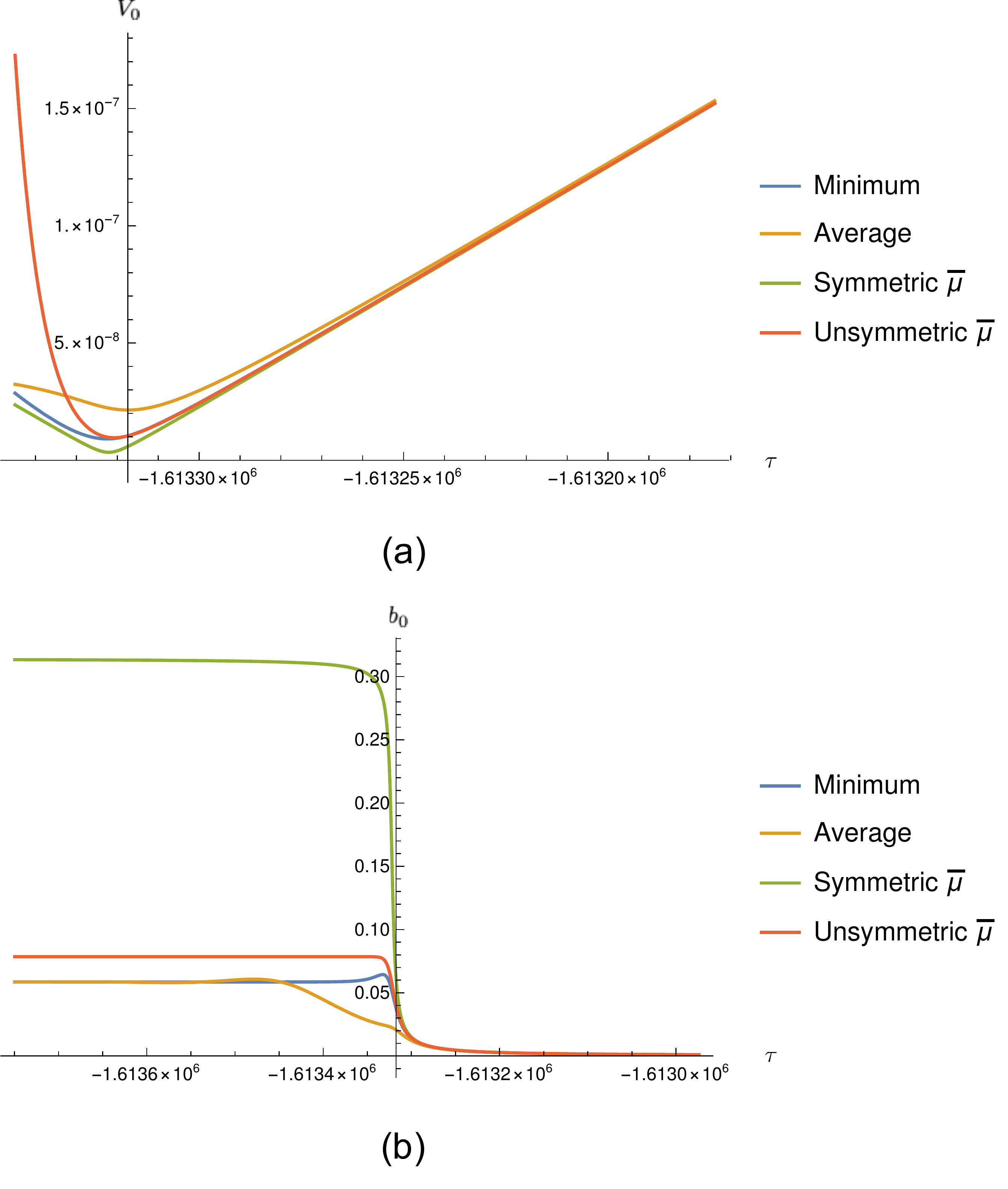}
    \caption{The plot of the average effective dynamics given by EOMs \eqref{bp11} - \eqref{bp33} (orange), and comparison with $\mu_{min}$-scheme effective dynamics  (blue), the $\bar{\mu}$-scheme LQC effective dynamics in \cite{Yang:2009fp,Assanioussi:2019iye} with unsymmetric bounce (red), and the traditional $\bar{\mu}$-scheme LQC effective dynamics in \cite{Ashtekar:2006wn} with symmetric bounce (green). The upper panel plots $V_0(\t)=P_0(\t)^{3/2}$, and the lower panel plots $b_0(\t)$. We set the pivot time to be $\t_{pivot}=0$. The solution is determined by the values at $\t_{pivot}$: $b_0(\t_{pivot})=1.21\times 10^{-6}m_P$, $V_0(\t_{pivot})=1$, $\phi_0(\t_{pivot})=1.07 m_P$, $\pi_0(\t_{pivot})=-5.03\times 10^{-9}m_P^2$. The parameters take values $m=2.44\times10^{-6}m_P$, $\L=0$, $\sqrt{\Delta}=10 l_P$, and $\b=1$.  }
  \label{bounce}
  \end{center}
\end{figure}

\begin{figure}[t]
\begin{center}
  \includegraphics[width = 0.8\textwidth]{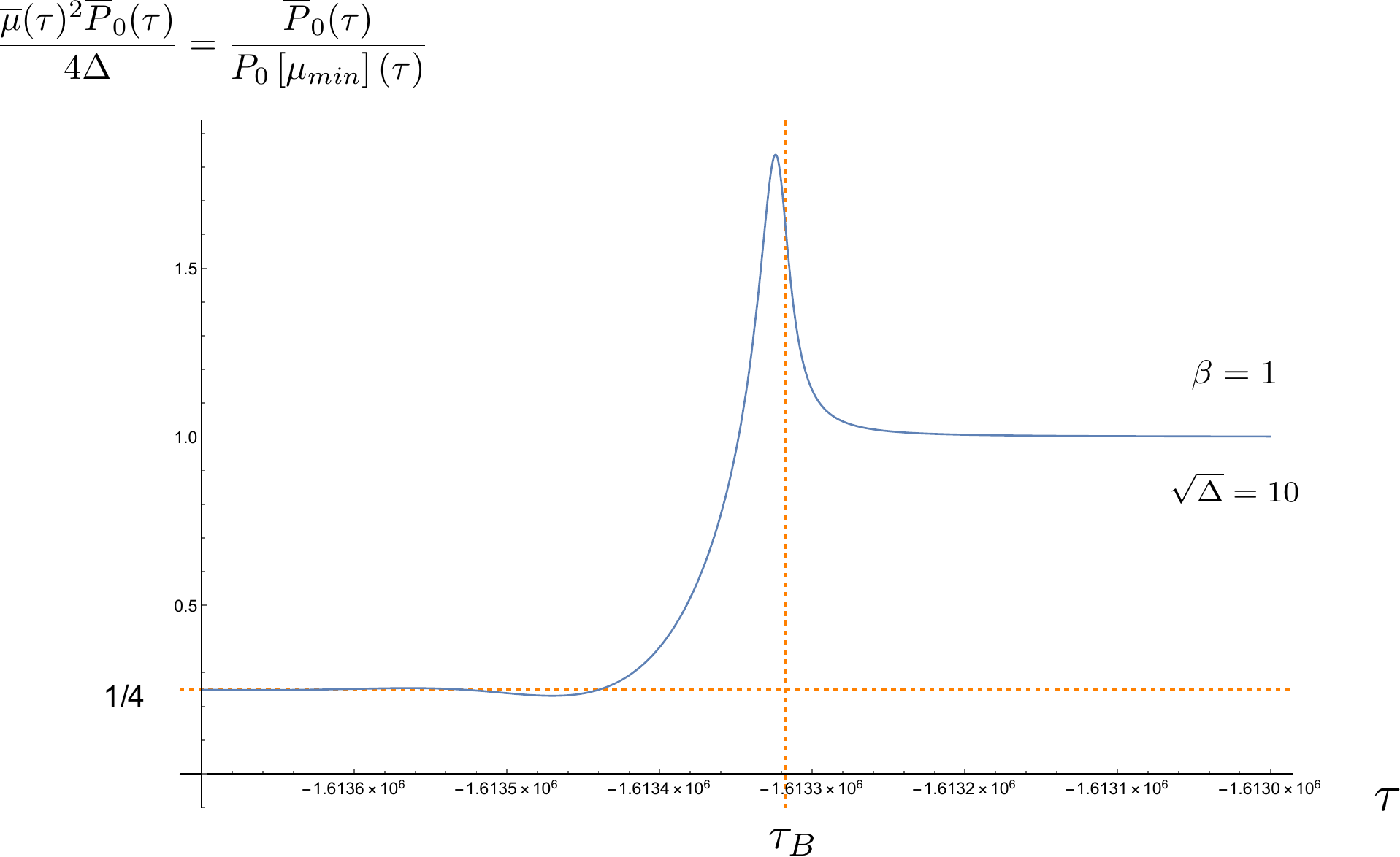}
    \caption{The Plot of $\overline{P}_0(\t)/P_0[\mu_{min}](\t)$ at $\b=1,\ \sqrt{\Delta}=10 l_P$. The vertical dashed line is at $\t_B$ of the bounce. The horizontal dashed line is at $\overline{\mu}^2 \overline{P}_0=\Delta$}
  \label{mu2P01}
  \end{center}
\end{figure}

\begin{figure}[t]
\begin{center}
  \includegraphics[width = 0.8\textwidth]{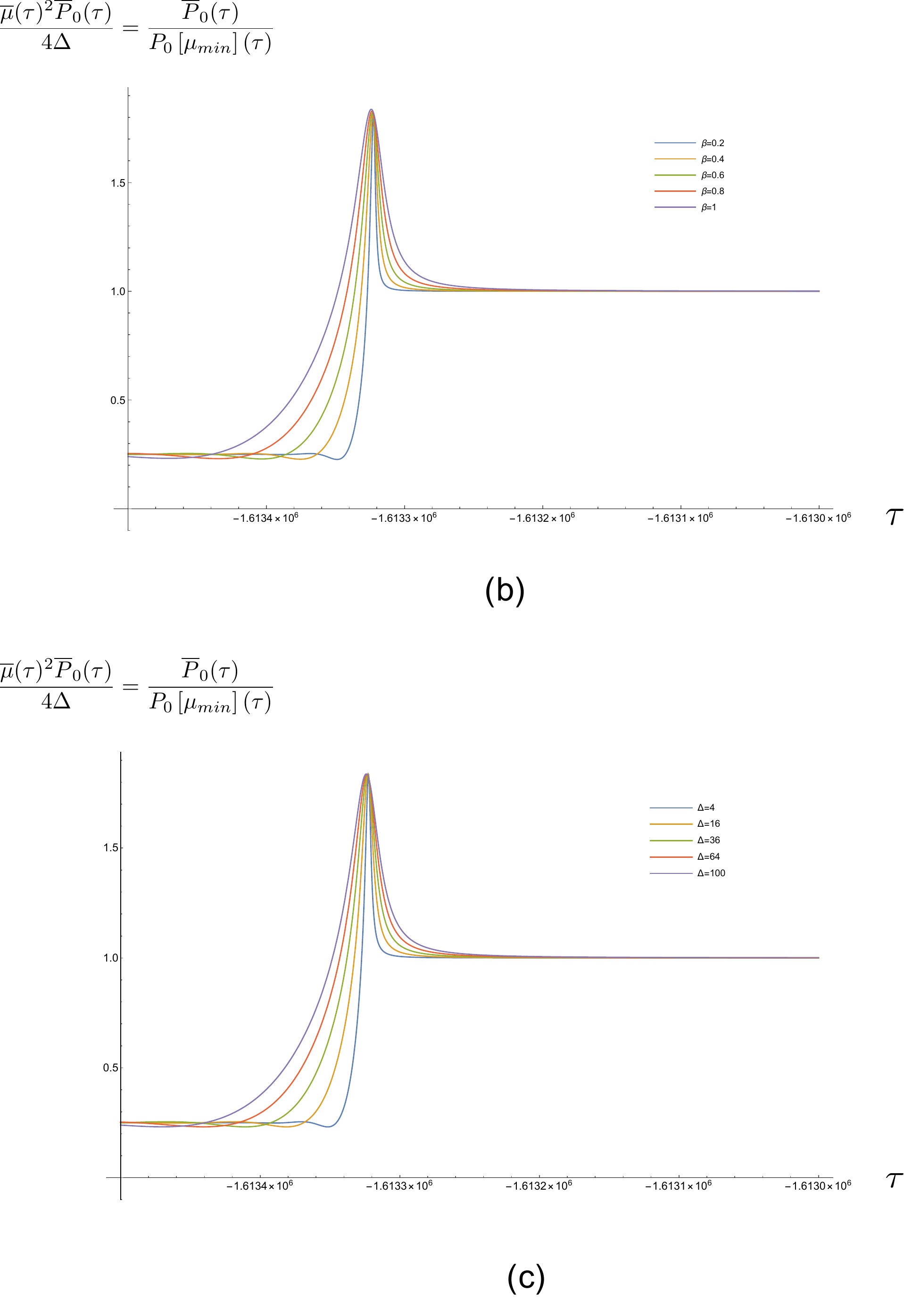}
    \caption{(a) The Plots of $\overline{P}_0(\t)/P_0[\mu_{min}](\t)$ for $\sqrt{\Delta}=10 l_P$ and several different values of $\b$, and (b) The Plots of $\overline{P}_0(\t)/P_0[\mu_{min}](\t)$ for $\b=1$ and several different values of $\Delta$. }
  \label{mu2P02}
  \end{center}
\end{figure}

\begin{figure}[h]
\begin{center}
  \includegraphics[width = 0.58\textwidth]{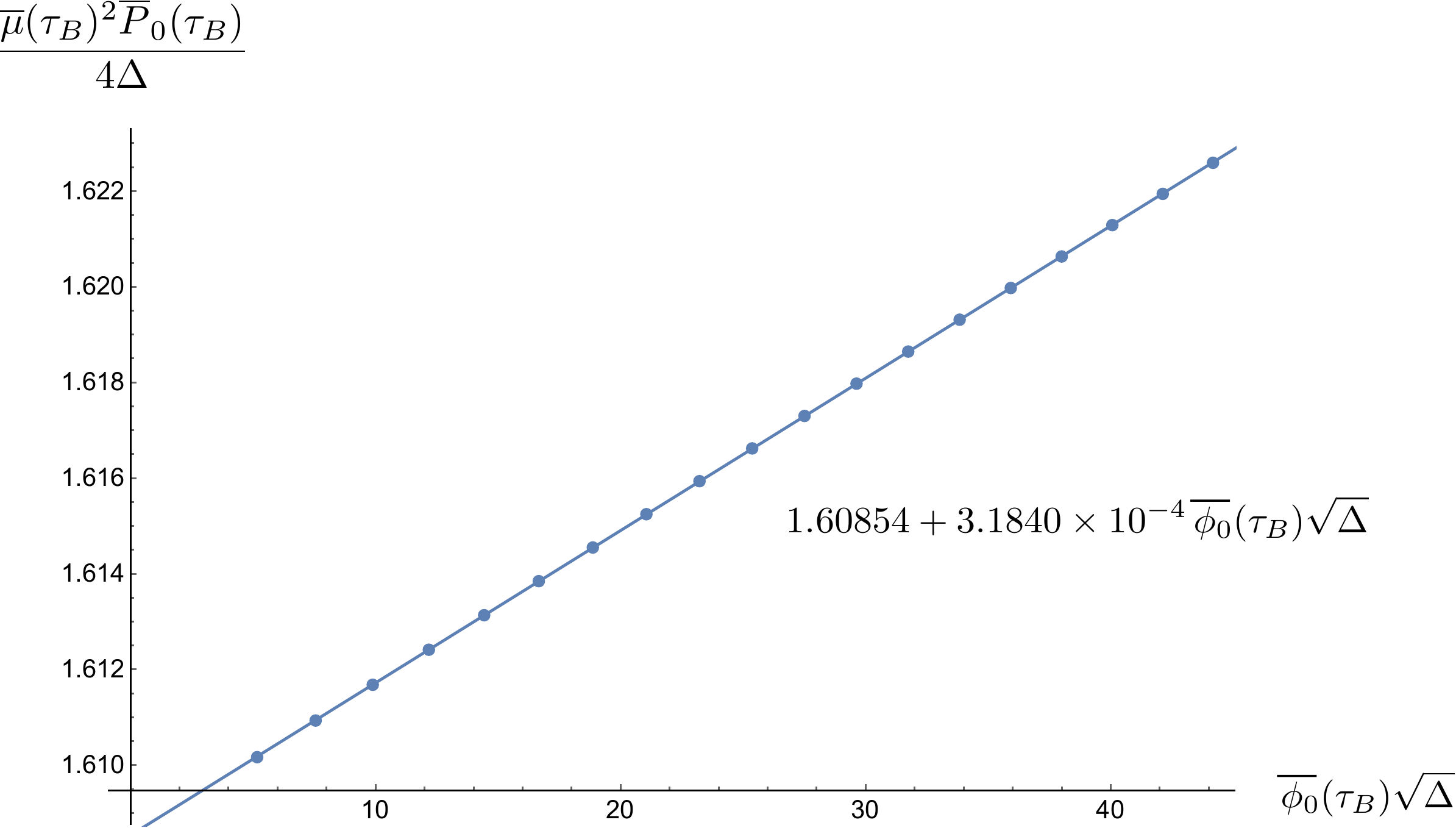}
    \caption{The Plot of $\overline{\mu}(\t_B)^2 \overline{P}_0(\t_B)/(4\Delta)$ versus $\overline{\phi}_0(\t_B)\sqrt{\Delta}$ with $\b=1$ and different values of $\Delta$. }
  \label{mu2P0B}
  \end{center}
\end{figure}

\begin{figure}[h]
\begin{center}
  \includegraphics[width = 0.5\textwidth]{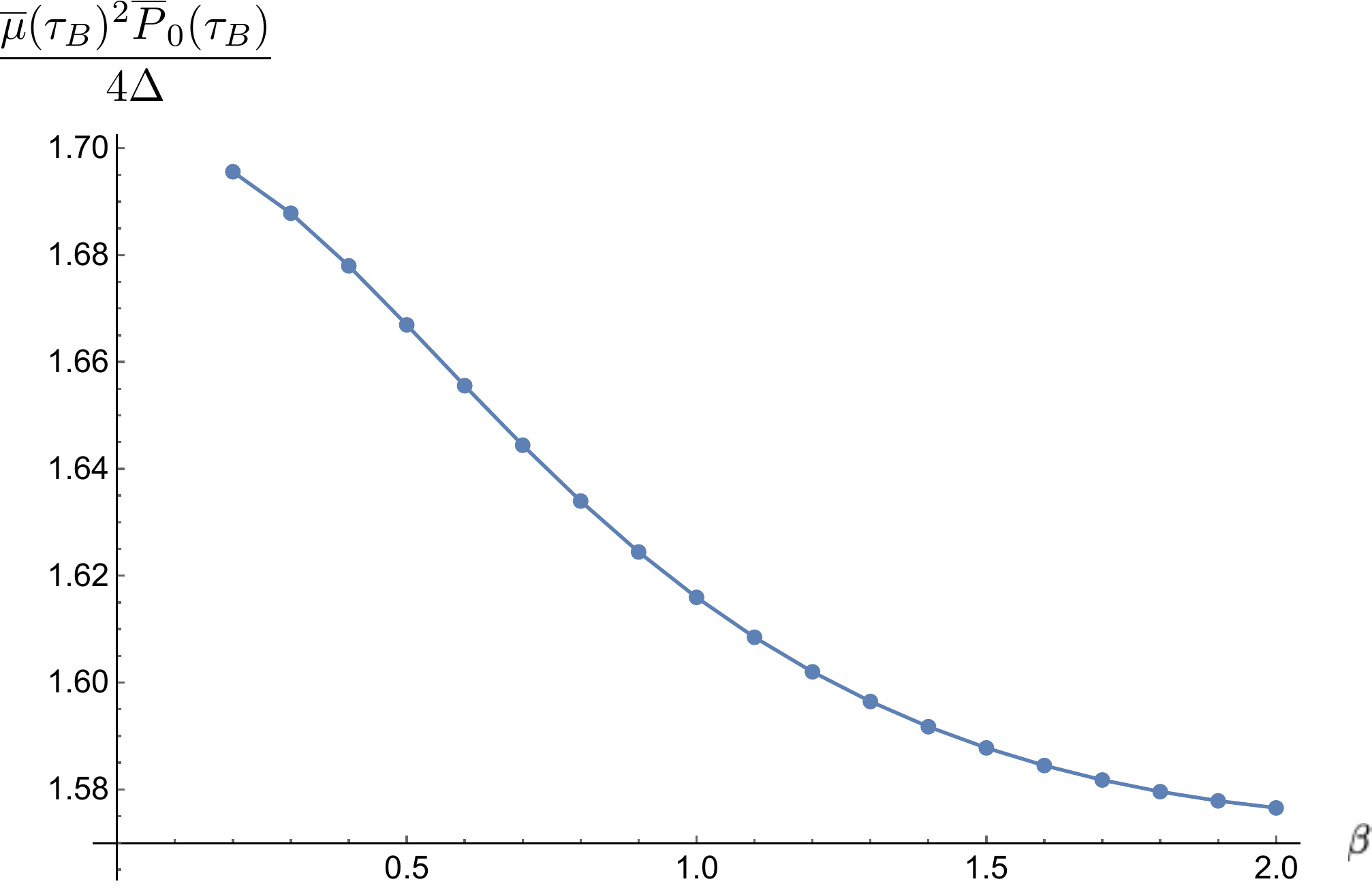}
    \caption{The Plot of $\overline{\mu}(\t_B)^2 \overline{P}_0(\t_B)/(4\Delta)$ versus $\b$ with $\sqrt{\Delta}=10 l_P$. }
  \label{mu2P0beta}
  \end{center}
\end{figure}

\begin{figure}[h]
\begin{center}
\includegraphics[width = 1\textwidth]{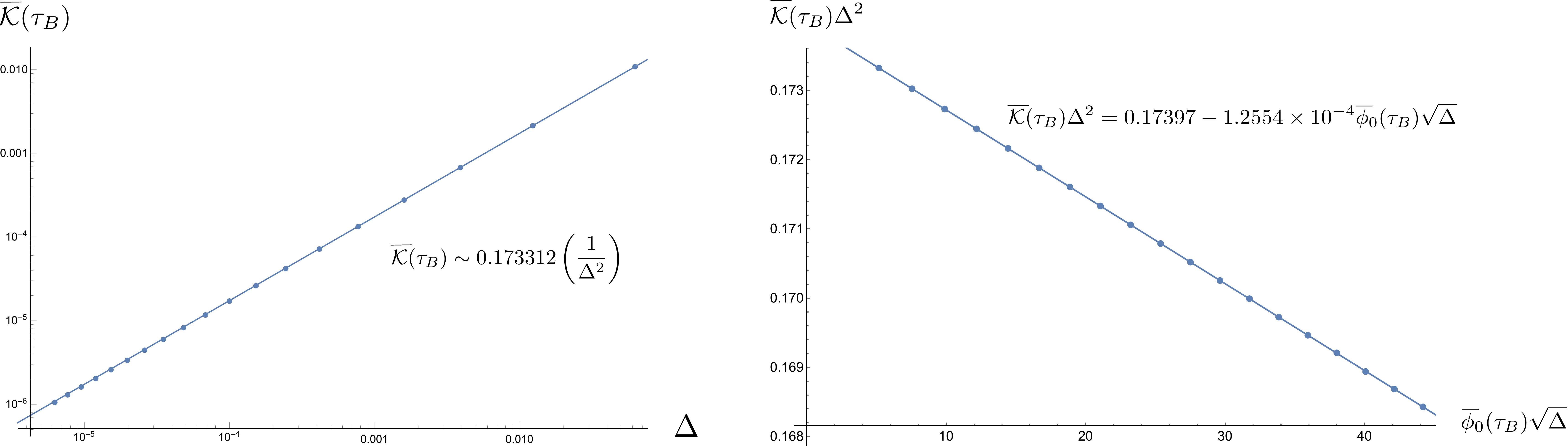}
    \caption{The left panel plots the Kretschmann scalar $\ck(\t_B)$ versus $\Delta$ and finds the leading order behavior $\ck\sim \Delta^{-2}$ (with $\b=1$). The right panel plots $\ck(\t_B)\Delta^2$ versus $\phi_0(\t_B)\sqrt{\Delta}$ and finds the subleading correction.}
  \label{krets_bounce}
  \end{center}
\end{figure}

\begin{figure}[h]
\begin{center}
\includegraphics[width = 1\textwidth]{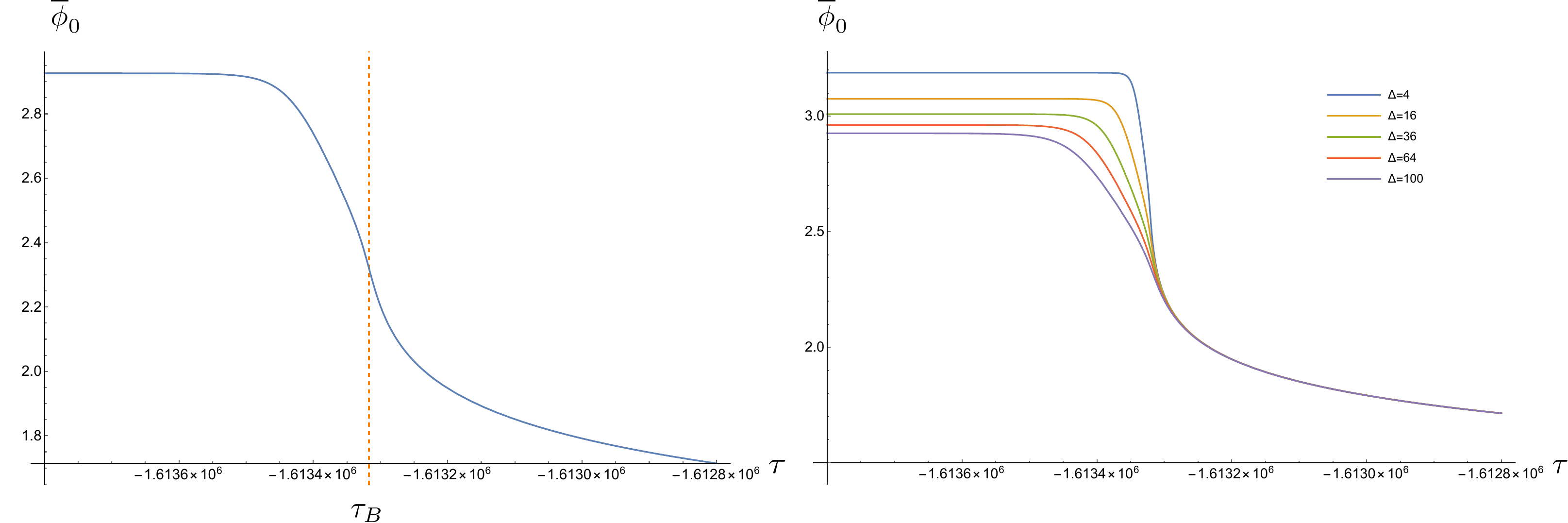}
    \caption{The left panel plots $\phi_0(\t_B)$ in the case of $\b=1,\ \sqrt{\Delta}=10 l_P$. The right panel plots $\phi_0(\t_B)$ in the case of $\b=1$ and different values of $\Delta$.}
  \label{scalarevo}
  \end{center}
\end{figure}

\begin{figure}[h]
\begin{center}
\includegraphics[width = 0.8\textwidth]{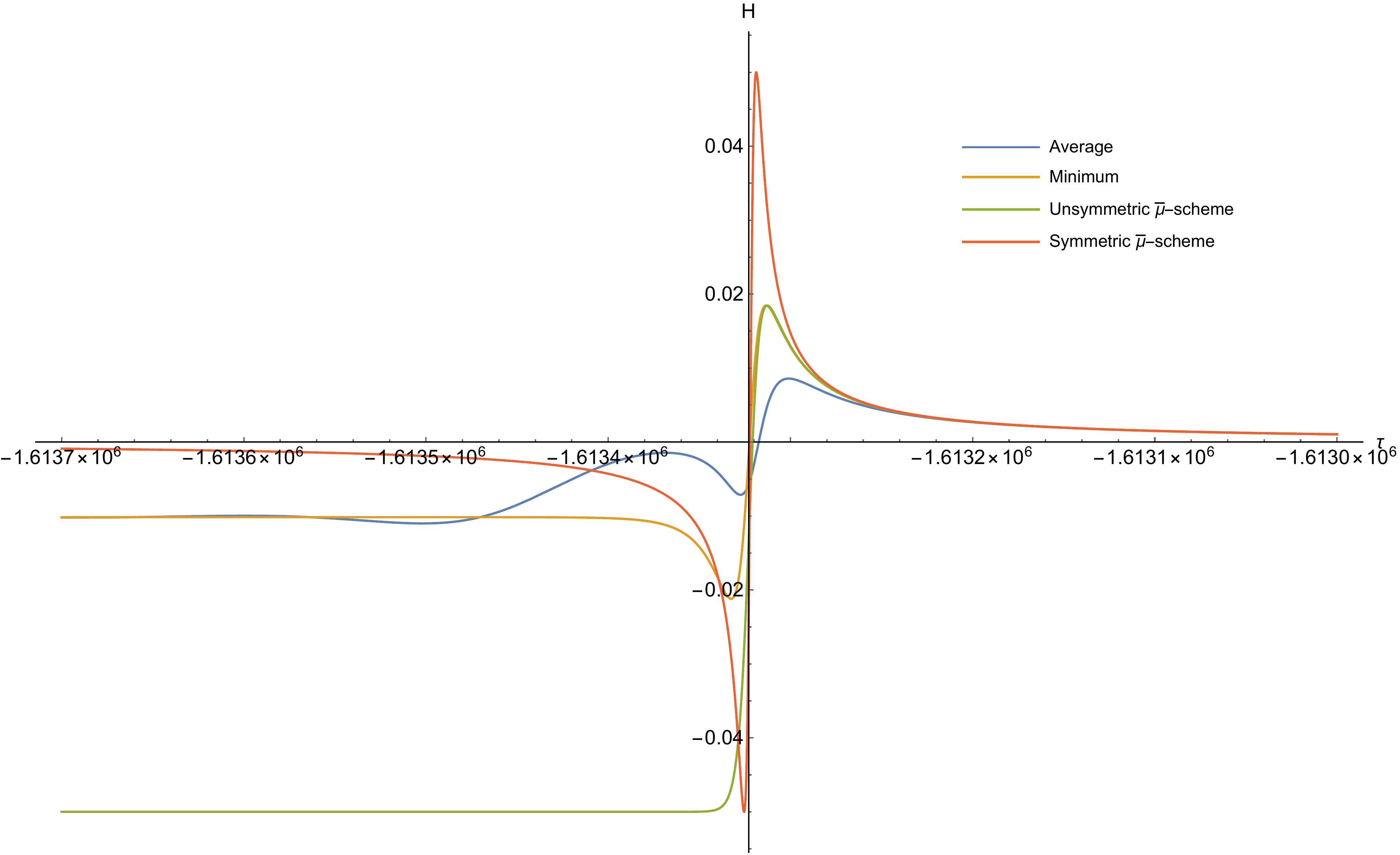}
    \caption{The Hubble parameter $H$ of the averaged dynamics (Average) and comparison with the dynamics with $\mu_{min}$ (Minimum) and the LQC $\bar{\mu}$-schemes with unsymmetric and symmetric bounces. }
  \label{hubble_compare}
  \end{center}
\end{figure}

\begin{figure}[h]
\begin{center}
\includegraphics[width = 0.8\textwidth]{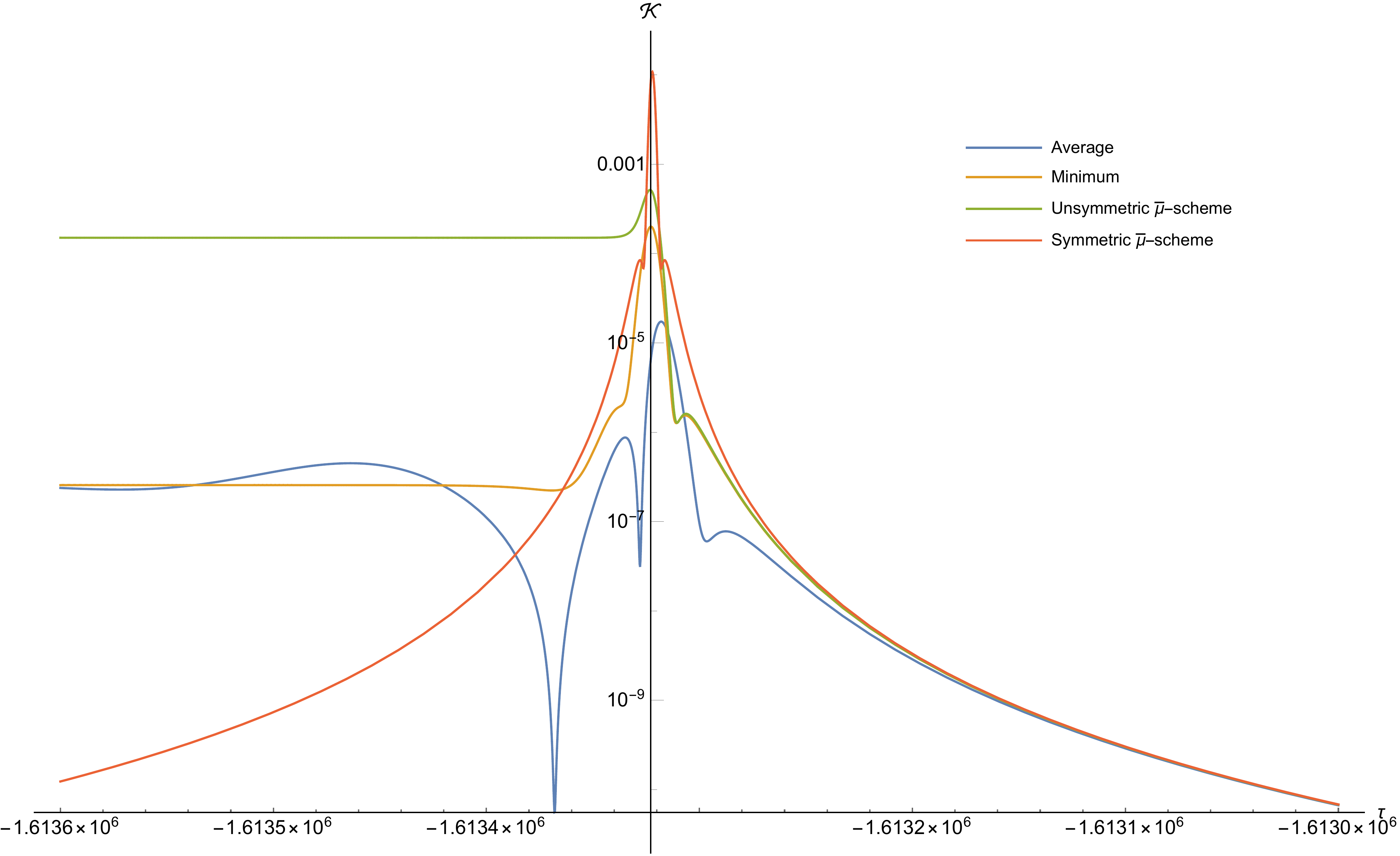}
    \caption{The Kretschmann scalar $\ck$ of the averaged dynamics and comparison with the dynamics with $\mu_{min}$ and the LQC $\bar{\mu}$-schemes with unsymmetric and symmetric bounces. }
  \label{krest_compare}
  \end{center}
\end{figure}

\begin{figure}[h]
\begin{center}
\includegraphics[width = 0.8\textwidth]{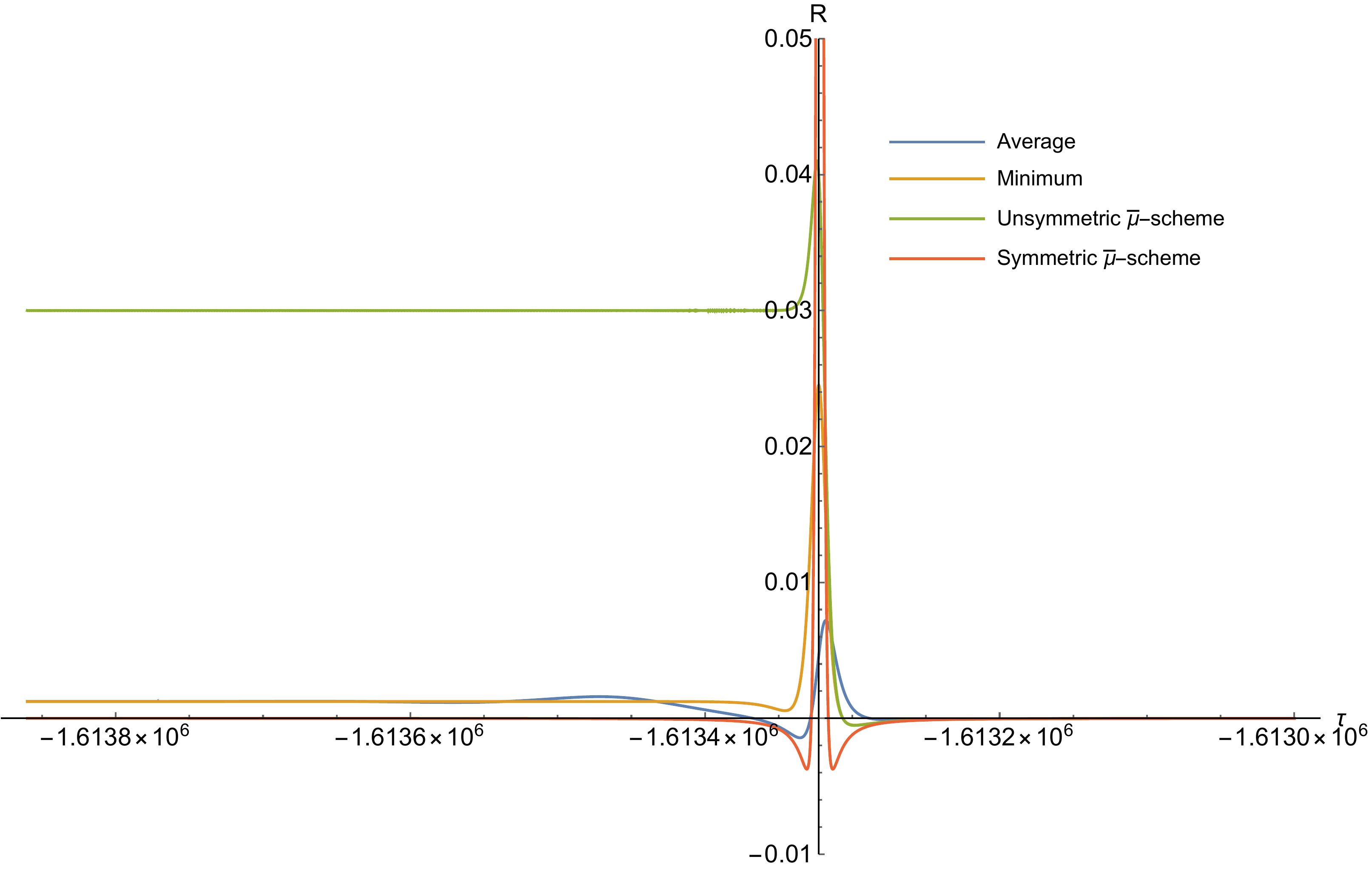}
    \caption{Plots of the scalar curvature $R$ of the averaged dynamics and comparison with the dynamics with $\mu_{min}$ and the LQC $\bar{\mu}$-schemes with unsymmetric and symmetric bounces. }
  \label{ricci_compare}
  \end{center}
\end{figure}

\begin{figure}[h]
\begin{center}
  \includegraphics[width = 0.6\textwidth]{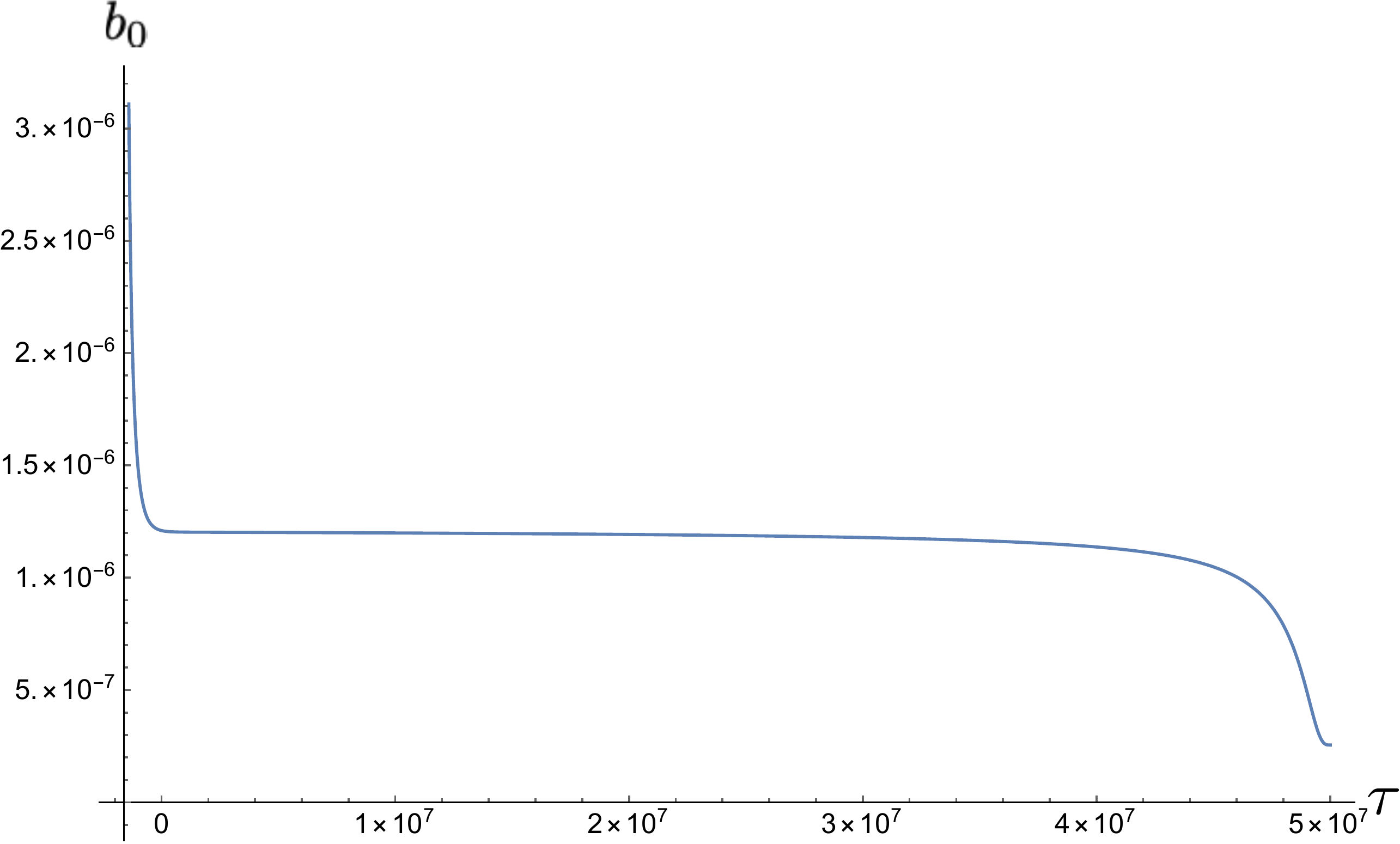}
    \caption{The plot of $b_0$ during the inflation. $\t=0$ is the pivot time. During the inflation $b_0\sim 10^{-6}m_P$, so $\b\sqrt{\Delta} b_0\sim 10^{-5}$ (we set $\b=1$ and $\sqrt{\Delta} = 10 l_P$ in this solution) in sines and cosines in Eqs.\eqref{bp11}, \eqref{bp22}, and \eqref{bp33}. }
  \label{b0inflation}
  \end{center}
\end{figure}

The Plot of $\mu_{min}$-scheme and average effective dynamics and comparison with $\bar{\mu}$-schemes of LQC are given in Fig.\ref{bounce} (we set $\L=0$ in the figures in this section). All effective dynamics converge at late time while behaving differently near and on the other side of the bounce. In Fig.\ref{bounce}, we have identify our UV cut-off $\Delta$ to the parameter $\Delta$ (usually set to be the minimal area gap) in LQC.

Let's focus on the bounce in the $\mu_{min}$-scheme effective dynamics, $\dot{P}_0(\t^{(min)}_B)=0$ at the bounce and Eq.\eqref{min2} gives
\be
\beta  \overline{b}_0(\t^{(min)}_B) \sqrt{\Delta} =\half\cos^{-1}\lt(\frac{\beta ^2}{\beta ^2+1}\rt)
\ee
where $\t^{(min)}_B$ is the instance when the bounce occurs in the $\mu_{min}$-scheme effective dynamics. Recall Eqs.\eqref{H=|C|} and \eqref{rhodustandH} for the Hamiltonian ${\bf H}$, the total matter density (including cosmological constant) $\rho=\rho_{dust}+\rho_s+\rho_\L$ at the bounce is given by the critical density
\be
\rho_c[\mu_{min}]=\frac{3}{2 \b^2\left(\beta ^2+1\right) \kappa  \Delta}
\ee 
which is constant. This expression is the same as in \cite{Yang:2009fp,Assanioussi:2019iye}.

For the bounce in the average effective dynamics, $\dot{P}_0(\t_B)=0$ at the bounce and Eq.\eqref{bp22} gives
\be
\beta  \overline{b}_0(\t_B) \sqrt{\overline{P}_0(\t_B)} \overline{\mu} (\t_B)=\half\cos^{-1}\lt(\frac{\beta ^2}{\beta ^2+1}\rt)
\ee
where $\t_B$ is the instance when the bounce occurs in the average effective dynamics. The critical density is given by
\be
\overline{\rho}_c=\rho(\t_B)=\frac{3}{2 \left(\beta ^4+\beta ^2\right) \kappa  \overline{P}_0(\t_B) \overline{\mu} (\t_B)^2}.
\ee
Note that here $\overline{\mu}(\t)=2\mu_{min}(\t)$, and $\overline{P_0}(\t)$ shares the same final condition $P_0(T)$ as the one used for solving $\mu_{min}(\t)$. Recall that $P_0(T)$ is the squared scale factor and only defined up to a scaling. We are going to show that $\overline{\rho}_c$ does not change by rescaling $P_0(T)$. Indeed, if we rescale $P_0(T)\to \a P_0(T)$ and $\pi_0(T)\to\a^{3/2}\pi_0(T)$ of the final condition ($b_0,\dot{\phi}_0,\phi_0$ are left invariant), the solution of the average effective dynamics is rescaled by
\be
&\overline{P}_0(\t)\to \a \overline{P}_0(\t),\quad \overline{\pi}_0(\t)\to\a^{3/2}\overline{\pi}_0(\t)\,&\\
& \overline{b}_0(\t)\to \overline{b}_0(\t),\quad \overline{\phi}_0(\t)\to\overline{\phi}_0(\t),&
\ee 
by the discussion below Eqs.\eqref{bp11} - \eqref{bp33}. This indicates that the average effective dynamics of $H$ and $\phi_0$ is free of the ambiguity of $P_0(T)$, and the following quantity is invariant under the rescaling
\be
\overline{\mu}(\t)^2\overline{P}_0(\t)=4\mu_{min}(\t)^2\overline{P}_0(\t)=4\Delta \overline{P}_0(\t)/P_0[\mu_{min}](\t),
\ee 
whose values are plotted in Fig.\ref{mu2P01} with $\b=1,\sqrt{\Delta}=10 l_P$. Therefore
\be
\overline{\rho}_c=\frac{3}{8 \left(\beta ^4+\beta ^2\right) \kappa  \Delta \overline{P}_0(\t_B)/P_0[\mu_{min}](\t_B)}.
\ee
is not affected by the rescaling thus is ambiguity free. $\overline{P}_0(\t_B)/P_0[\mu_{min}](\t_B)\simeq 1.6 $ when $\b=1,\sqrt{\Delta}=10 l_P$, and it has mild dependence on $\b$ and $\Delta$ (see Figs.\ref{mu2P02} - \ref{mu2P0beta}). More precisely, Fig.\ref{mu2P0B} shows that at $\b=1$
\be
\overline{\mu}(\t_B)^2\overline{P}_0(\t_B)\simeq 4\Delta\lt(1.60854+3.1840\times 10^{-4}\, \overline{\phi}_0(\tau_B)\sqrt{\Delta}\rt),
\ee
$\overline{\phi}_0(\t_B)$ is the value of scalar field at the bounce (of the averaged dynamics). $\overline{\phi}_0(\t_B)\sim O(1)$ (see Fig.\ref{scalarevo}) is determined by the final condition \eqref{finalconditionT} and relates to the subleading correction. It implies
\be
\overline{\rho}_c\simeq\frac{3}{16 \kappa  \Delta \lt(1.60854+3.1840\times 10^{-4}\, \overline{\phi}_0(\tau_B)\sqrt{\Delta}\rt)}.
\ee
for $\b=1$. $\overline{\rho}_c$ is close to be Planckian when $\Delta$ is close to be Planckian. The numerics shows that the Kretschmann scalar at the bounce $\overline{\ck}(\t_B)\sim \Delta^{-2}$ and more precisely as shown in Fig.\ref{krets_bounce},
\be
\overline{\ck}(\t_B)\simeq\frac{0.17397-1.2554 \times 10^{-4} \overline{\phi}_{0}\left(\tau_{B}\right) \sqrt{\Delta}}{\Delta^2}.
\ee
which includes the subleading correction.

On the other hand, as illustrated by Figs.\ref{mu2P01} and \ref{mu2P02}, $\overline{P}_0(\t)/P_0[\mu_{min}](\t)\simeq 1$ after the bounce (approximately $\overline{\mu}(\t)\simeq 2\sqrt{\Delta/\overline{P}_0(\t)}$ ), the average dynamics converges to $\mu_{min}$-scheme dynamics. $b_0(\t)\to 0$ at late time (see Fig.\ref{bounce}(b)), so at late time the quantities inside sine functions in Eqs.\eqref{bp11} - \eqref{bp22} are small enough to validate $\sin(x)\simeq x$, which reduce Eqs.\eqref{bp11} - \eqref{bp22} to the classical cosmology. The classical limit at late time is also illustrated in Fig.\ref{bounce}(a). Particularly, the deviation from classical cosmology is negligible during the inflation. The plot of slow-roll parameter $\eps_H$ has negligible difference from Figure \ref{inflation0}. The plot of $b_0$ during the inflation is given in Figure \ref{b0inflation} (the difference between the average and $\mu_{min}$-scheme dynamics is negligible). $\b\sqrt{\Delta} b_0\ll1$ and $\overline{P}_0(\t)/P_0[\mu_{min}](\t)\simeq 1$ guarantee that the cosmological dynamics is semiclassical during the inflation (for both the average and $\mu_{min}$-scheme dynamics), as promised at the end of Section \ref{Motivation Lattice Refinement}.

We notice that on the other side of the bounce, there exists a short time period having $\overline{\mu}(\t)^2 \overline{P}_0(\t)/\Delta$ slightly smaller than 1. $\overline{\mu}(\t)^2 \overline{P}_0(\t)$ even below the UV cut-off $\Delta$ might seems to be problematic for the effective dynamics on the other side of the bounce. However, this issue happens in the universe on the other side of the bounce, so it does not affect our predictions at and after the bounce, given that we have fixed the final condition at $T$ and evolved back in time. Secondly, $\overline{\mu}(\t)^2 \overline{P}_0(\t)/\Delta$ is only slightly below 1, and we always have the averaged $\overline{\mu} = 2\mu_{min}$ all the time. It may be still acceptable if we view the UV cut-off \eqref{saturate} to be not restrictive but approximate.

As is demonstrated in Fig.\ref{bounce}, $\overline{b}_0$ approaches to constant value on the other side of the bounce as $\t\to-\infty$. Fig.\ref{mu2P01} shows that ${\bar{P}_{0}(\tau)}/{P_{0}\left[\mu_{min }\right](\tau)}\sim 1/4$ as $\t\to-\infty$. Then Eqs.\eqref{bp22} and \eqref{min2} implies that the Hubble parameter ${H}=\dot{{P}}_0/(2{P}_0)$ approaches to a constant in both the average and $\mu_{min}$-scheme effective dynamics. $H$ in both cases can be shown to be negative and coincide by numerics (see Fig.\ref{hubble_compare}). The effective spacetime is asymptotically de Sitter (dS) in the infinitely past to the bounce
\be
\rmd s^2=-\rmd\t^2+e^{2 {H}\t}(\rmd x^2+\rmd y^2+\rmd z^2).
\ee 
Here $H$ is negative and $\t$ is running to the past in the emergent de-sitter spacetime. Figs.\ref{hubble_compare}, \ref{krest_compare}, and \ref{ricci_compare} plot the Hubble parameter $H$, the Kretschmann scalar $\ck$, and the scalar curvature $R$ of the averaged and $\mu_{min}$-scheme dynamics, and compare with the $\bar{\mu}$-schemes in LQC. By Eqs.\eqref{bp22} and \eqref{min2} and the facts that $\overline{b}_0$ approaches to constant and $\overline{P}_0/P_0[\mu_{min}]\sim 1/4$ in the dS phase, the emergent cosmological constant in both the averaged and $\mu_{min}$-scheme dynamics
\be
\L_{eff}= 3 {H}^2 \sim \Delta^{-1}
\ee
is an effect from the UV cut-off $\Delta$. The emergent dS phase and cosmological constant $\L_{eff}$ are consequences from including the Lorentzian term in $\hat{\bf H}$ (see Eq.\eqref{operatorK}), similar to the situation in \cite{Assanioussi:2019iye}.

Table \ref{Comparing effective dynamics} summarizes some key properties of the average and $\mu_{min}$-scheme effective dynamics and compare with two $\bar{\mu}$-scheme effective dynamics in LQC (the effective dynamics in \cite{Yang:2009fp,Assanioussi:2019iye} with unsymmetric bounce, and the traditional LQC effective dynamics in \cite{Ashtekar:2006wn} with symmetric bounce). We observe that the $\mu_{min}$-scheme effective dynamics share the same features as the $\bar{\mu}$-scheme LQC with unsymmetric bounce, although $\L_{eff}$ in their dS phases take different values.

\begin{table}[h]
\caption{Comparing effective dynamics}
\begin{center}
\begin{tabular}{|c|c|c|c|c|}
\hline
  & average & $\mu_{min}$ & unsymmetric $\bar{\mu}$ & symmetric $\bar{\mu}$ \\
 \hline
Asymptotic FRW & & &  & \\
at late time & Yes & Yes & Yes & Yes \\
\hline
Singularity resolution & & &  & \\
and bounce  & Yes & Yes & Yes & Yes \\
\hline
Critical density &$\frac{3}{16 \kappa  \Delta \lt(1.6+3\times 10^{-4}\, \overline{\phi}_0(\tau_B)\sqrt{\Delta}\rt)}$ & &  & \\
at the bounce &  (for $\b=1$) & $\frac{3}{2 \b^2\left(\beta ^2+1\right) \kappa  \Delta}$ & $\frac{3}{2 \b^2\left(\beta ^2+1\right) \kappa  \Delta}$ & $\frac{16}{\beta^{2} \Delta \kappa}$\\
\hline
dS phase in the & & &  & \\
past to the bounce & Yes & Yes & Yes & No\\
\hline
\end{tabular}
\end{center}
\label{Comparing effective dynamics}
\end{table}%

\begin{figure}[h]
\begin{center}
  \includegraphics[width = 1\textwidth]{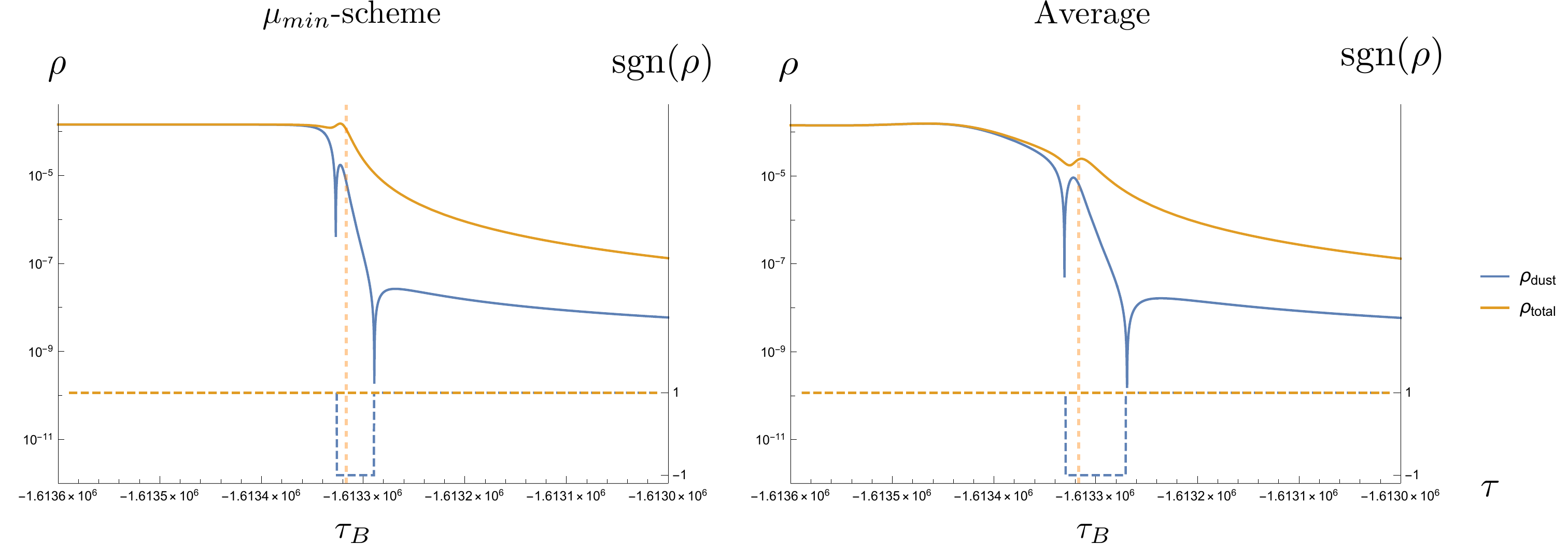}
    \caption{The evolution of $\rho$ and $\rho_{dust}$ in both the $\mu_{min}$-scheme and average effective dynamics. The solid curves plot $\rho$ while the dashed horizontal lines plot $\sgn(\rho)$. The vertical orange dashed lines label the instance of bounce $\t_B$. }
  \label{plotdensity}
  \end{center}
\end{figure}

For both the $\mu_{min}$-scheme and average effective dynamics, $\rho_{dust}$ becomes negative near the bounce, but the total density $\rho$ is positive. Thus the energy density of the scalar field plays the dominant role in the critical density $\rho_c$. In the dS phase, both $\rho$ and $\rho_{dust}$ are positive and approximately coincide, while the energy density of the scalar field becomes negligible. Fig.\ref{plotdensity} plots the evolution of $\rho$ and $\rho_{dust}$ in both the $\mu_{min}$-scheme and average effective dynamics.


\section{Effective Hamiltonian and Poisson Bracket}\label{Effective Hamiltonian and Poisson Bracket}

The effective dynamics of the LQC $\bar{\mu}$-scheme are given by the Hamilton's equations from the LQC Hamiltonian which replacing ${\mu}$ by $\sqrt{\Delta/P_0(\t)}$ at the level of Hamiltonian. The average and $\mu_{min}$-scheme effective dynamics analyzed here are given by imposing certain dynamical $\mu(\t)$ at the level of EOMs \eqref{bp1} - \eqref{bp3}, so they give different dynamics comparing to the LQC $\bar{\mu}$-scheme. However, we can extract the effective Hamiltonian $H_{eff}$ and effective Poisson bracket $\{\cdot,\cdot\}_{eff}$ for the $\mu_{min}$-scheme, such that the $\bar{\mu}$-scheme dynamics is equivalent to the Hamilton's equations from $H_{eff}$ and $\{\cdot,\cdot\}_{eff}$.  

We turn off the scalar fields $\phi_0,\pi_0$ and cosmological constant $\L$ for simplicity. It turns out to be convenient to use $b_0$ and $V_0=P_0^{3/2}$ to express $H_{eff}$. Our aim is to find $\{V_0,\,b_0\}_{eff},\, H_{eff}(b_0, V_0)$ to write Eqs.\eqref{min1} and \eqref{min2} as
\be
\dot{V}_0= \{V_0,\,b_0\}_{eff} \partial_{b_0} H_{eff}, \qquad \dot{b}_0= - \{V_0,\,b_0\}_{eff} \partial_{V_0} H_{eff}
\ee
These equations imply that
\be
\dot{b}_0 \partial_{b_0} H_{eff} +\dot{V}_0 \partial_{V_0} H_{eff}=0
\ee
which is the conservation of $H_{eff}$. Here $\dot{b}_0,\dot{V}_0$ are given by Eqs.\eqref{min1} and \eqref{min2}. The general solution of this equation is
\be
&H_{eff} =\cf\lt( V_0\exp\lt[ \int_1^{b_0} dx\, \xi(\b,\Delta,x)\rt] \rt),\\
&\xi=\frac{6 \sqrt{\Delta} \beta \cos(\beta \sqrt{\Delta} x)(\beta^2 -(1+\beta^2)\cos(2 \beta \sqrt{\Delta} x))}{\beta (\beta^2-1) \sqrt{\Delta} \cos(\beta \sqrt{\Delta} x)x - \beta (1+\beta^2) \sqrt{\Delta} \cos(3 \beta \sqrt{\Delta} x)x -\cos(\beta \sqrt{\Delta} x)^2 \sin(\beta \sqrt{\Delta} x ) +\beta^2 \sin(\beta \sqrt{\Delta} x)^3 }\nonumber
\ee
where $\cf$ is an arbitrary single-variable function. To determine $\cf$, we take the limit $\Delta\to0$
\be
H_{eff} \to\cf(b_0^2 V_0)
\ee
Comparing to the classical Hamiltonian of FRW cosmology we obtain $\cf(b_0^2 V_0)=\frac{6}{\kappa}b_0^2 V_0$ therefore
\be
H_{eff} =\frac{6}{\kappa} V_0\exp\lt[ \int_1^{b_0} dx\, \xi(\b,\Delta,x)\rt] 
\ee
Its expansion in ${\Delta}$ gives
\be
H_{eff} &=& \frac{6}{\kappa}V_0\Big[ b_0^2 -\frac{1}{9} \Delta \beta ^2 b_0^2 \left(b_0^2-1\right) \left(3 \beta ^2+4\right)\\
&&-\frac{1}{405} \Delta^2 \beta ^4 b_0^2 \left(b_0^2-1\right) \left(135 \left(\beta ^2+2\right) \beta ^2+2 \left(45 \beta ^4+75 \beta ^2+32\right) b_0^2+144\right)+O\left({\Delta} ^3\right)\Big], \nonumber
\ee
which can be compared with the $\bar{\mu}$-scheme Hamiltonian in LQC
\be
H_{LQC}&=&\frac{6}{\kappa}{V_0}\frac{  \sin ^2\left(\beta  \sqrt{\Delta } b_0 \right) \left(-\beta ^2+\left(\beta ^2+1\right) \cos \left(2 \beta  \sqrt{\Delta } b_0 \right)+1\right)}{2\beta ^2 \Delta   }\\
&=&\frac{6}{\kappa}{V_0}\lt[b_0^2-\frac{1}{3} \Delta  \beta ^2 \left(3 \beta ^2+4\right) b_0^4+\frac{2}{45} \beta ^4 \left(15 \beta ^2+16\right) b_0^6 \Delta ^2+O\left(\Delta ^{3}\right)\rt].
\ee
$H_{eff}$ and $H_{LQC}$ shares the same classical limit as $\Delta\to0$, while having different $O(\Delta)$ corrections.

The effective Poisson bracket is given by
\be
\{V_0,\,b_0\}_{eff}&=& \frac{\dot{V}_0}{ \partial_{b_0} H_{eff}} = \frac{\dot{V}_0}{  \frac{6}{\kappa}V_0\xi(\b,\Delta,b_0)}\exp\lt[ -\int_1^{b_0} dx\, \xi(\b,\Delta,x)\rt]\\
\frac{\dot{V}_0}{  \frac{6}{\kappa} V_0\xi(\b,\Delta,b_0)}&=&-\frac{\kappa }{96 \beta ^2 \Delta } \Bigg(3 \beta ^2+4 b_0 \sqrt{\Delta } \left[2 \beta ^3 \sin \left(2 \beta  b_0 \sqrt{\Delta }\right)-\left(\beta ^3+\beta \right) \sin \left(4 \beta  b_0 \sqrt{\Delta }\right)\right]\nonumber\\
&&-4 \beta ^2 \cos \left(2 \beta  b_0 \sqrt{\Delta }\right)+\left(\beta ^2+1\right) \cos \left(4 \beta  b_0 \sqrt{\Delta }\right)-1\Bigg).
\ee
We can expand $\{V_0,\,b_0\}_{eff}$ in $\Delta$,
\be
\{V_0,\,b_0\}_{eff}&=&\frac{\kappa }{4}-\frac{\kappa}{36} \Delta  \beta ^2 \left(3 \beta ^2+4\right) \left(4 b_0^2+1\right) \nonumber\\
&&-\frac{\kappa}{810} \Delta ^2 \beta ^4  \left(45 \beta ^4+75 \beta ^2+\left(45 \beta ^4-150 \beta ^2-208\right) b_0^4-10 \left(3 \beta ^2+4\right)^2 b_0^2+32\right)\nonumber\\
&&+O\left(\Delta ^{3}\right), 
\ee
which reduces to the classical limit (equivalent to $\{P_0,K_0\}_{classical}=\frac{\kappa}{6}$) as $\Delta\to0$.

The above $H_{eff}$ is for the $\mu_{min}$-scheme effective dynamics. Unfortunately we are not able to obtain the effective Hamiltonian for the average effective dynamics since $(P_0[\mu_{min}],b_0[\mu_{min}])$ are complicated function of $(\overline{P}_0,\overline{b}_0)$ \footnote{$(P_0[\mu_{min}],b_0[\mu_{min}])$ and $(\overline{P}_0,\overline{b}_0)$ evolve from the same final condition with different EOMs, so $(P_0[\mu_{min}],b_0[\mu_{min}])$ can be seen as functions of $(\overline{P}_0,\overline{b}_0)$}. However Fig.\ref{mu2P01} indicates that the average effective dynamics coincides with the $\mu_{min}$-scheme in both the dS and FRW phases, so the above $H_{eff}$ should approximate the effective Hamiltonian of the averaged effective dynamics in these 2 phases.

\section{Cosmological Perturbations}\label{Perturbations}

\subsection{Linearization of EOMs}\label{Linearization of EOMs}

We insert perturbations Eqs.\eqref{perturb1} and \eqref{perturb2} in the EOMs \eqref{hamilton} and linearize, followed by the Fourier transform \eqref{fourierksig} on the fixed lattice (with fixed $\mu$). We consider both situations of the  average and $\mu_{min}$ scheme cosmological dynamics as the background. These 2 situations correspond to the replacements $\mu\to\overline{\mu}(\t)=2\sqrt{\Delta/P_0[\mu_{min}](\t)}$ and $\mu\to{\mu}_{min}(\t)=\sqrt{\Delta/P_0[\mu_{min}](\t)}$  respectively. Again due to that all intervals $[\t_i,\t_{i+1}]$ are very small so that we approximate $V^\rho(\t,\vec{k})$ as smooth function (allowed by Eq.\eqref{VVidentification1} ) and 
\be
\dot{\tilde V}^\rho(\t_i,\vec{k})\simeq \frac{{\tilde V}^\rho(\t_{i+1},\vec{k})_{\g_i}-{\tilde V}^\rho(\t_i,\vec{k})_{\g_i}}{\t_{i+1}-\t_i},
\ee
similar to the approximation for Eqs.\eqref{mueom1} - \eqref{mueom3}. We obtain the following linearized EOMs for each mode $\vec{k}$:
\be
\dot{{\tilde V}}^\rho(\t,\vec{k})={\bf U}^\rho_{\ \nu}(\Delta,\t,\vec{k})\,{\tilde V}^\nu(\t,\vec{k}).\label{lineareom}
\ee
where ${\bf U}^\rho_{\ \nu}$ depends on $\t$ through the background fields $P_0(\t),K_0(\t),\phi_0(\t),\pi_0(\t)$.
For simplicity we are going to assume that $\vec{k}$ has only one nonzero component $k^x=k$, i.e.
\be
\vec{k}=(k,0,0).
\ee 
The derivation of Eq.\eqref{lineareom} is carried out by expanding $S[Z,u]$ up to quadratic order in perturbations followed by variations. The Mathematica code of the derivation can be downloaded in \cite{github}, where one can find the explicit expression of the $20\times 20$ matrix ${\bf U}^\rho_{\ \nu}(\Delta,\t,\vec{k})$.


The path integral \eqref{cagg} needs to integrate over SU(2) gauge transformation $u^{(i)}$ at every $\t_i$ of changing the lattice. The variation of $u^{(i)}$ gives the closure condition \eqref{closure0} at $\t_i$. When $[\t_i,\t_{i+1}]$ are small, we make the continuous time approximation as the above. Then the closure condition is imposed approximately at all time through out the evolution. Due to the spatial homogeneity, the closure condition is satisfied exactly for both the $\mu_{min}$-scheme and average effective cosmological backgrounds. For cosmological perturbations, the linearized closure condition \eqref{closure0} reads
\be
0&=&P_0\big[({\tilde V}^{15}-{\tilde V}^{18}) \sin (\b^2\mu  K_0)-({\tilde V}^{16}+{\tilde V}^{17}) (\cos (\b^2\mu  K_0)-1)\big]\nonumber\\
&&+\ \beta  K_0 \big[-i {\tilde V}^{1} \sin (k \b\mu )+{\tilde V}^{1} \cos (k \b\mu )-{\tilde V}^{6} \sin (\b^2\mu  K_0)+{\tilde V}^{9} \sin (\b^2\mu  K_0)\nonumber\\
&&+\ {\tilde V}^{7} \cos (\b^2\mu  K_0)+{\tilde V}^{8} \cos (\b^2\mu  K_0)-{\tilde V}^{1}-{\tilde V}^{7}-{\tilde V}^{8},\nonumber\\
0&=&P_0\big[\cos (k \b\mu ) ({\tilde V}^{14} \sin (\b^2\mu  K_0)+{\tilde V}^{13} \cos (\b^2\mu  K_0)-{\tilde V}^{13})-i \sin (k \b\mu ) ({\tilde V}^{14} \sin (\b^2\mu  K_0)\nonumber\\
&&+\ {\tilde V}^{13} \cos (\b^2\mu  K_0)-{\tilde V}^{13})-{\tilde V}^{17} \sin (\b^2\mu  K_0)+{\tilde V}^{18} \cos (\b^2\mu  K_0)-{\tilde V}^{18}\big]\nonumber\\
&&+\ \beta  K_0 \big[i {\tilde V}^{5} \sin (k \b\mu ) \sin (\b^2\mu  K_0)-\cos (k \b\mu ) ({\tilde V}^{5} \sin (\b^2\mu  K_0)+{\tilde V}^{4} \cos (\b^2\mu  K_0))\nonumber\\
&&+\ (-{\tilde V}^{9}+i {\tilde V}^{4} \sin (k \b\mu )) \cos (\b^2\mu  K_0)+{\tilde V}^{8} \sin (\b^2\mu  K_0)+{\tilde V}^{4}+{\tilde V}^{9},\nonumber\\
0&=&P_0\big[-\cos (k \b\mu ) ({\tilde V}^{13} \sin (\b^2\mu  K_0)-{\tilde V}^{14} (\cos (\b^2\mu  K_0)-1))+i \sin (k \b\mu ) ({\tilde V}^{13} \sin (\b^2\mu  K_0)\nonumber\\
&&-\ {\tilde V}^{14} \cos (\b^2\mu  K_0)+{\tilde V}^{14})+{\tilde V}^{16} \sin (\b^2\mu  K_0)+{\tilde V}^{15} \cos (\b^2\mu  K_0)-{\tilde V}^{15}\big]\nonumber\\
&&+\ \beta  K_0 \big[\cos (k \b\mu ) ({\tilde V}^{4} \sin (\b^2\mu  K_0)-{\tilde V}^{5} \cos (\b^2\mu  K_0))-i \sin (k \b\mu ) ({\tilde V}^{4} \sin (\b^2\mu  K_0)\nonumber\\
&&-\ {\tilde V}^{5} \cos (\b^2\mu  K_0))-{\tilde V}^{7} \sin (\b^2\mu  K_0)-{\tilde V}^{6} \cos (\b^2\mu  K_0)+{\tilde V}^{5}+{\tilde V}^{6},\label{linearclosure}
\ee
where ${\tilde V}^\rho={\tilde V}^\rho(\t,k)$. $\mu$ is $\overline{\mu}$ or $\mu_{min}$ for the average or $\mu_{min}$-scheme backgrounds. In both Eqs.\eqref{lineareom} and \eqref{linearclosure}, $\mu$ appears in two types of combinations in sines and cosines 
\be
&{\mu}_{min}  {K}_0=\sqrt{\Delta}\,{b}_0[\mu_{min}](\t),\quad {\mu}_{min}  k = \sqrt{\Delta}\,k\sqrt{\frac{1}{P_0[\mu_{min}](\t)}}\\
&\overline{\mu}  {K}_0=2\sqrt{\Delta}\,\overline{b}_0(\t)\sqrt{\frac{\overline{P}_0(\t)}{P_0[\mu_{min}](\t)}},\quad \overline{\mu}  k = 2\sqrt{\Delta}\,k\sqrt{\frac{1}{P_0[\mu_{min}](\t)}}
\ee
For the average or $\mu_{min}$-scheme backgrounds respectively, Eqs.\eqref{lineareom} and \eqref{linearclosure} are invariant under the rescaling \eqref{rescaleoverline} or \eqref{rescalemin} complemented by 
\be
k\to \a^{1/2} k,\quad \delta \pi \to \a^{3/2} \delta\pi
\ee 
In particular the momentum $k$ is rescaled when the initial/final condition of $P_0$ is rescaled, and can be seen from the expression of the pivot mode $k_{pivot}=\sqrt{P_0(\t_{pivot})}H(\t_{pivot})$.

Eqs.\eqref{lineareom} and \eqref{linearclosure}, derived from the full LQG, govern the dynamics of cosmological perturbations. Given initial conditions of ${\tilde V}^{\rho=1,\cdots, 20}$ satisfying the closure condition \eqref{linearclosure}, the $\t$-evolution of ${\tilde V}^\rho$'s can be computed by numerically solving Eqs.\eqref{lineareom}.

\begin{figure}[h]
\begin{center}
  \includegraphics[width = 0.6\textwidth]{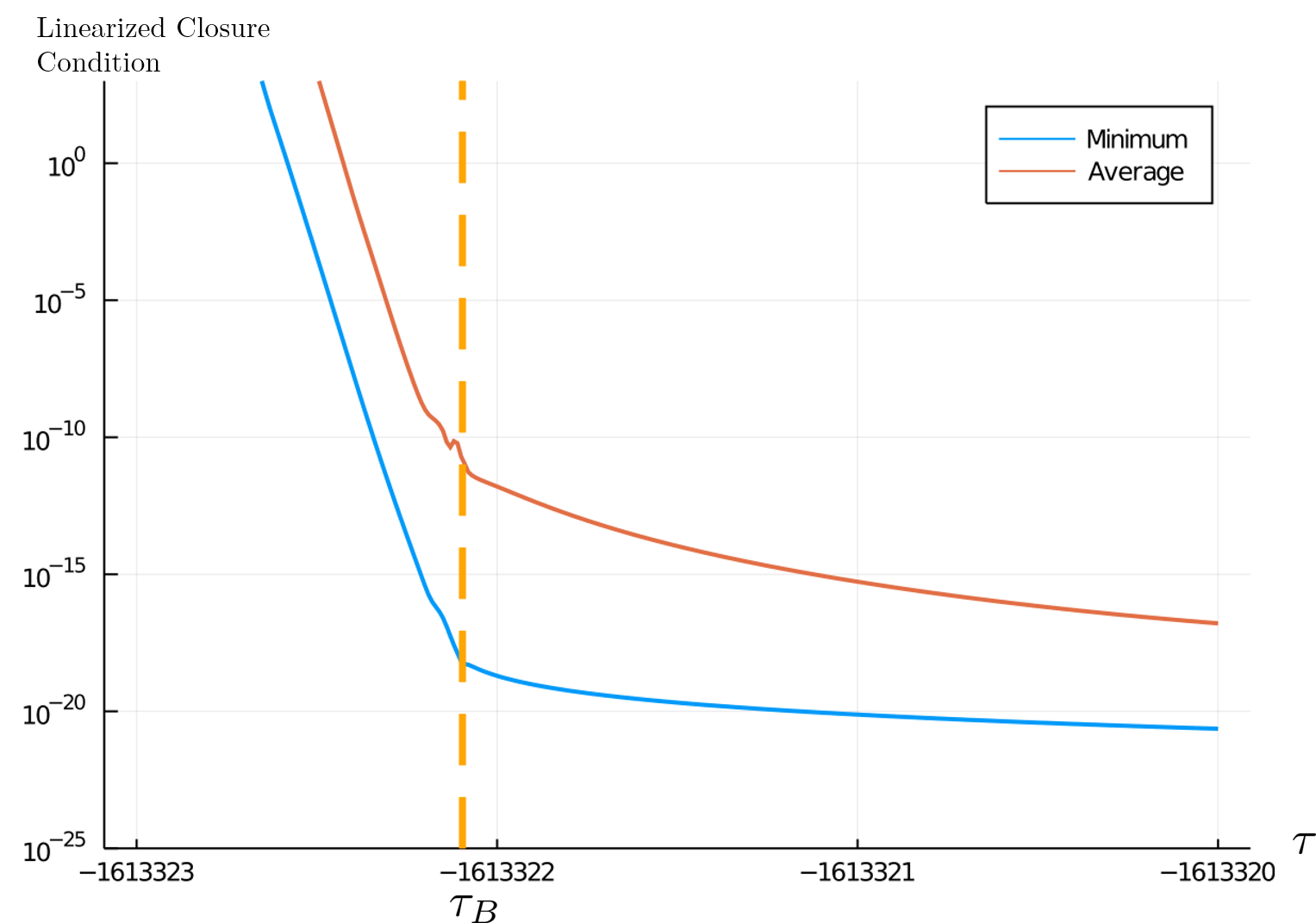}
   \caption{Writing the linearized closure condition Eqs.\eqref{linearclosure} as $0=G_{j=1,2,3}$, this figure plots the average of absolute values, $\frac{1}{3}\sum_{j=1}^3|G_j|$, in the time evolution. The green vertical line is the instance of the bounce $\t_B\simeq-1.6133221\times 10^6$. The initial data determining this solution is set at $\t_0=0$. Nonzero values of perturbations at $\t_0$ are $V^{10}=-6.8565\times 10^{-32} -6.4871\times 10^{-32} i, V^{11}=V^{12}=-8.0409\times 10^{-34}+8.1062\times 10^{-34} i$, $\delta\pi=-1.2980\times 10^{-27}-1.2568\times 10^{-27} i $, $\delta\varphi=-1.2723\times 10^{-23}-1.2826\times 10^{-23} i$. The values of parameters are $\sqrt{\Delta}= \ell_P$, $\b=10^{-3}$, $k=10^{-4}l_P^{-1}$. The average and $\mu_{min}$-scheme cosmological background are the same as in Fig.\ref{bounce}. }
  \label{closureevo}
  \end{center}
\end{figure}

The linearized closure condition is not exactly satisfied due to the dynamical lattice refinement (note that $\{G^a_v,\,{\bf H}\}=0$ is only satisfied on the fixed lattice). But the numerics demonstrates that the linearized closure condition is approximately satisfied with high accuracy near and in the future of the bounce (see Fig.\ref{closureevo} for illustration). On the other side of the bounce the perturbation grows significantly, which cause the linearized closure condition to be violated. For the initial condition used in plotting Fig.\ref{closureevo}, we have to exclude from the path integral the part of the evolution which violates the closure condition, in order that the path integral is not exponentially suppressed.

\subsection{Late-Time Behavior: Classical Limit}

Particularly at late time, the large $P_0[\mu_{min}]$ causes ${\mu}_{min},\overline{\mu}\to0$ so that the continuum limit gives a good approximation to the dynamics on the lattice. We focus on the long wavelength modes with $|k|\leq 10^3 k_{pivot}$ within the observational range. The pivot scale $k_{pivot}=\sqrt{P_0(\t_{pivot})}H(\t_{pivot})$ where $H(\t_{pivot})=1.21\times 10^{-6}l_P^{-1}=9.37\times10^{-6}\ell_P^{-1}$, 
\be
&{\mu}_{min}(\t_{pivot})| k|\leq 10^3k_{pivot}{\mu}_{min}(\t_{pivot})=9.37\times10^{-3}\ell_P^{-1}\sqrt{\Delta},\label{kbound1}\\
&\overline{\mu}(\t_{pivot})| k|\leq 10^3k_{pivot}\overline{\mu}(\t_{pivot})=18.74\times10^{-3}\ell_P^{-1}\sqrt{\Delta},\label{kbound2}
\ee
where we have used $\overline{\mu}=2\mu_{min}$. Recall that $\Delta^2\gg\frac{1}{4}\b^2\ell_P^4/t$, it is possible to have $\Delta\sim \ell_P^2$ when we set $\frac{1}{4}\b^2/ t\ll1$, e.g. we can set $\b=10^{-3}$, $t=10^{-4}$. $\Delta\sim \ell_P^2$ implies that for both the average and $\mu_{min}$-scheme,
\be
{\mu}_{min}(\t_{pivot})| k|\leq 9.37\times 10^{-3}\ll1,\quad \text{and}\quad \overline{\mu}(\t_{pivot}) |k|\leq18.74\times 10^{-3}\ll1.
\ee

The linearized EOMs \eqref{lineareom} and closure condition \eqref{linearclosure} depend on $k$ through $\sin({\mu} k)$ and $\cos({\mu}k)$. Given the above bound of $k$ in the observational range, they can be approximated by the expansion
\be
\sin({\mu} k)\simeq {\mu} k+\frac{1}{6}(\mu k)^3+O(\mu^5),\quad \cos(\mu k)\simeq 1+\frac{1}{2}(\mu k)^2+O(\mu^4),\label{approxcon1}
\ee 
at the pivot time ($\mu=\overline{\mu}$ or $\mu_{min}$). $\mu k\leq \mu(\t_{pivot}) k$ is even smaller after the pivot time. On the other hand, Recall that the background $b_0=K_0/\sqrt{P_0}$ is small at and after the pivot time, so that 
\be
\sin(\b \mu K_0)\simeq \b \mu K_0+\frac{1}{6}(\b \mu K_0)^3+O(\mu^5),\quad \cos(\b \mu K_0)\simeq 1+\frac{1}{2}(\b \mu K_0)^2+O(\mu^4),\label{approxcon2}
\ee 
and the cosmological background is approximately classical. The small $\mu k$ and $\b \mu K_0$ permit us to make power series expansion in $\mu$ of the linearized EOMs and closure condition, whose leading order approximation give the continuum limit of the effective dynamics, at and after the pivot time.

By \eqref{kbound1} and \eqref{kbound2} and $\Delta^2\gg\frac{1}{4}\b^2\ell_P^4/t$, a small $\b$ is needed in order that that the semiclassical approximation is valid for all $k$ within the observational range $|k|\leq 10^3 k_{pivot}$.

The continuum limit allows us to organize perturbations of holonomies and fluxes, $V^{\rho=1,\cdots,18}$, into scalar, tensor, and vector modes \cite{Han:2020iwk}, consistent with the scalar-tensor-vector (STV) decomposition in the standard cosmological perturbation theory. These 3 different modes are decoupled in the linearized EOMs in the limit that \eqref{approxcon1} and \eqref{approxcon2} are valid.

In the following, we only focus on the sector of scalar mode perturbations which contain the scalar field $\delta\varphi,\delta\pi$ different from the discussion in \cite{Han:2020iwk}. The discussion of tensor and vector modes is identical to \cite{Han:2020iwk}.

The scalar mode contains 4 DOFs ${\tilde V}^1,\ {\tilde V}^2={\tilde V}^3, {\tilde V}^6=-{\tilde V}^9,\ \delta \tilde{\varphi}\equiv {\tilde V}^{20}$, while ${\tilde V}^{\rho=4,5,7,8}$ are set to vanish. $\tilde{V}^{\rho=10,\cdots,19}$ can be eliminated by algebraically solving 10 linear EOMs in \eqref{lineareom}. It reduces \eqref{lineareom} to 10 differential equations of second order in $\t$. Following the standard cosmological perturbation theory, we define the following variables
\be
\psi &=&\frac{1}{2 } {\tilde V}^{1} \\
\mathcal{E} &=&-\frac{-2 {\tilde V}^{1}+({\tilde V}^{2}+{\tilde V}^{3})}{2 k^{2}}\\
B&=&-\frac{2 P_{0}^{3 / 2}\lt(\kappa \delta \tilde{\varphi}\dot{\phi_0}+4\dot{\psi}\rt)}{\lt[4\Lambda+\kappa \lt(U(\phi_0)+\dot{\phi}_0^2\rt)\rt] P_{0}^{2}-3 \dot{P}_{0}^{2}}
\ee
and Bardeen potentials
\be
\Phi=-\left(\mathcal{H}\left(B-\mathcal{E}^{\prime}\right)+\left(B-\mathcal{E}^{\prime}\right)^{\prime}\right)=\mathcal{H} \mathcal{E}^{\prime}+\mathcal{E}^{\prime \prime}, \quad \Psi=\psi+\mathcal{H}\left(B-\mathcal{E}^{\prime}\right).
\ee
Note that all above perturbative variables are defined at a given momentum $k$ which is conserved in the evolution. $\ch=P_0'/(2 P_0)$ is the Hubble parameter with respect to the conformal time $\eta$, and $\mathcal{E}^{\prime}$ is the derivative with respect to $\eta$.

By using the background EOMs and the conservation of $\cc_j$ and $h$, It is straight-forward to check that the continuum limit of the linearized EOMs implies the following equations
\be
\Phi-\Psi&=&0,\label{potential1} \\
\left[2 \Psi^{\prime \prime}+2\left(2 \mathcal{H}^{\prime}+\mathcal{H}^{2}\right) \Phi+\mathcal{H}(2 \Psi+\Phi)^{\prime}\right] &=&-\frac{\kappa}{4 \lambda}\left[2 \phi_0^{\prime}\left[\Phi+Z^{\prime}\right]-U^{\prime}(\phi_0) P_0 Z\right], \label{potential2} 
\ee
where
\be
Z=\delta\tilde{\varphi}+ \phi_0^{\prime} \left(B-\ce^{\prime}\right)
\ee
This result coincides with the classical dynamics of the scalar mode perturbations, see e.g. (3.48) and (3.49) in \cite{Giesel:2007wk}. 

{On the other hand, the linearized closure condition gives only one nontrivial equation 
\be
\frac{\mathrm{d}}{\mathrm{d} \tau}\tilde{V}^{9}=0.\label{linearclo}
\ee
}

\begin{figure}[h]
\begin{center}
  \includegraphics[width = 0.8\textwidth]{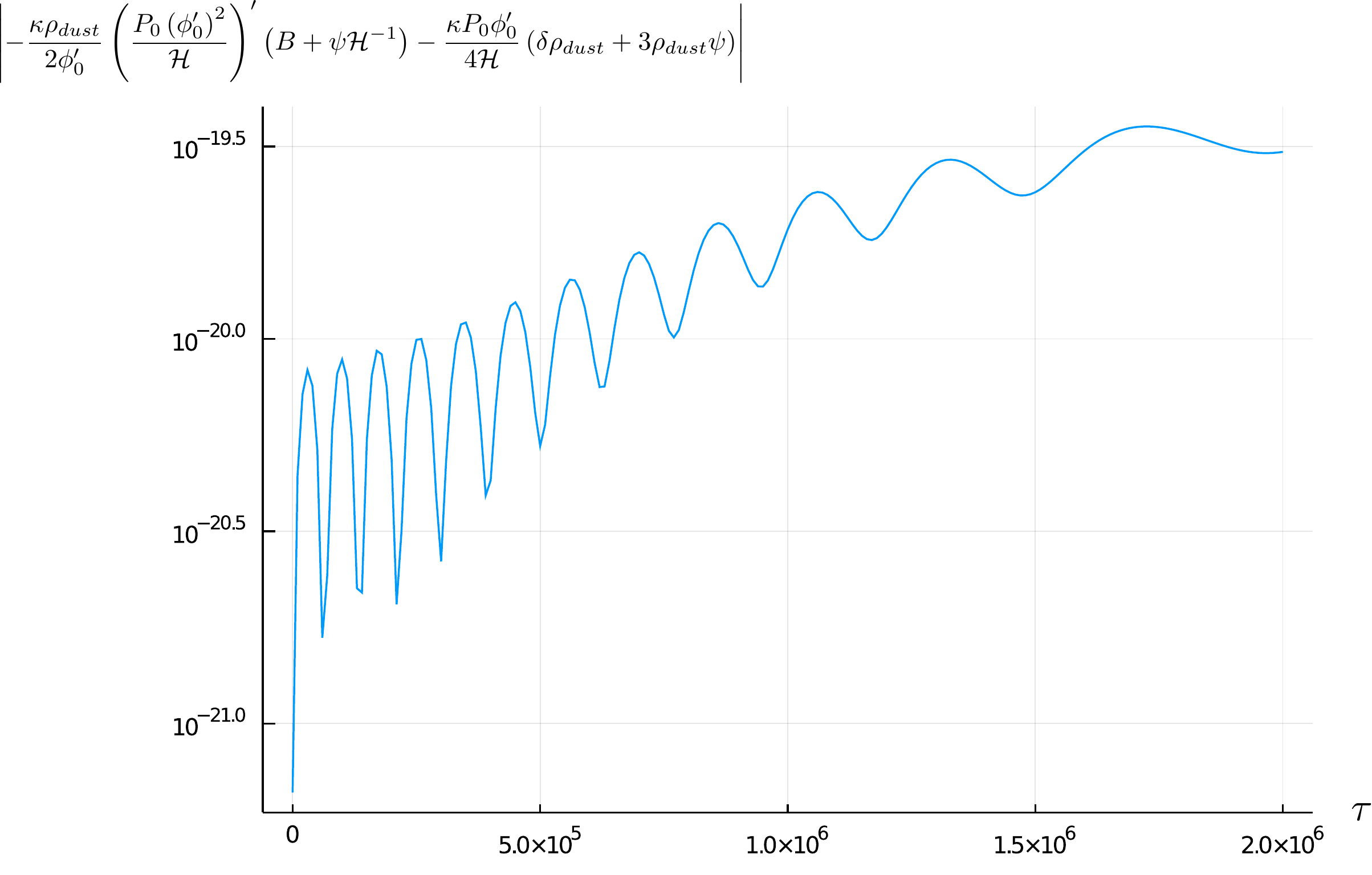}
    \caption{Plot of the right-hand side of Eq.\eqref{MSeqn} in the late-time evolution: $\t=0$ is the pivot time. The nonzero initial perturbations at $\t=0$ are $\tilde{V}^{10}=8.7864\times 10^{-15}+\left(3.7087\times 10^{-13}\right) i,\ \tilde{V}^{11}=\tilde{V}^{12}=4.4689\times 10^{-15},\ \tilde{V}^{19}=8.5206\times 10^{-11}+\left(7.0711\times 10^{-9}\right) i,\ \tilde{V}^{20}=0.000070711 $ (pure scalar-mode perturbation with $\delta\rho_{dust}=B=0$). The values of parameters are $\Delta=\ell_P^2$, $\b=10^{-3}$, $k=10^{-4}l_P^{-1}$.}
  \label{reminderQ}
  \end{center}
\end{figure}

The dynamics of the scalar field $\delta\varphi$ can be conveniently studied with the Mukhanov-Sasaki variable \cite{montani2011primordial}
\be
Q :=  \delta \tilde{\varphi}-\phi_{0}^{\prime} \frac{\psi}{\mathcal{H}}
\ee
A closely related quantity is the perturbation of the scalar curvature:
\be
\delta\mathcal{R}=\Psi-\frac{\mathcal{H}\left(\mathcal{H} \Phi+\Psi^{\prime}\right)}{\mathcal{H}^{\prime}-\mathcal{H}^{2}}
\ee
which relates to $Q$ by 
\be
{(\delta\mathcal{R}-\psi){P_0 \rho_{dust} }} +{\delta\mathcal{R}\phi_0^{\prime}{}^2}  = - \mathcal{H} Q\phi_0^{\prime}
\ee
It recovers the standard relation between $Q$ and $\delta\calr$ when $\rho_{dust}\to0$. The linear EOMs implies the following modified Mukhanov-Sasaki equation:
\be
&&Q''+ 2 \mathcal{H} Q' +k^2 Q + \left[ \frac{1}{2}P_0 U''(\phi_0) - \frac{\kappa}{2 P_0} \left( \frac{P_0 (\phi_0')^2}{\mathcal{H}} \right)' \right] Q \nonumber\\
& =&  - \frac{ \kappa \rho_{dust}}{2 \phi_0'} \left( \frac{P_0 (\phi_0')^2}{\mathcal{H}}\right)' \left( B + \psi \mathcal{H}^{-1} \right) - \frac{\kappa P_0 \phi_0'}{4 \mathcal{H}} \left( \delta \rho_{dust} + 3 \rho_{dust} \psi  \right),\label{MSeqn}
\ee
The modification from the standard Mukhanov-Sasaki equation is the right hand side (see also \cite{Giesel:2020xnb,Giesel:2020bht} for Mukhanov-Sasaki equation in presence of dusts). Eq.\eqref{MSeqn} reduces to the standard Mukhanov-Sasaki equation when the right-hand side of Eq.\eqref{MSeqn} is negligible. Indeed we show that with the suitable initial condition, the evolution by Eq.\eqref{lineareom} gives the negligible right-hand side of Eq.\eqref{MSeqn} after the pivot time: The right-hand side of Eq.\eqref{MSeqn} is $\sim 10^{-18}$ (see Figure.\ref{reminderQ}), while terms in the left-hand side are much larger than $10^{-18}$. Therefore in the late-time evolution after the pivot time Eq.\eqref{lineareom} can be well approximated by the standard Mukhanov-Sasaki equation with vanishing right-hand side in Eq.\eqref{MSeqn}.

\section{Outlook}

Finally we would like to mention a few future perspectives of our program: Firstly, given our proposal of the LQG evolution with dynamical lattice, the map $\ci_{\g_i,\g_{i-1}}$ transforming between different spatial lattices should be better understood in the future, especially in the aspect of the relation with coarse-grain and renormalization. As mentioned in Section \ref{The Linear Map}, the transformation between different lattices might relate to the Wilsonian renormalization procedure of integrating out high frequency modes on the lattice. This procedure should provide the correction to the action in the lattice path integral. We expect that this procedure should closely relate to the Hamiltonian renormalization program in \cite{Lang:2017beo,Thiemann:2020cuq}. The study of $\ci_{\g_i,\g_{i-1}}$ is expected to understand how to restore the unitarity in the the path integral \eqref{cagg}.

Secondly, on the cosmological perturbation theory, some future research should be spent on understanding the initial state of cosmology. One advantage of our result is the dS phase on the other side of the bounce. There is the preferred Bunch-Davies vacuum state from the viewpoint of quantum field theories on curved spacetime. There has been studies of applying the Bunch-Davies vacuum as the initial state of LQC \cite{Agullo:2018wbf}. We have to understand how to translate the Bunch-Davies vacuum state to the framework of the full LQG, in order to apply to the initial state in our program. A related perspective is the $O(\ell_P^2)$ correction to the cosmological perturbation theory. The recent work in \cite{Zhang:2020mld} computes the $O(\ell_P^2)$ correction to the expectation value of the physical Hamiltonian at the coherent state peaked at the homogeneous and isotropic data. The next step is to generalize the $O(\ell_P^2)$ computation to include cosmological perturbations.


\section*{Acknowledgements}

This work receives support from the National Science Foundation through grant PHY-1912278. 
	

\appendix

\section{Discretizing and Quantizating Scalar-Field Contributions of Constraints}\label{Discretizations and Quantizations of Matter Constraints}

The following discretization of cotriad $e^a_i$ is frequently used in this appendix:
\begin{align}
e_{i} & =e_{i}^{a}\frac{\tau^{a}}{2}=\frac{1}{2}\mathrm{sgn}(e)\frac{\epsilon^{abc}\epsilon_{ijk}E_{b}^{j}E_{c}^{k}}{\sqrt{\det(q)}}\frac{\tau^{a}}{2}=\frac{1}{2}\mathrm{sgn}(e)\epsilon_{ijk}\frac{\left[E^{j},E^{k}\right]}{\sqrt{\det(q)}}\nonumber\\
\int_{e_{v;is_i}}dx^{i}e_{i} & \simeq\frac{1}{2}\mathrm{sgn}(e)\frac{\epsilon^{abc}\epsilon_{ijk}\int dx^{i}dx^{k}E_{b}^{j}\int dx^{i}dx^{j}E_{c}^{k}}{\int dx^{i}dx^{j}dx^{k}\sqrt{\det(q)}}\frac{\tau^{a}}{2}\\
&\simeq\mathrm{sgn}(Q)\frac{a^{4}\beta^{2}s}{2}\frac{\epsilon^{abc}\epsilon_{ijk}s_{i}s_{k}\frac{p^{b}\left(e_{v;j+}\right)-p^{b}\left(e_{v;j-}\right)}{4}s_{i}s_{j}\frac{p^{b}\left(e_{v;k+}\right)-p^{b}\left(e_{v;k-}\right)}{4}}{s_{i}s_{j}s_{k}V_{v}}\frac{\tau^{a}}{2}\nonumber\\
 & =\mathrm{sgn}(Q)\frac{a^{4}\beta^{2}s_{i}}{2V_{v}}\epsilon^{abc}\epsilon_{ijk}\frac{\tau^{a}}{2}\frac{p^{b}\left(e_{v;j+}\right)-p^{b}\left(e_{v;j-}\right)}{4}\frac{p^{b}\left(e_{v;k+}\right)-p^{b}\left(e_{v;k-}\right)}{4}\nonumber\\
 & =\frac{8}{\beta\kappa}h(e_{v;is_i})\left\{ h(e_{v;is_i})^{-1},V_{v}\right\}.
\end{align}
In this appendix we denote by $x^i$ the coordinate along $e_{v;i,s_i}$. Moreover we have following relations of $\t^a$:
\be
&&\left[\frac{\tau^{a}}{2},\frac{\tau^{b}}{2}\right]=-\frac{1}{4}\left[\sigma^{a},\sigma^{b}\right]=-i\varepsilon^{abc}\sigma^{c}/2=\varepsilon^{abc}\frac{\tau^{c}}{2},\quad\operatorname{Tr}\left(\frac{\tau^{a}}{2}\frac{\tau^{b}}{2}\right)=-\frac{1}{2}\delta^{ab},\quad\operatorname{Tr}\left(\tau^{a}\frac{\tau^{b}}{2}\right)=-\delta^{ab}\nonumber\\
&&\mathrm{Tr}\left(\frac{\tau^{a}}{2}\frac{\tau^{b}}{2}\frac{\tau^{c}}{2}\right)=\frac{i}{8}\mathrm{Tr}\left(\sigma^{a}\sigma^{b}\sigma^{c}\right)=\frac{i}{8}2i\epsilon_{abc}=-\frac{1}{4}\epsilon_{abc}
\ee

What to be quantized is not $\mathcal{C}^{S}$ but $\sgn(e)\mathcal{C}^{S}$. $\sgn(e)$ can be discretized in the cube $\Box_{s_1s_2s_3}$ bounded by $e_{v;1,s_1}$, $e_{v;2,s_2}$, $e_{v;3,s_3}$: 
\begin{align*}
&\sgn(e)_v  =\frac{\int_{\Box_{s_1s_2s_3}} d^{3}x\det e_{j}^{a}}{\int_{\Box_{s_1s_2s_3}} d^{3}x\sqrt{\det(q)}}=\frac{\int_{\Box_{s_1s_2s_3}} d^{3}x\frac{1}{3!}\epsilon^{ijk}\epsilon_{abc}e_{i}^{a}e_{j}^{b}e_{k}^{c}}{\int_{\Box_{s_1s_2s_3}} d^{3}x\sqrt{\det(q)}}=-\frac{4}{3!}\frac{\int_{\Box_{s_1s_2s_3}} d^{3}x\epsilon^{ijk}\mathrm{Tr}\left(e_{i}e_{j}e_{k}\right)}{\int_{\Box_{s_1s_2s_3}} d^{3}x\sqrt{\det(q)}}\\
 & =-\frac{2}{3}\left(\frac{8}{\kappa\beta}\right)^{3}\frac{\epsilon^{ijk}\mathrm{Tr}\left(h(e_{v;is_i})\left\{ h(e_{v;is_i})^{-1},V_{v}\right\} h(e_{v,js_j})\left\{ h(e_{v;js_j})^{-1},V_{v}\right\} h(e_{v;ks_k})\left\{ h(e_{k})^{-1},V_{v}\right\} \right)}{s_1s_2s_3V_{v}^{1/3}V_{v}^{1/3}V_{v}^{1/3}}\\
 & =-\frac{2}{3}\left(\frac{8}{\kappa\beta}\right)^{3}\left(\frac{3}{2}\right)^{3}s_1s_2s_3\epsilon^{ijk}\\
 &\qquad\mathrm{Tr}\left(h(e_{v;is_i})\left\{ h(e_{v;is_i})^{-1},V_{v}^{2/3}\right\} h(e_{v,js_j})\left\{ h(e_{v,js_j})^{-1},V_{v}^{2/3}\right\} h(e_{v;ks_k})\left\{ h(e_{v;ks_k})^{-1},V_{v}^{2/3}\right\} \right)\\
 & =-8\left(\frac{9\times16}{\kappa^{3}\beta^{3}}\right)s_1s_2s_3\epsilon^{ijk}\\
 &\qquad\mathrm{Tr}\left(h(e_{v;is_i})\left\{ h(e_{v;is_i})^{-1},V_{v}^{2/3}\right\} h(e_{v,js_j})\left\{ h(e_{v,js_j})^{-1},V_{v}^{2/3}\right\} h(e_{v;ks_k})\left\{ h(e_{v;ks_k})^{-1},V_{v}^{2/3}\right\} \right)
\end{align*}
We average the above result over all 8 cubes at $v$ (summing over $s_1,s_2,s_3$ and divide by $8$) and quantize: 
\begin{align*}
\widehat{\sgn(e)}_v & =\left(\frac{9\times16}{i\ell_{P}^{6}\beta^{3}}\right)\sum_{s_{1}s_{2}s_{3}}s_1s_2s_3\epsilon^{ijk}\nonumber\\
&\qquad \mathrm{Tr}\left(\hat{h}(e_{v;is_{1}})\left[\hat{h}(e_{v;is_{1}})^{-1},\hat{V}_{v}^{2/3}\right]\hat{h}(e_{v;js_{2}})\left[\hat{h}(e_{v;js_{2}})^{-1},\hat{V}_{v}^{2/3}\right]\hat{h}(e_{v;ks_{3}})\left[\hat{h}(e_{v;ks_{3}})^{-1},\hat{V}_{v}^{2/3}\right]\right)\\
 & =-\left(\frac{9\times16}{\ell_{P}^{6}\beta^{3}}\right)\sum_{s_{1}s_{2}s_{3}}s_1s_2s_3\epsilon^{ijk}\mathrm{Tr}\left(\hat{\mathcal{Q}}_{2/3}(e_{v;is_{1}})\hat{\mathcal{Q}}_{2/3}(e_{v;js_{2}})\hat{\mathcal{Q}}_{2/3}(e_{v;ks_{3}})\right),
\end{align*}
where
\be
\hat{\cq}_{r}^{a}(e)=i\mathrm{Tr}\left(\tau^{a}\hat{h}(e)\left[\hat{h}(e)^{-1},\hat{V}_{v}^{r}\right]\right),\quad \hat{\mathcal{Q}}_{r}(e)=\hat{\mathcal{Q}}_{r}^{a}(e)\frac{\tau^{a}}{2}=-i\hat{h}(e)\left[\hat{h}(e)^{-1},\hat{V}_{v}^{r}\right].
\ee

The scalar Hamiltonian constraint reads
\[
\mathcal{C}^{S}=\frac{1}{2\sqrt{\det(q)}}\pi\pi^{T}+\frac{1}{2}\sqrt{\det(q)}q^{jk}\left(\mathcal{D}_{j}\phi\right)^{T}\mathcal{D}_{k}\phi+\sqrt{\det(q)}\lt[U_1(\phi)+\sgn(e) U_2(\phi)\rt].
\]
Discretization of $\sgn(e)\mathcal{C}^{S}$ gives
\begin{align*}
&\int_{\Box_{s_1s_2s_3}} d^{3}x\,\mathrm{sgn}(e)\mathcal{C}^{S} \\
& =\int_{\Box_{s_1s_2s_3}} d^{3}x\left[\frac{\mathrm{sgn}(e)}{2\sqrt{\det(q)}}\pi\pi^{T}+\frac{\mathrm{sgn}(e)}{2}\frac{E_{a}^{j}E_{a}^{k}}{\sqrt{\det(q)}}\left(\mathcal{D}_{j}\phi\right)^{T}\mathcal{D}_{k}\phi+\mathrm{sgn}(e)\sqrt{\det(q)}U_{1}(\phi)+\sqrt{\det(q)}U_{2}(\phi)\right]\\
 & =\frac{\mathrm{sgn}(e)}{2\int_{\Box_{s_1s_2s_3}} d^{3}x\sqrt{det(q)}}\int_{\Box_{s_1s_2s_3}} d^{3}x\pi\int_{\Box_{s_1s_2s_3}} d^{3}x\pi^{T}\\
 &\quad +\frac{\mathrm{sgn}(e)}{2\int_{\Box_{s_1s_2s_3}} d^{3}x\sqrt{det(q)}}\int dx^idx^kE_{a}^{j}\int dx^idx^jE_{a}^{k}\left(\int dx^{j}\mathcal{D}_{j}\phi\right)^{T}\int dx^{k}\mathcal{D}_{k}\phi\\
 &\quad +\left[-\frac{4}{3!}\int_{\Box_{s_1s_2s_3}} d^{3}x\epsilon^{ijk}\mathrm{Tr}\left(e_{i}e_{j}e_{k}\right)\right]U_{1}(\phi)+s_1s_2s_3 V_{v}U_{2}(\phi)\\
 & =\frac{\mathrm{sgn}(e)}{2V}s_1s_2s_3\pi(v)\pi(v)^{T}+s_1s_2s_3\frac{\mathrm{sgn}(e)}{2V}\left({a^{2}\beta}\right)^{2}s_is_kX^j_{a}(v)s_is_jX^k_{a}(v)\left(\delta_{j,s_j}^{(R_{s})}\phi(v)\right)^{T}\delta_{k,s_k}^{(R_{s})}\phi(v)\\
 & \quad+\left[-\frac{2}{3}\left(\frac{8}{\kappa\beta}\right)^{3}\epsilon^{ijk}\mathrm{Tr}\left(h(e_{v;is_i})\left\{ h(e_{v;is_i})^{-1},V_{v}\right\} h(e_{v;js_j})\left\{ h(e_{v;js_j})^{-1},V_{v}\right\} h(e_{v;ks_k})\left\{ h(e_{v;ks_k})^{-1},V_{v}\right\} \right)\right]U_{1}(\phi)\nonumber\\
 &\quad +s_1s_2s_3V_{v}U_{2}(\phi)
\end{align*}
where $\int_{\Box_{s_1s_2s_3}} d^{3}x\,\pi=s_1s_2s_3\pi(v)$ and 
\[
\int dx^{j}\mathcal{D}_{j}\phi\simeq R_{s}\left(\underline{h}(e_{v;js_j})\right)\phi\left(t(e_{v;js_j})\right)-\phi\left(v\right)\equiv\delta_{j}^{(R_{s})}\phi(v)
\]
Averaging $\frac{1}{8}\sum_{s_1s_2s_3}s_1s_2s_3\cdots$ followed by quantization, we obtain
\begin{align*}
\hat{C}_{v}^{S} & =\half\widehat{\left(\frac{\mathrm{sgn}(e)}{V}\right)}\pi(v)\pi(v)^{T}+\half\widehat{\left(\frac{\mathrm{sgn}(e)}{V}\right)}\left({a^{2}\beta}\right)^{2}\frac{1}{8}\sum_{s_{1}s_{2}s_{3}}\sum_{j,k}s_jX^j_{a}(v)s_kX^k_{a}(v)\left(\delta_{j,s_{j}}^{(R_{s})}\phi(v)\right)^{T}\delta_{k,s_{k}}^{(R_{s})}\phi(v)\\
 & \quad-\frac{2}{3}\left(\frac{8}{i\hbar\kappa\beta}\right)^{3}\frac{1}{8}\sum_{s_{1}s_{2}s_{3}}s_{1}s_{2}s_{3}\epsilon^{ijk}\nonumber\\
 &\quad \mathrm{Tr}\left(h(e_{v;is_{1}})\left[h(e_{v;is_{1}})^{-1},V_{v}\right]h(e_{v;js_{2}})\left[h(e_{v;js_{2}})^{-1},V_{v}\right]h(e_{v;ks_{3}})\left[h(e_{v;ks_{3}})^{-1},V_{v}\right]\right)U_{1}(\phi)+V_{v}U_{2}(\phi)\\
 & =\half\widehat{\left(\frac{\mathrm{sgn}(e)}{V}\right)}\hat{\pi}(v)\hat{\pi}(v)^{T}+\half\widehat{\left(\frac{\mathrm{sgn}(e)}{V}\right)}\frac{a^{4}\beta^{2}}{8}\sum_{s_{1}s_{2}s_{3}}\sum_{j,k}s_jX^j_{a}(v)s_kX^k_{a}(v)\left(\delta_{j,s_{j}}^{(R_{s})}\hat{\phi}(v)\right)^{T}\delta_{k,s_{k}}^{(R_{s})}\hat{\phi}(v)\\
 & \quad-\frac{2}{3}\frac{8^{2}}{(i\ell_{P}^{2}\beta)^{3}}\sum_{s_{1}s_{2}s_{3}}s_{1}s_{2}s_{3}\epsilon^{ijk}\mathrm{Tr}\left(i\mathcal{Q}_{1}(e_{v;is_{1}})i\mathcal{Q}_{1}(e_{v;js_{2}})i\mathcal{Q}_{1}(e_{v;ks_{3}})\right)U_{1}(\phi)+V_{v}U_{2}(\phi)\\
 & =\half\widehat{\left(\frac{\mathrm{sgn}(e)}{V}\right)}\hat{\pi}(v)\hat{\pi}(v)^{T}+\half\widehat{\left(\frac{\mathrm{sgn}(e)}{V}\right)}\frac{a^{4}\beta^{2}}{8}\sum_{s_{1}s_{2}s_{3}}\sum_{j,k}s_jX^j_{a}(v)s_kX^k_{a}(v)\left(\delta_{j,s_{j}}^{(R_{s})}\hat{\phi}(v)\right)^{T}\delta_{k,s_{k}}^{(R_{s})}\hat{\phi}(v)\\
 & \quad-\frac{2}{3}\frac{8^{2}}{(\ell_{P}^{2}\beta)^{3}}\sum_{s_{1}s_{2}s_{3}}s_{1}s_{2}s_{3}\epsilon^{ijk}\mathrm{Tr}\left[\hat{\mathcal{Q}}_{1}(e_{v;is_{1}})\hat{\mathcal{Q}}_{1}(e_{v;js_{2}})\mathcal{\hat{Q}}_{1}(e_{v;ks_{3}})\right]U_{1}(\hat{\phi})+\hat{V}_{v}U_{2}(\hat{\phi})
\end{align*}

\section{Corrections of $\tilde{\eps}_{i+1,i}\lt(\frac{\delta\t}{\hbar}\rt)$ in EOMs}\label{Corrections of eps in EOMs}

We show below the variation of $\tilde{\varepsilon}_{i+1,i}\left(\frac{\delta\tau}{\hbar}\right)$ vanishes in the time continuous limit $\delta\tau\to0$. We denote by $\delta_{Z_i}$ the holomorphic derivative $\partial_{\eps^a_i(e)}$ or $\partial_{z_i(v)}$ (the anti-holomorphic derivative $\delta_{\bar{Z}_i}=\partial_{\bar{\eps}^a_i(e)}$ or $\partial_{\bar{z}_i(v)}$ can be derived similarly).
\be
&&\delta_{Z_{i}}\tilde{\varepsilon}_{i+1,i}\left(\frac{\delta\tau}{\hbar}\right)\nonumber\\
&	=&\frac{\hbar}{\delta\tau}\delta_{Z_{i}}\ln\left[1-\frac{i\delta\tau}{\hbar}\frac{\langle\psi_{Z_{i+1}}^{\hbar}|\hat{\mathbf{H}}|\psi_{Z_{i}}^{\hbar}\rangle}{\langle\psi_{Z_{i+1}}^{\hbar}\mid\psi_{Z_{i}}^{\hbar}\rangle}+\frac{\delta\tau}{\hbar}\frac{\langle\psi_{Z_{i+1}}^{\hbar}\left|\hat{\varepsilon}\left(\frac{\delta\tau}{\hbar}\right)\right|\psi_{Z_{i}}^{\hbar}\rangle}{\langle\psi_{Z_{i+1}}^{\hbar}\mid\psi_{Z_{i}}^{\hbar}\rangle}\right]+i\delta_{Z_{i}}\frac{\langle\psi_{Z_{i+1}}^{\hbar}|\hat{\mathbf{H}}|\psi_{Z_{i}}^{\hbar}\rangle}{\langle\psi_{Z_{i+1}}^{\hbar}\mid\psi_{Z_{i}}^{\hbar}\rangle},\nonumber\\
	&=&\frac{-i\delta_{Z_{i}}\frac{\langle\psi_{Z_{i+1}}^{\hbar}|\hat{\mathbf{H}}|\psi_{Z_{i}}^{\hbar}\rangle}{\langle\psi_{Z_{i+1}}^{\hbar}\mid\psi_{Z_{i}}^{\hbar}\rangle}+\delta_{Z_{i}}\frac{\langle\psi_{Z_{i+1}}^{\hbar}\left|\hat{\varepsilon}\left(\frac{\delta\tau}{\hbar}\right)\right|\psi_{Z_{i}}^{\hbar}\rangle}{\langle\psi_{Z_{i+1}}^{\hbar}\mid\psi_{Z_{i}}^{\hbar}\rangle}}{1-\frac{i\delta\tau}{\hbar}\frac{\langle\psi_{Z_{i+1}}^{\hbar}|\hat{\mathbf{H}}|\psi_{Z_{i}}^{\hbar}\rangle}{\langle\psi_{Z_{i+1}}^{\hbar}\mid\psi_{Z_{i}}^{\hbar}\rangle}+\frac{\delta\tau}{\hbar}\frac{\langle\psi_{Z_{i+1}}^{\hbar}\left|\hat{\varepsilon}\left(\frac{\delta\tau}{\hbar}\right)\right|\psi_{Z_{i}}^{\hbar}\rangle}{\langle\psi_{Z_{i+1}}^{\hbar}\mid\psi_{Z_{i}}^{\hbar}\rangle}}+i\delta_{Z_{i}}\frac{\langle\psi_{Z_{i+1}}^{\hbar}|\hat{\mathbf{H}}|\psi_{Z_{i}}^{\hbar}\rangle}{\langle\psi_{Z_{i+1}}^{\hbar}\mid\psi_{Z_{i}}^{\hbar}\rangle},\nonumber\\
	&\to&-i\delta_{Z}\frac{\langle\psi_{Z}^{\hbar}|\hat{\mathbf{H}}|\psi_{Z}^{\hbar}\rangle}{\langle\psi_{Z}^{\hbar}\mid\psi_{Z}^{\hbar}\rangle}+\lim_{\delta\tau\to0}\delta_{Z_{i}}\frac{\langle\psi_{Z_{i+1}}^{\hbar}\left|\hat{\varepsilon}\left(\frac{\delta\tau}{\hbar}\right)\right|\psi_{Z_{i}}^{\hbar}\rangle}{\langle\psi_{Z_{i+1}}^{\hbar}\mid\psi_{Z_{i}}^{\hbar}\rangle}+i\delta_{Z}\frac{\langle\psi_{Z}^{\hbar}|\hat{\mathbf{H}}|\psi_{Z}^{\hbar}\rangle}{\langle\psi_{Z}^{\hbar}\mid\psi_{Z}^{\hbar}\rangle},\quad\left(\delta\tau\to0\right)\label{ZtoZlimit}\\
	&=&\lim_{\delta\tau\to0}\delta_{Z_{i}}\frac{\langle\psi_{Z_{i+1}}^{\hbar}\left|\hat{\varepsilon}\left(\frac{\delta\tau}{\hbar}\right)\right|\psi_{Z_{i}}^{\hbar}\rangle}{\langle\psi_{Z_{i+1}}^{\hbar}\mid\psi_{Z_{i}}^{\hbar}\rangle}\nonumber\\
	&=&\lim_{\delta\tau\to0}\frac{\langle\psi_{Z_{i+1}}^{\hbar}\left|\hat{\varepsilon}\left(\frac{\delta\tau}{\hbar}\right)\right|\delta_{Z_{i}}\psi_{Z_{i}}^{\hbar}\rangle\langle\psi_{Z_{i+1}}^{\hbar}\mid\psi_{Z_{i}}^{\hbar}\rangle-\langle\psi_{Z_{i+1}}^{\hbar}\left|\hat{\varepsilon}\left(\frac{\delta\tau}{\hbar}\right)\right|\psi_{Z_{i}}^{\hbar}\rangle\langle\psi_{Z_{i+1}}^{\hbar}\mid\delta_{Z_{i}}\psi_{Z_{i}}^{\hbar}\rangle}{\langle\psi_{Z_{i+1}}^{\hbar}\mid\psi_{Z_{i}}^{\hbar}\rangle^{2}}\nonumber\\
	&=&0,
\ee
where in the step \eqref{ZtoZlimit} we denote $Z_{i+1}\to Z_{i}\equiv Z$ and apply Eqs.\eqref{limitZZZZ1}, \eqref{limitZZZZ2} for $z(v)$ and analogs for $g(e)$ in \cite{Han:2019vpw}. We use the strong limit $\hat{\varepsilon}\left(\frac{\delta \tau}{\hbar}\right)|\psi\rangle \rightarrow 0 \text { as } \delta \tau \rightarrow 0$ (for all $\psi$ in the domain of $\hat{\bf H}$) in the last step.

\section{Properties of $\ci_{\g_i\g_{i-1}}$: I. Dense Domain of Definition}\label{Properties of I Isometry}

Given $f_1,f_2\in \ch_{\g_{i-1}}$, the inner product of their images by $\ci_{\g_i,\g_{i-1}}$ is given by the following integral
\be
&&\langle \ci_{\g_i,\g_{i-1}} f_1| \ci_{\g_i,\g_{i-1}}f_2\rangle_{\ch_{\g_i}}\nonumber\\
&=&\int\rmd Z(\g_{i-1})\rmd Z'(\g_{i-1})\rmd u\rmd u_1\rmd u_2\lag\tilde{\psi}^\hbar_{Z'(\g_i)}\Big|\tilde{\psi}^\hbar_{Z^u(\g_i)}\rag\nonumber\\
&&\frac{\big\langle f_1\big|{\psi}^\hbar_{Z'{}^{u_1}(\g_{i-1})}\big\rangle}{\big\|{\psi}^\hbar_{Z'(\g_{i-1})}\big\|}\frac{\big\langle{\psi}^\hbar_{Z^{u_2}(\g_{i-1})}\big|f_2\big\rangle}{\big\|{\psi}^\hbar_{Z(\g_{i-1})}\big\|}.\label{IfIf}
\ee
where $\rmd u$, $\rmd u_1$, and $\rmd u_2$ are Haar integrals of SU(2) gauge transformations. When $f_1=f_2$ is any coherent state $\tilde{\psi}^\hbar_{Z_0(\g_{i-1})}\in \ch_{\g_{i-1}}$, Eq.\eqref{IfIf} integrates 3 Gaussian-like functions peaked at $Z(\g_{i-1})=Z'(\g_{i-1})=Z_0(\g_{i-1})$ up to gauge transformations, so Eq.\eqref{IfIf} is finite. The coherent states and their finite linear combinations are dense in $\ch_{\g_{i-1}}$, so $\ci_{\g_i,\g_{i-1}}$ is densely defined on $\ch_{\g_{i-1}}$.

\section{Properties of $\ci_{\g_i\g_{i-1}}$: II. Equations of Motion}\label{Properties of I Equations of Motion}

Recall the definition of the linear map $\ci_{\g_i,\g_{i-1}}$: $\ch_{\g_{i-1}}\to\ch_{\g_i}$
\be
\ci_{\g_i,\g_{i-1}}&=&\int\rmd Z_0(\g_{i-1})\frac{\big|{\Psi}^\hbar_{[Z_0(\g_i)]}\big\rangle \big\langle{\Psi}^\hbar_{[Z_0(\g_{i-1})]}\big|}{\big\|{\psi}^\hbar_{Z_0(\g_i)}\big\|\,\big\|{\psi}^\hbar_{Z_0(\g_{i-1})}\big\|},\\
Z_0(\g_i)&=&Z(\mu_i,\,\mathscr{F}_{\g_i}\tilde{\Phi}_{\g_i}),\quad Z_0(\g_{i-1})=Z(\mu_{i-1},\,\mathscr{F}_{\g_{i-1}}\tilde{\Phi}_{\g_{i-1}})
\ee
where $\tilde{\Phi}_{\g_i}$ and $\tilde{\Phi}_{\g_{i-1}}$ are constrained by 
\be
\tilde{\Phi}^{\rho}(\tau_{i}, \vec{m})_{\g_i}&=&\tilde{\Phi}^{\rho}(\tau_{i}, \vec{m})_{\g_{i-1}},\quad \vec{m}\in\Z(N_{i-1})^3,\label{PhiPhiconstraint1}\\
\tilde{\Phi}^{\rho}(\tau_{i}, \vec{m})_{\g_i}&=&0,\qquad \vec{m}\in\Z(N_{i})^3\setminus\Z(N_{i-1})^3. \label{PhiPhiconstraint2}
\ee
When inserting $\ci_{\g_{i},\g_{i-1}}$ in $\ca_{[Z],[Z']}$ and consider the variation with respect to $Z_0(\g_{i})$, the integral of $Z_0(\g_{i})$ involves (recall Eq.\eqref{smallsteps})
\be
&&\int\rmd Z_0(\g_{i-1})\cdots\frac{\big\langle\tilde{\psi}^\hbar_{Z_1(\g_i)}\big|{\Psi}^\hbar_{[Z_0(\g_i)]}\big\rangle \big\langle{\Psi}^\hbar_{[Z_0(\g_{i-1})]}\big|\tilde{\psi}^\hbar_{Z_N(\g_{i-1})}\big\rangle}{\big\|{\psi}^\hbar_{Z_0(\g_i)}\big\|\,\big\|{\psi}^\hbar_{Z_0(\g_{i-1})}\big\|}\cdots\nonumber\\
&=&\int\rmd Z_0(\g_{i-1})\rmd u_1\rmd u_2\,\nu[Z]\cdots e^{\ck(Z_1(\g_{i}),Z_0(\g_i))/t+\ck(Z_0(\g_{i-1}),Z_N(\g_{i-1}))/t}\cdots,\label{ciintinca}
\ee 
where $\ck(Z,Z')$ is expressed in Eq.\eqref{ckZZ}. The exponent contains two $\ck$'s on $\g_i$ and $\g_{i-1}$ respectively.





We focus on cosmological perturbations \eqref{perturb1} and $\eqref{perturb2}$ applied to $Z_1(\g_{i}),Z_0(\g_i), Z_0(\g_{i-1}),Z_N(\g_{i-1})$, and expand $\ck(Z_1(\g_{i}),Z_0(\g_i))+\ck(Z_0(\g_{i-1}),Z_N(\g_{i-1}))$ to quadratic order in $V^\rho$. The expansion to quadratic order is sufficient for studying the linear perturbation theory.
\be
&&\ck(Z_1(\g_{i}),Z_0(\g_i))+\ck(Z_0(\g_{i-1}),Z_N(\g_{i-1}))\nonumber\\
&=&\ck_{0,\g_i}+\ck_{0,\g_{i-1}}+\sum_{v\in V(\g_i)}\ck^\rho_{1}{}_{\g_i} V^\rho(v)_{\g_i}+\sum_{v\in V(\g_{i-1})}\ck^\rho_{1}{}_{\g_{i-1}} V^\rho(v)_{\g_{i-1}}+\cdots\nonumber\\
&&+\sum_{v\in V(\g_i)}\ck^{\rho\sig}_{2}{}_{\g_i} V^\rho(v)_{\g_i}V^\sig(v)_{\g_i}+\sum_{v\in V(\g_{i-1})}\ck^{\rho\sig}_{2}{}_{\g_{i-1}} V^\rho(v)_{\g_{i-1}}V^\sig(v)_{\g_{i-1}}+\cdots\nonumber\\
&=&\ck_{0,\g_i}+\ck_{0,\g_{i-1}}+\lt(\ck^\rho_{1}{}_{\g_i} +\ck^\rho_{1}{}_{\g_{i-1}}\rt)\tilde{V}^\rho(0)+\cdots\nonumber\\
&&+\sum_{\vec{m}\in \Z(N_{i-1})^3}\lt(\ck^{\rho\sig}_{2}{}_{\g_i}+\ck^{\rho\sig}_{2}{}_{\g_{i-1}}\rt) \tilde{V}^\rho(-\vec{m})\tilde{V}^\sig(\vec{m})+\cdots,\label{KKexpand}
\ee
where $\ck_{0,1,2}$ depend on $\mu_i,\mu_{i-1}$ and the homogeneous-isotropic backgrounds in $Z_1(\g_{i}),Z_0(\g_i),Z_0(\g_{i-1}),Z_N(\g_{i-1})$. $ V^\sig(v)_{\g_{i-1}},V^\sig(v)_{\g_{i}}$ are perturbations in $Z_0(\g_i),Z_0(\g_{i-1})$, and $\cdots$ contains linear and quadratic terms involving perturbations in $Z_1(\g_i),Z_N(\g_{i-1})$. $Z_0(\g_i),Z_0(\g_{i-1})$ have the same background $P_0,K_0,\phi_0,\pi_0$ and nonzero Fourier modes $\tilde{V}^\sig(\vec{m})=\tilde{V}^\sig(\vec{m})_{\g_i}=\tilde{V}^\sig(\vec{m})_{\g_{i-1}}$ of perturbations on $\g_i$ and $\g_{i-1}$ by the definition of $\ci_{\g_i,\g_{i-1}}$.

Considering $t\to0$ and the stationary phase approximation of the integral in Eq.\eqref{ciintinca}. The variations with respect to the background $\delta\cb=(\delta P_0,\delta K_0,\delta\phi_0,\delta\pi_0)$ and perturbations $\delta \tilde{V}^\rho(\vec{m})$, $\vec{m}\in \Z(N_{i-1})$ of $Z_0(\g_{i-1})$ give the EOMs
\be
&&\frac{\delta}{\delta\cb}\lt(\ck_{0,\g_i}+\ck_{0,\g_{i-1}}\rt),\label{eomVapp0}\\
&&\lt(\ck^\rho_{1}{}_{\g_i} +\ck^\rho_{1}{}_{\g_{i-1}}\rt)\delta_{\vec{m},0}+2\lt(\ck^{\rho\sig}_{2}{}_{\g_i}+\ck^{\rho\sig}_{2}{}_{\g_{i-1}}\rt) \tilde{V}^\sig(-\vec{m})+\cdots.\label{eomVapp}
\ee
The variational principle for integrals over $Z_1(\g_{i})$ and $Z_N(\g_{i-1})$ in $\ca_{[Z],[Z']}$ gives $Z_1(\g_{i})=Z_0(\g_{i})$ and $Z_N(\g_{i-1})=Z_0(\g_{i-1})$ (as the initial and final conditions for the Hamiltonian evolutions after and before $\t_i$ \cite{Han:2019vpw}). Applying this result to \eqref{eomVapp0} gives
\be
\left(\mu_{i}^3-\mu_{i-1}^3\right)\lt(0,\frac{2 i  {P_0}}{a^2 \beta },-\frac{i \kappa  {\pi_0}}{2 a^2},\frac{i \kappa   {\phi_0}}{2 a^2}\right).\label{backgroundvar}
\ee 
$( P_0, K_0,\phi_0,\pi_0)$ are the background data in $Z_0(\g_{i-1}),Z_0(\g_{i})$ and are the final/initial data of the evolution of the background before/after $\t_i$. Although \eqref{eomVapp0} is not precisely zero due to $\mu_i\neq\mu_{i-1}$, it is arbitrarily small when $N_i,N_{i-1}\gg1$, and $N_{i}-N_{i-1}\sim O(1)$, since $\mu_{i}^3-\mu_{i-1}^3=\mu_i^3(1-N_i^3/N_{i-1}^3)\sim \mu_i^3 O(1/N_i)$. The assumptions $N_i,N_{i-1}\gg1$, and $N_{i}-N_{i-1}\sim O(1)$ also qualifies the continuous approximation in e.g. Eqs.\eqref{mueom1} - \eqref{mueom3}).

In \eqref{eomVapp}, $\lt(\ck^\rho_{1}{}_{\g_i} +\ck^\rho_{1}{}_{\g_{i-1}}\rt)$ is nonvanishing only for 4 $\rho$'s correspond to perturbation components preserving the homogeneity and isotropy. By applying $Z_1(\g_{i})=Z_0(\g_{i})$ and $Z_N(\g_{i-1})=Z_0(\g_{i-1})$, the nonzero components in $\lt(\ck^\rho_{1}{}_{\g_i} +\ck^\rho_{1}{}_{\g_{i-1}}\rt)$ are the same as \eqref{backgroundvar}.

$2\lt(\ck^{\rho\sig}_{2}{}_{\g_i}+\ck^{\rho\sig}_{2}{}_{\g_{i-1}}\rt) \tilde{V}^\sig(-\vec{m})+\cdots$ in \eqref{eomVapp} is reduced to the following by $Z_1(\g_{i})=Z_0(\g_{i})$ and $Z_N(\g_{i-1})=Z_0(\g_{i-1})$:
\be
&(\mu_i^3-\mu_{i-1}^3)\,\tilde{V}(-\vec{m})^{T}\cdot\nonumber\\
&{\tiny\left(
\begin{array}{cccccccc}
 0 & 0 & 0 & \frac{4 i}{a^2 \beta } & 0 & 0 & 0 & 0 \\
 0 & 0 & 0 & 0 & \frac{4 i \left(\mu_{i-1}^2 \sin (\mu_{i-1} \b K_0)-\mu_{i}^2 \sin (\mu_{i} \b K_0)\right)}{a^2 \b^2 K_0 \left(\mu_{i-1}^3-\mu_{i}^3\right)} & \frac{4 i \left(\mu_{i-1}^2 \cos (\mu_{i-1} \b K_0)-\mu_{i}^2 \cos (\mu_{i} \b K_0)-\mu_{i-1}^2+\mu_{i}^2\right)}{a^2 \b^2 K_0 \left(\mu_{i-1}^3-\mu_{i}^3\right)} & 0 & 0 \\
 0 & 0 & 0 & 0 & \frac{4 i \left(\mu_{i-1}^2 \cos (\mu_{i-1} \b K_0)-\mu_{i}^2 \cos (\mu_{i} \b K_0)-\mu_{i-1}^2+\mu_{i}^2\right)}{a^2 \b^2 K_0 \left(\mu_{i}^3-\mu_{i-1}^3\right)} & \frac{4 i \left(\mu_{i-1}^2 \sin (\mu_{i-1} \b K_0)-\mu_{i}^2 \sin (\mu_{i} \b K_0)\right)}{a^2 \b^2 K_0 \left(\mu_{i-1}^3-\mu_{i}^3\right)} & 0 & 0 \\
 0 & 0 & 0 & 0 & 0 & 0 & 0 & 0 \\
 0 & 0 & 0 & 0 & \frac{4 i P_0 \left(\b K_0 \left(\mu_{i-1}^3-\mu_{i}^3\right)+\mu_{i-1}^2 (-\sin (\mu_{i-1} \b K_0))+\mu_{i}^2 \sin (\mu_{i} \b K_0)\right)}{a^2 \b^2 K_0^2 \left(\mu_{i-1}^3-\mu_{i}^3\right)} & \frac{4 i P_0 \left(\mu_{i-1}^2 \cos (\mu_{i-1} \b K_0)-\mu_{i}^2 \cos (\mu_{i} \b K_0)-\mu_{i-1}^2+\mu_{i}^2\right)}{a^2 \b^2 K_0^2 \left(\mu_{i}^3-\mu_{i-1}^3\right)} & 0 & 0 \\
 0 & 0 & 0 & 0 & \frac{4 i P_0 \left(\mu_{i-1}^2 \cos (\mu_{i-1} \b K_0)-\mu_{i}^2 \cos (\mu_{i} \b K_0)-\mu_{i-1}^2+\mu_{i}^2\right)}{a^2 \b^2 K_0^2 \left(\mu_{i-1}^3-\mu_{i}^3\right)} & \frac{4 i P_0 \left(\b K_0 \left(\mu_{i-1}^3-\mu_{i}^3\right)+\mu_{i-1}^2 (-\sin (\mu_{i-1} \b K_0))+\mu_{i}^2 \sin (\mu_{i} \b K_0)\right)}{a^2 \b^2 K_0^2 \left(\mu_{i-1}^3-\mu_{i}^3\right)} & 0 & 0 \\
 0 & 0 & 0 & 0 & 0 & 0 & 0 & -\frac{i \kappa }{a^2} \\
 0 & 0 & 0 & 0 & 0 & 0 & \frac{i \kappa }{a^2} & 0 \\
\end{array}
\right)}\nonumber\\
&\sim \mu_i^3\,O(1/N_i).
\ee
where $\tilde{V}^\rho(\vec{m})$ are the final/inital data of the Hamiltonian evolution before/after $\t_i$.

Given the Hamiltonian evolutions in $[\t_{i-1},\t_i]$ and $[\t_i,\t_{i+1}]$ and their solutions which are connected by identifying the final and initial data at $\t_i$, the variations \eqref{eomVapp0} and \eqref{eomVapp} vanishes approximately up to errors bounded by $\mu_i^3\,O(1/N_i)$. These errors can be arbitrarily small if sizes of lattices are arbitrarily large. Connecting solutions from the Hamiltonian evolutions on different lattices gives the approximate solutions satisfying the variational principle of the path integral $\ca_{[Z],[Z']}(\ck)$, up to $\mu_i^3\,O(1/N_i)$.

\bibliographystyle{jhep}

\bibliography{muxin}

\end{document}